\documentclass[12pt, a4paper]{article}
\usepackage[a4paper, left=3cm,right=2cm]{geometry}
\usepackage{hyperref}
\usepackage{amsfonts} 
\usepackage{amsmath}
\usepackage{amssymb}
\usepackage{setspace}
\usepackage{color}

\usepackage[utf8,applemac]{inputenc}
\usepackage{tensor}
\usepackage{cite}
\usepackage{tikz}
\usepackage{graphicx,subfigure}% Include figure files
\usepackage[font=small, font+=it, format=hang, margin={1cm, 1cm}]{caption}
\graphicspath{{figure/}}
\usepackage{dcolumn}% Align table columns on decimal point
\usepackage{bm}% bold math

\numberwithin{equation}{section}

\newcommand{\bea}{\begin{eqnarray}}
\newcommand{\eea}{\end{eqnarray}}
\newcommand{\be}{\begin{equation}}
\newcommand{\ee}{\end{equation}}
\def\nn{\nonumber}

\def\p{\partial}
\def\eps{\epsilon}

\newcommand{\cR}{\mathcal{R}}
\newcommand{\cS}{\mathcal{S}}

\newcommand{\cM}{\mathcal{M}}
\newcommand{\cW}{\mathcal{W}}
\newcommand{\cX}{\mathcal{X}}
\newcommand{\cK}{\mathcal{K}}
\newcommand{\cG}{\mathcal{G}}
\newcommand{\cZ}{\mathcal{Z}}

\newcommand{\cE}{\mathcal{E}}

\newcommand{\cY}{\mathcal{Y}}

\renewcommand{\d}{\textrm{d}}

\begin{document}

%\maketitle
%\abstract{[...]}

\setcounter{tocdepth}{2}

\begin{titlepage}

\begin{flushright}\vspace{-3cm}
{\small
%{\tt arXiv:yymm.nnnn} \\
\today }\end{flushright}
\vspace{0.5cm}

\begin{center}

{{ \LARGE{\bf{Gravitational waves from plunges\\ }}}}
{{ \LARGE{\bf{into Gargantua\\ }}}} 
\vspace{5mm}

\bigskip

\centerline{\large{\bf{Geoffrey Comp\`{e}re$^\dagger$\footnote{email: gcompere@ulb.ac.be}, Kwinten Fransen$^*$\footnote{email: kwinten.fransen@kuleuven.be}, Thomas Hertog$^*$\footnote{email:
thomas.hertog@kuleuven.be}}}}\vspace{2pt}
\centerline{\large{\bf{
and Jiang Long$^\dagger$\footnote{email: jiang.long@ulb.ac.be}}}}

\vspace{2mm}
\normalsize
\bigskip\medskip
\textit{{}$^\dagger$ Centre for Gravitational Waves, Universit\'{e} Libre de Bruxelles, \\
 International Solvay Institutes, CP 231, B-1050 Brussels, Belgium}\\
\textit{{}$^*$ Centre for Gravitational Waves, Institute for Theoretical Physics, \\
KU Leuven, Celestijnenlaan 200D, 3001 Leuven, Belgium}

\vspace{15mm}

\begin{abstract}
\noindent
{We analytically compute time domain gravitational waveforms produced in the final stages of extreme mass ratio inspirals of non-spinning compact objects into supermassive nearly extremal Kerr black holes. Conformal symmetry relates all corotating equatorial orbits in the geodesic approximation to circular orbits through complex conformal transformations. We use this to obtain the time domain Teukolsky perturbations for generic equatorial corotating plunges in closed form. The resulting gravitational waveforms consist of an intermediate polynomial ringdown phase in which the decay rate depends on the impact parameters, followed by an exponential quasi-normal mode decay. The waveform amplitude exhibits critical behavior when the orbital angular momentum tends to a minimal value determined by the innermost stable circular orbit. We show that either near-critical or large angular momentum leads to a significant extension of the LISA observable volume of gravitational wave sources of this kind. 
}

\end{abstract}

%\pacs{04.65.+e,04.70.-s,11.30.-j,12.10.-g}
%PACS: 04.20.-q, 04.20.Ha, 11.25.Tq, 11.30.Cp

\end{center}
%%%%%%%%%%%%%%%%%%%%%%%%%%%%%%%%%%%%%%%%%%%%%%%%%%%%%%%%%%%%%%%%%%%%%%%%%%%%%%%%%%%%%%%%

\end{titlepage}

\tableofcontents

\section{Introduction}

Extreme Mass Ratio Inspirals (EMRIs) describe the long-lasting inspiral -- from months to a few years -- and final plunge of stellar mass black holes into supermassive black holes in the centre of galaxies. The gravitational wave (GW) pattern emitted by EMRIs encodes detailed and rich information about the strong gravity region in the neighbourhood of the central massive object. The observation of EMRIs is therefore one of the central goals of LISA, the planned space-based gravitational wave observatory \cite{Audley:2017drz,AmaroSeoane:2007aw,Babak:2017tow}. 

To fully exploit the scientific potential of EMRI observations requires an extremely accurate modelling of the relativistic dynamics over a large number of cycles. However the construction of precise gravitational waveform templates needed to interpret LISA observations of EMRIs is a challenge at several levels. First, the orbits are highly relativistic which makes the post-Newtonian methods commonly used in the LIGO/VIRGO data analysis inadequate. Second, the large number of observable cycles and the separation of scales pose severe difficulties for numerical simulations. Finally, self-force effects which are relevant corrections in perturbation methods still require further development \cite{Detweiler:2005kq,Barack:2009ux,Blanchet:2011zz,Poisson:2011nh,Amaro-Seoane:2014ela,Wardell:2015kea,Pound:2015tma}.

In this paper we identify new corners in the parameter space of EMRIs where semi-analytical waveforms can be computed from first principles. Specifically we consider EMRIs involving a nearly maximally spinning supermassive black hole, which we refer to as Gargantua, following the terminology of \cite{Gralla:2016qfw}\footnote{Gargantua refers to the black hole featured in Nolan's science-fiction movie \emph{Interstellar}, which according to Thorne \cite{Thorne:2014} must be very rapidly spinning, with $1-J/M^2 < 10^{-14}$ to allow key pieces of the narrative.} The comparison between the analytic leading order near-horizon results and numerical results for gravitational emission on the ISCO orbit allows to test the precision of the leading order near-horizon methods. For $1-J/M^2 \sim 10^{-4}$ one obtains $\sim 10\%$ precision, and $1-J/M^2 \sim 10^{-6}$ gives $\sim 1\%$ precision\footnote{See Table I of \cite{Gralla:2015rpa}. For the precision of near-horizon methods in the context of accretion disks, see \cite{Compere:2017zkn}.}. According to recent X-ray observing campaigns, 7 out of the 22 Active Galactic Nuclei (AGNs) analyzed are candidates for being high spin supermassive black holes\footnote{See Table 1 of \cite{Brenneman:2013oba}. These are MCG-6-30-15 \cite{Brenneman:2006hw}, 1H 0707-495 \cite{2009Natur.459..540F}, NGC 3783, Mrk 110, RBS 1124, IRAS 13224-3809 and NGC 4051.}. Accretion is the main mechanism to spin up black holes. For geometrically thin disk models this leads to the Thorne bound $J/M^2 <99.8\%$ \cite{1974ApJ...191..507T}. Spins near the upper bound are reached according to the black hole spin evolution model of \cite{Berti:2008af} for supermassive black holes of $10^9$ solar masses or higher, if the accretion gas is anisotropic. The Thorne bound can however be exceeded in the presence of magnetic fields, as shown e.g. in the magneto-hydrodynamic simulations of \cite{2011A&A...532A..41S}. The number of EMRI events expected to be detected by LISA ranges from a few to a few thousands per year due to astrophysical uncertainties \cite{Babak:2017tow}. The currently preferred population model for LISA uses a Gaussian distribution of spins centered at $98\%$ and capped off at the Thorne bound \cite{Babak:2017tow}. However in light of \cite{2011A&A...532A..41S} EMRIs involving a high spin supermassive black holes with $J/M^2 =0.9999$ are candidate events for the LISA observatory.

The physics of such black holes in the near-horizon region is controlled by $SL(2,\mathbb R)$ conformal symmetry \cite{Bardeen:1999px,Amsel:2009ev,Dias:2009ex,Bredberg:2009pv} and, at the quantum level, by infinite-dimensional conformal symmetries \cite{Guica:2008mu,Bredberg:2011hp,Compere:2012jk}. The combination of a small near-extremality parameter and near-horizon $SL(2,\mathbb R)$ conformal symmetry allows one to use unique analytical methods, including matched asymptotic expansions and conformal representation theory methods, which in turn enables one to obtain gravitational waveforms in closed form. This program was initiated in \cite{Porfyriadis:2014fja,Hadar:2014dpa,Hadar:2015xpa,Gralla:2015rpa,Hadar:2016vmk}. Here we extend this to generic corotating equatorial orbits into supermassive, nearly extremal Kerr black holes. All known previous results for gravitational wave emission \cite{Porfyriadis:2014fja,Hadar:2014dpa,Gralla:2015rpa} are particular instances of our general analysis, which also generalizes to spin 2 the known analogous spin 0 results \cite{Hadar:2015xpa,Hadar:2016vmk}. Moreover, the semi-analytic waveforms we obtain contain novel remarkable ``smoking gun'' signatures associated with near extremality. They should also be useful to calibrate numerical simulations and effective models, and serve as a basis for further analytical endeavor.

The opening up of an extended near-horizon region at high redshift around rapidly spinning black holes gives rise to very specific strong gravity physics. For instance the Aretakis instability occurs in the extremal limit \cite{Aretakis:2011gz} and leads to the amplification of signals emerging from the near-horizon due to high-energy interaction with near-horizon modes \cite{1975ApJ...196L.107P,Banados:2009pr,Gralla:2016sxp}. In \cite{Gralla:2016qfw} it was found that the EMRI GW signal obtained by adiabatically evolving the ISCO into Gargantua develops a characteristic tail associated with GW emission during the late, near-horizon phase of the inspiral. Here we obtain new gravitational wave signatures specific to plunges of non-spinning compact objects along general corotating geodesic orbits into Gargantua. Importantly, the spectrum of quasi-normal modes splits in the near-extremal limit into damped modes in the asymptotically flat region, and zero-damped modes in the near-horizon region \cite{Yang:2012pj}. This leads to polynomial quasi-normal mode ringing due to harmonic stacking of overtones \cite{Yang:2013uba}. Polynomial ringing was described so far without reference to the source of gravitational waves. Here we show that this effect is a manifest feature or a ``smoking gun'' signature of the gravitational wave emission from a plunging source into Gargantua. We also discuss the implications and the observability of the resulting signal for LISA. 

As mentioned already, a distinctive feature of EMRIs into Gargantua is the presence of spontaneously broken conformal symmetry. The existence of the asymptotic extremal system with exact conformal symmetry leads to critical behavior, as already demonstrated for various physical processes probing nearly maximally spinning black holes \cite{Gralla:2016jfc,Compere:2017zkn,Gralla:2017lto}. Here we find a new kind of critical behavior, directly associated with the GW emission from EMRIs, which occurs when the orbital angular momentum of the plunging body approaches a minimal angular momentum bound set by the innermost stable circular orbit. We find that in this critical limit, the amplitude of the GW signal in the geodesic approximation diverges. Hence one expects that non-linear backreaction becomes important. On a more technical point we note that the existence of conformal symmetry is the key feature that allows us to explicitly evaluate the Fourier integrals and to resum the zero-damped quasi-normal modes which ultimately leads to the closed form, analytical, late time domain waveform for all equatorial plunging orbits. 

The paper is organized as follows. In Section \ref{app:NHEK}, we recall the three distinct spacetime regions that arise from the Kerr metric in the near-extremal limit. In Section \ref{taxo} we perform a systematic classification of conjugacy classes of equatorial timelike geodesics under \emph{complexified} conformal symmetry, which consists of $SL(2,\mathbb C) \times U(1)$ transformations combined with PT symmetry (flip of time and axial angle). We will derive that there are only two conjugacy classes whose representatives can be chosen to be the circular orbit in either NHEK or near-NHEK. All other trajectories can be obtained from conformal symmetry by employing complex transformations. In Section \ref{sec:Perturbation} we obtain the general solution (up to an important coefficient that depends on the source) of the Teukolsky master equation governing gravitational perturbations of nearly-extremal Kerr, using the method of matched asymptotic expansions and with appropriate boundary conditions. Selecting the zero-damped QNM contribution we also convert the frequency-based solution for the perturbation into a waveform in the time domain at late times. In Section \ref{Emission} we obtain the spectrum of GW emission of a body moving on a circular geodesic in (near-)NHEK at first order in the asymptotically matched expansion. We use this in Section \ref{Generic} to derive the gravitational waveforms for all other equatorial (corotating) orbits in (near-)NHEK by applying the conformal transformations of Section \ref{taxo} which relate the waveforms associated with generic equatorial plunges in (near-)NHEK to one of the two sets of circular ``seed orbits''. We obtain the analytical critical behavior of all resulting waveforms in Section \ref{Seccrit} in the limit of minimal angular momentum compatible with the existence of (near-)NHEK orbits. We also obtain the analytical large orbital angular momentum behavior. We discuss the observability of such signals by LISA in Section \ref{sec:LISA}. We conclude in Section \ref{sec:ccl}.

\section{The three extremal limits of Kerr}
\label{app:NHEK}

Given a metric depending on a parameter $\lambda$, the limiting spacetime when $\lambda \rightarrow 0$ might depend on the coordinates in which the limit was taken \cite{Geroch:1969ca}.
For a near-extremal Kerr black hole with
\bea
\lambda \equiv \sqrt{1-\frac{J^2}{M^4}} \ll 1, \label{lambda}
\eea
it turns out that there exist 3 distinct coordinate systems that admit a distinct spacetime limit when $\lambda \rightarrow 0$, as we will now review. In turn, this implies that a near-extremal Kerr black hole is composed of three patches glued together that each resolve the physics in a distinct scaling limit:  the extremal Kerr region, the near-horizon extremal Kerr (NHEK) region and the very near-horizon region (near-NHEK). 
%When approaching extremality, the NHEK region first opens up while the near-NHEK region becomes relevant further closer to extremality. 

The extremal Kerr metric is obtained by taking the limit $\lambda \rightarrow 0$ in Boyer-Linquist coordinates $(\hat{t},\hat{r},\theta,\hat{\phi})$. We recall that the Kerr metric in Boyer-Linquist coordinates is given by 
\bea
ds^2&=&-(1-\frac{2M \hat r}{\Sigma})d\hat t^2+\frac{\Sigma}{\Delta}d\hat r^2+\Sigma d\theta^2 + (\hat r^2+a^2 + \frac{2M a^2 \hat r \sin^2\theta}{\Sigma})\sin^2\theta d\hat \phi^2\nn\\
&&
-\frac{4M a \hat r \sin^2 \theta}{\Sigma} d\hat t d\hat \phi \label{eqn:kerrmetric}
\eea
where
\bea
\Delta \equiv \hat r^2-2M \hat r+a^2,\qquad \Sigma \equiv \hat r^2+a^2 \cos^2\theta. \label{defD}
\eea
The inner and outer horizons are denoted as $\hat r_\pm = M \pm \sqrt{M^2-a^2}$ or $\hat r_\pm = M(1 \pm \lambda)$. The angular velocity and Hawking temperature are denoted as $\Omega_H=a/(2M r_+)$ and $T_H=(r_+-M)/(4\pi M r_+)$. The radial tortoise coordinate $\hat r^*$ of the Kerr metric is defined as 
\be
\hat r^{*} = \hat{r} +  \frac{2M \hat r_{+}}{\hat r_{+}-\hat r_{-}}\ln{\frac{\hat{r}-\hat r_{+}}{2M}} -  \frac{2M \hat r_{-}}{\hat r_{+}-\hat r_{-}}\ln{\frac{\hat{r}-\hat r_{-}}{2M}}.\label{rstar}
\ee
The limit $\lambda \rightarrow 0$ in these coordinates gives the extremal Kerr metric. 

It was first observed by Bardeen, Press and Teukolsky \cite{Bardeen:1972fi} that in the extremal limit the innermost stable circular orbit (ISCO) and the innermost bound circular orbit (IBCO) (sometimes called the marginally bound circular orbit) admit distinct scaling  limits: $\hat r_{ISCO}=M +2^{1/3}\lambda^{2/3}M+O(\lambda)$ while $\hat r_{IBCO} = M + 2^{1/2} \lambda M + o(\lambda)$. These timelike orbits degenerate with the null event horizon in Boyer-Linquist coordinates. Therefore,  two additional separate coordinate systems are required to resolve the physics around these orbits, which turn out to lead to distinct limiting spacetimes.  

The near-horizon extremal Kerr geometry (NHEK) \cite{Bardeen:1999px}
\bea
ds^2= 2M^2 \Gamma(\theta) \left( -R^2 dT^2 + \frac{dR^2}{R^2}+d\theta^2 + \Lambda^2(\theta) (d\Phi + R dT)^2 \right)\label{NHEKGEO}
\eea
where 
\bea
\Gamma(\theta)=\frac{1+\cos^2\theta}{2},\qquad \Lambda(\theta) = \frac{2\sin\theta}{1+\cos^2\theta}
\eea
is obtained from the Kerr geometry in the limit $\lambda \rightarrow 0$ after a zoom close to the horizon in corotating frame, 
\bea
T&=& \frac{\hat t}{2M}\lambda^{2/3}, \nn\\
R &=&\frac{\hat r-\hat r_+}{M}\lambda^{-2/3}, \label{NHEKlimit}\\
\Phi &=& \hat \phi - \Omega_{ext} \hat t, \qquad \Omega_{ext} \equiv \frac{1}{2M}.\nn
\eea
We call the coordinates $(T,R,\theta,\Phi)$ the Poincar\'e coordinates of NHEK.  The event horizon is mapped in the limit to the NHEK Poincar\'e horizon $R=0$. This spacetime admits an enhanced $SL(2,\mathbb R) \times U(1)$ Killing symmetry. By construction, the ISCO is mapped into the NHEK orbit $R=2^{1/3}$ in the near-extremal limit but because of the emerging scale invariance, the orbit $R=2^{1/3}$ takes no specific role within the NHEK spacetime alone. The location $R=2^{1/3}$ takes the meaning to be the ISCO only after specifying how the asymptotically flat region is glued to the NHEK region through the specific coordinate change \eqref{NHEKlimit}. This spacetime is relevant for Boyer-Linquist coordinates in the range $\hat r \sim \hat r_+ + M \times O(\lambda^{2/3})$\footnote{In fact $\hat r \sim \hat r_+ + M \times O(\lambda^{p})$ for $ 0 < p< 1$ but the ISCO limit is defined only for $p=\frac{2}{3}$.\label{footn}}. 

The very near-horizon extremal Kerr geometry (near-NHEK) \cite{Amsel:2009ev,Bredberg:2009pv}
\bea
ds^2= 2M^2 \Gamma(\theta) \left( -r(r+2\kappa) dt^2 + \frac{dr^2}{r(r+2\kappa)}+d\theta^2 + \Lambda^2(\theta) (d\phi + (r+\kappa) dt)^2 \right)\label{nearNHEKGEO}
\eea
is obtained from the Kerr geometry in the limit $\lambda \rightarrow 0$ via the change of coordinates
\bea
t&=& \frac{\hat t}{2M\kappa} \lambda , \nn\\
r &=&\kappa \frac{\hat r-\hat r_+}{M\lambda} , \label{chgtnear}\\
\phi &=& \hat \phi - \frac{\hat t}{2M},\nn
\eea
where $\kappa > 0$ is arbitrary as a consequence of emerging scale invariance. The event horizon is mapped in the limit to the NHEK black hole horizon $r=0$. By definition, this spacetime is relevant for Boyer-Linquist coordinates in the range $\hat r \sim \hat r_+ + M \times O(\lambda)$ which is closer to the horizon than the NHEK patch. 

The NHEK and near-NHEK spacetimes are diffeomorphic to each other. The near-NHEK coordinate patch $(t,r,\phi )$ associated with the parameter $\kappa$ 
that lies at the lowest corner of the NHEK Poincar\'e patch $(T,R,\Phi)$, is given by  
\bea
R &=& \frac{1}{\kappa} e^{\kappa  t} \sqrt{ r ( r + 2\kappa)},\nn \\
T &=& -e^{-\kappa  t} \frac{ r + \kappa}{\sqrt{ r ( r+2\kappa)}},\label{NHEKtonearNHEK} \\
\Phi &=&  \phi - \frac{1}{2} \log \frac{ r }{ r +2 \kappa}. \nn
\eea
While the global patch of NHEK does not appear as a limit of the Kerr metric, it will turn out to be useful in order to map the NHEK geodesic orbits and draw their Penrose diagrams. We use the standard embedding of the Poincar\'e coordinate patch $(T,R,\Phi)$ in the global coordinate patch $(\tau,y,\varphi)$ of the NHEK spacetime given by 
\bea
R &=& \sqrt{1+y^2}\cos\tau + y,\nn\\
T &=& \sqrt{1+y^2} \sin\tau \frac{1}{R},\label{glob}\\
\Phi &=& \varphi +  \log \frac{\cos\tau + y \sin \tau}{1+\sqrt{1+y^2}\sin\tau} , \nn
\eea
where the global metric ($-\infty < \tau <\infty$, $-\infty < y<\infty$) is
\bea
ds^2= 2M^2 \Gamma(\theta) \left( -(1+y^2) d\tau^2 + \frac{dy^2}{1+y^2}+d\theta^2 + \Lambda^2(\theta) (d\phi + yd\tau)^2 \right).\label{globalNHEKGEO}
\eea
Given the $2\pi$-periodicity of $\tau$, the conformal Penrose diagram can be drawn as a rectangle $-\pi < \tau < \pi$, $-\frac{\pi}{2} <\text{arctan}(y) < \frac{\pi}{2}$ where each point represents a two-sphere. 
For the convenience of the reader, we summarized our notation for the different coordinate systems and their relationship in Table 1.%\ref{table:conventions}. 
%We also include the barred notation that is used to distinguish coordinates of the target spaces of the conformal transformations in Section \ref{Generic}.

		\begin{table}[bht]\label{table:conventions}
	\centering
	\renewcommand{\arraystretch}{1.5}
	\begin{tabular}{l l  c  l l}
		\hline
		Coordinates & & Metric & Coordinate changes &  \\
		\hline
		Boyer-Linquist & $(\hat{t},\hat{r},\theta,\hat{\phi})$ & \eqref{eqn:kerrmetric} & $(\hat{t},\hat{r},\theta,\hat{\phi}) \to (T,R,\theta,\Phi)$ & \eqref{NHEKlimit} \\
		& & &  $(\hat{t},\hat{r},\theta,\hat{\phi}) \to (t,r,\theta,\phi)$ & \eqref{chgtnear} \\ \hline
	%	& & & & \\
		Poincar\'e NHEK & $(T,R,\theta,\Phi)$ & \eqref{NHEKGEO}& $(T,R,\theta,\Phi) \to (t,r,\theta,\phi)$ & \eqref{NHEKtonearNHEK} \\
		 &% $(\bar T,\bar R, \theta,\bar \Phi)$ 
		 & & $(T,R,\theta,\Phi) \to (\tau,y,\theta,\phi)$ & \eqref{glob}  \\ \hline
		% & & & & \\
		near-NHEK & $(t,r,\theta,\phi)$ & \eqref{nearNHEKGEO} & $(t,r,\theta,\phi) \to (T,R,\theta,\Phi)$ & \eqref{nearNHEKtoNHEK}  \\ \hline
	%	& %$(\bar t,\bar r,\theta,\bar \phi)$ 
	%	& & & \\
		global NHEK & $(\tau,y,\theta,\phi)$ & \eqref{globalNHEKGEO} & $(\tau,y,\theta,\phi) \to (T,R,\theta,\Phi)$ & \eqref{globinverse} \\
		\hline	
	\end{tabular}
	\caption{Summary of the different coordinates and their relations.}
\end{table}

\section{Taxonomy of timelike equatorial NHEK orbits}
\label{taxo}

The generic trajectory of a massive probe such as a star or a stellar black hole falling into a nearly extreme, spinning, supermassive black hole can be approximated in its vicinity by a null geodesic in one of the near-horizon geometries. This is due to the infinite redshift in the extremal limit which requires a rescaling of proper time. In the special case where the probe is nearly co-rotating with the central black hole, the trajectory can instead be approximated by a timelike geodesic in one of the near-horizon geometries. 

All previous studies of gravitational wave emission in the near-horizon Kerr geometry have concentrated on specific equatorial timelike geodesics and specific $SL(2,\mathbb R)$ transformations relating them \cite{Porfyriadis:2014fja,Hadar:2014dpa,Hadar:2015xpa,Gralla:2015rpa,Hadar:2016vmk,Gralla:2016qfw}. Here, we go further and perform a systematic classification of conjugacy classes of equatorial timelike geodesics under conformal symmetry, which consists of complexified $SL(2,\mathbb C) \times U(1)$ transformations combined with PT symmetry (flip of time and axial angle). We will see there are only two conjugacy classes whose representatives can be chosen to be the circular orbit in NHEK and the circular orbit in near-NHEK. All other (real) trajectories can be obtained from conformal symmetry by employing complex transformations. It therefore suffices to understand the physics of circular orbits in (near-)NHEK to understand all equatorial timelike orbits.

The energy per unit probe mass $\hat E$ associated with the asymptotically flat Killing vector $-\p_{\hat t}$ is conserved and positive. By contrast there is no global timelike Killing vector associated with the (near-)NHEK regions. In particular $\p_T$, $\p_t$ are spacelike, so the NHEK energy $E$ associated with $-\p_T$ and the near-NHEK energy $e$ associated with $-\p_t$ can be negative. We denote by $ \ell$ the angular momentum per unit probe mass associated with $\p_{\hat \phi}$, $\p_\Phi$, and $\p_\phi$. We will often call the quantity $\ell$ angular momentum in what follows by a slight abuse of language. These quantities are related as 
\bea
\hat E = \Omega_{Ext} \ell +\frac{\lambda^{2/3}}{2M} E = \Omega_{Ext} \ell+\frac{\lambda}{2} \frac{e}{M\kappa}\label{defE}
\eea
where $\Omega_{Ext}=1/(2M)$ and $\lambda$ is the near-extremal redshift factor \eqref{lambda}. 

\subsection{Timelike equatorial NHEK orbits}

We are interested in future-directed timelike orbits entering the NHEK geometry (which is equivalent to near-NHEK with $\kappa = 0$) 
from the asymptotically flat region. Accordingly we consider $dT/dR <0$ at $R \rightarrow \infty$.  We only look at orbits corotating with the black hole ($\ell\geq 0$). It turns out there are no orbits for which the conserved angular momentum per unit probe mass $\ell$ is lower than the critical angular momentum $\ell_*$, with
\bea
\ell_*=\frac{2}{\sqrt{3}}M.
\eea 
We distinguish critical $\ell = \ell_*$ from supercritical $\ell > \ell_*$ orbits. Plunges are defined as orbits that cross the horizon in finite affine time. Osculating orbits are defined as orbits that leave the NHEK geometry in finite affine time. In the marginal case, the orbit leaves the NHEK geometry at infinite time $T$. We find the following distinct orbits (where critical orbits are indicated with a $*$ subscript):

\bea
\begin{array}{llll}
\kappa = 0 & \ell=\ell_* & E < 0 & \text{Does not exist} \\
\kappa = 0 &\ell=\ell_* & E = 0 & \text{Circular$_*$ (ISCO)} \\
\kappa = 0 &\ell=\ell_* & E > 0 & \text{Plunging$_*$}(E) \\
\kappa = 0 & \ell>\ell_* & E < 0 &\text{Osculating}(E,\ell) \\
\kappa = 0 &\ell>\ell_* & E = 0 & \text{Marginal}(\ell) \\
\kappa = 0 &\ell>\ell_* & E > 0 & \text{Plunging}(E,\ell)  
\end{array}
\eea
All such orbits are detailed in Appendix \ref{formulae_NHEK} and illustrated in Figure \ref{fig:taxonomyNHEK}. The $\text{Plunging$_*$}(E)$ orbit is also known as the fast NHEK plunge \cite{Hadar:2015xpa}. 

\begin{figure}\vspace{-2cm}
\centering
\subfigure[\ Circular$_*$ (ISCO)]{
\includegraphics[width=.28\textwidth]{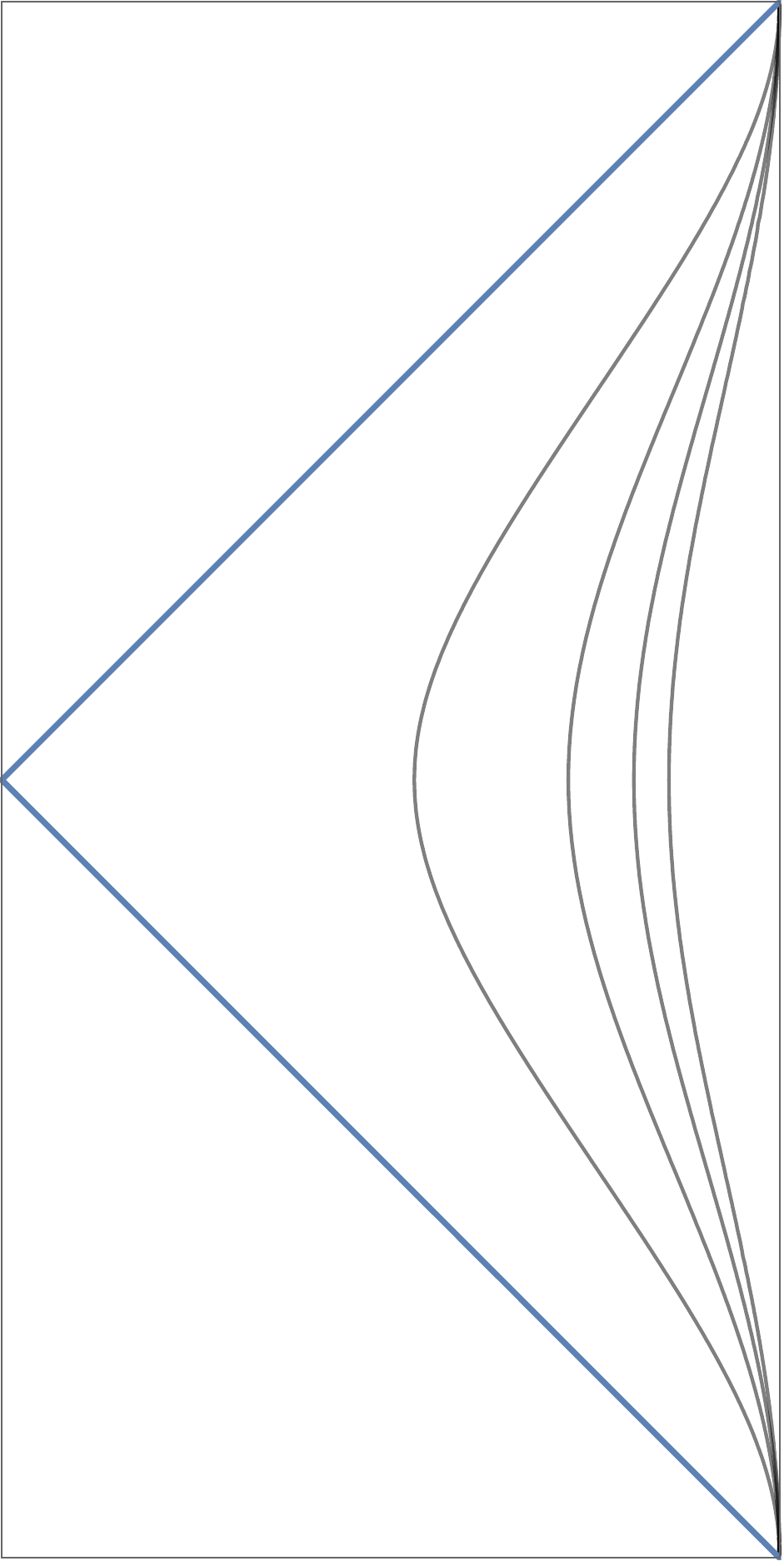}}
\qquad
\subfigure[\ Plunging$_*$$(E)$]{
\includegraphics[width=.28\textwidth]{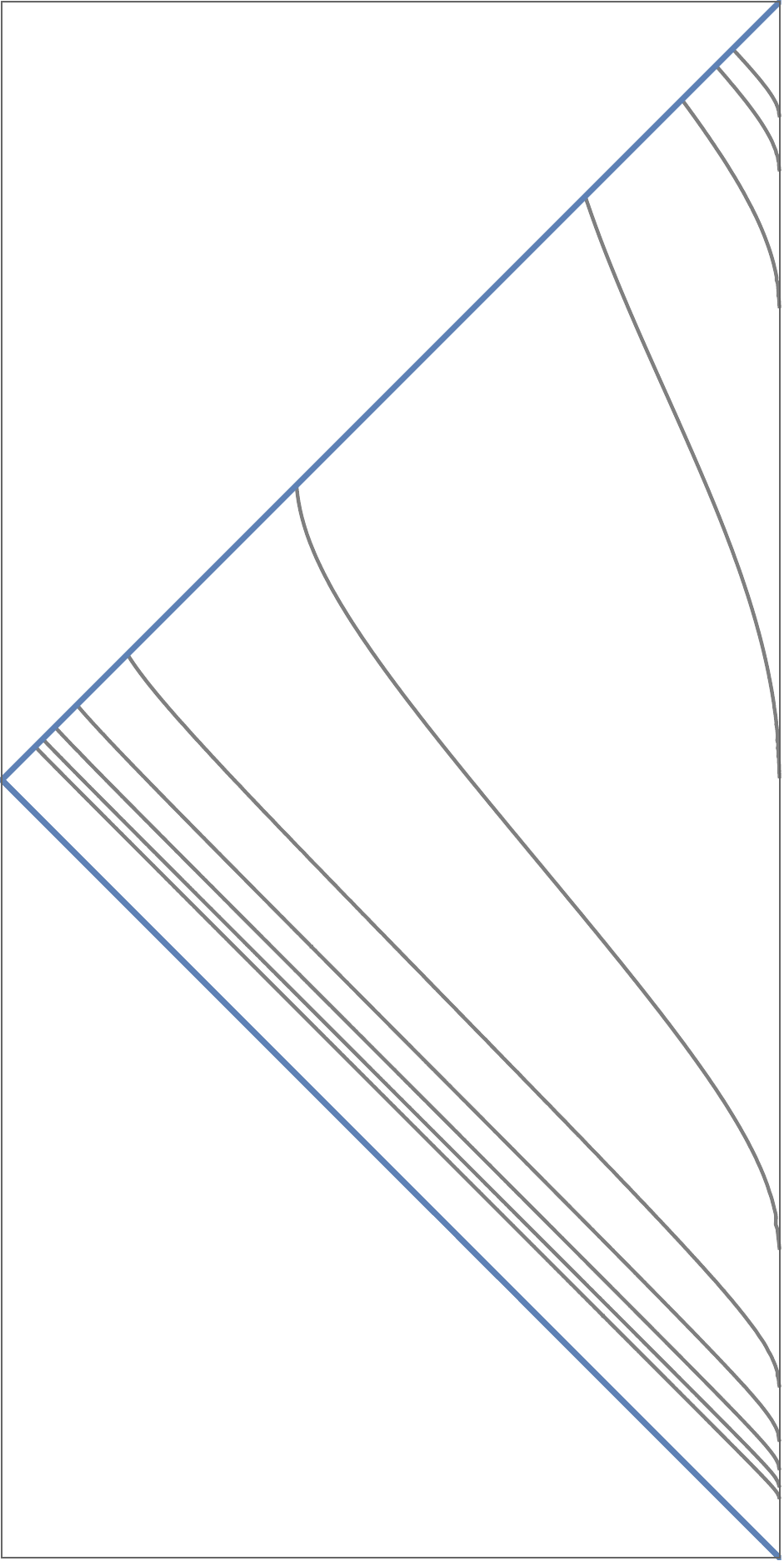}}
\qquad
\subfigure[\ Osculating$(E,\ell)$]{
\includegraphics[width=.28\textwidth]{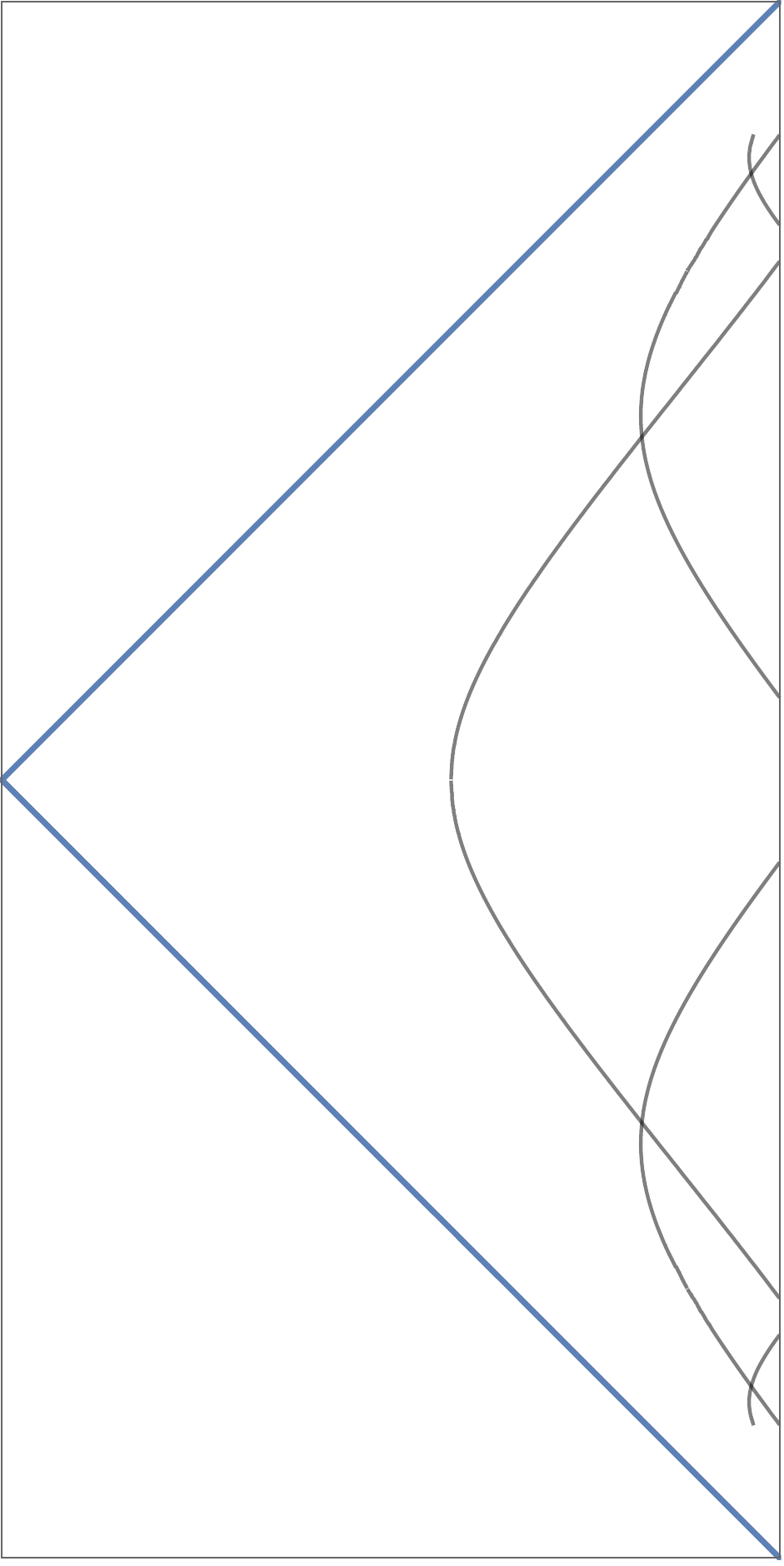}}\\
\subfigure[\ Marginal$(\ell)$]{
\includegraphics[width=.28\textwidth]{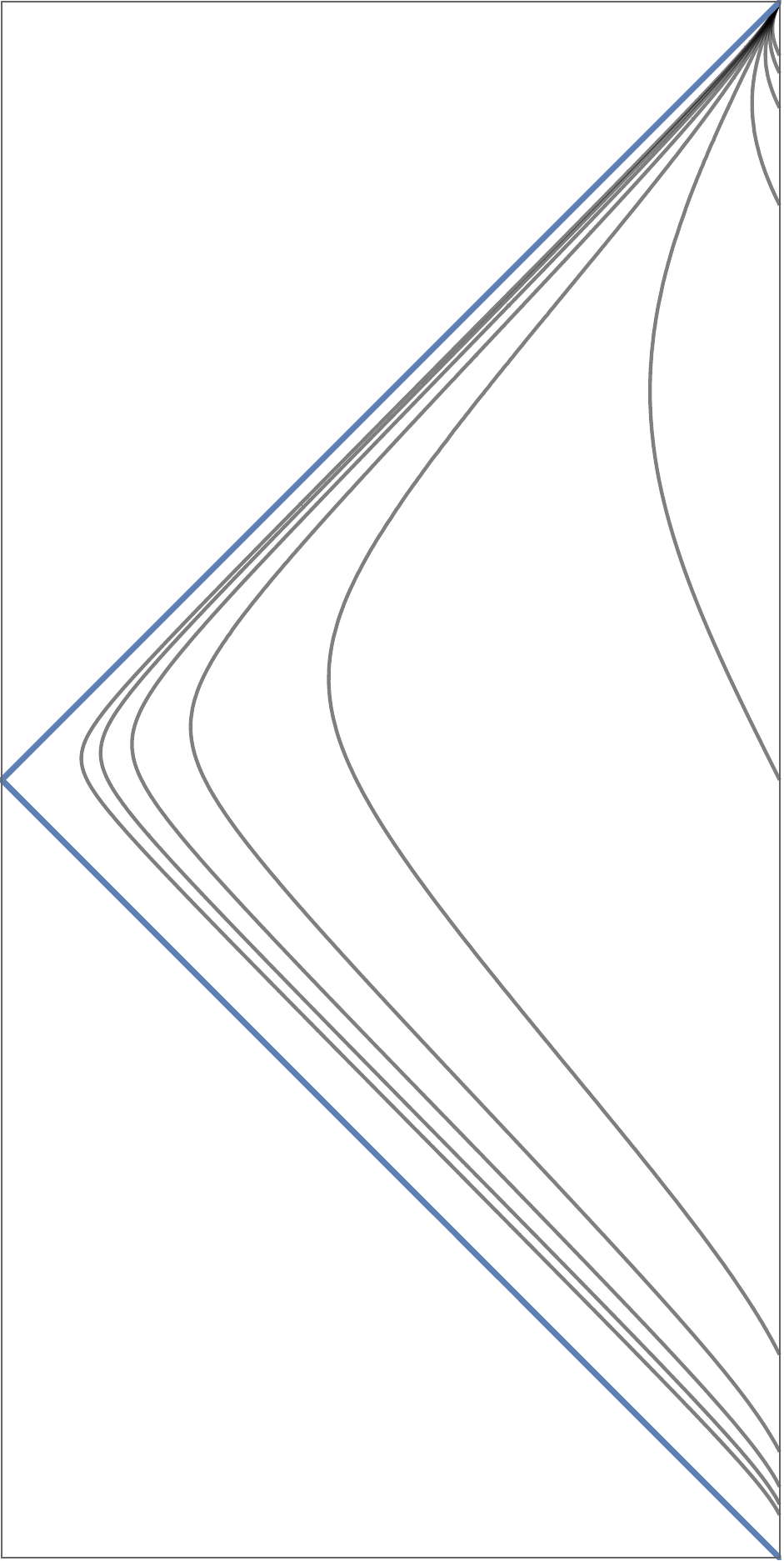}}
\qquad
\subfigure[\ Plunging$(E,\ell)$]{
\includegraphics[width=.28\textwidth]{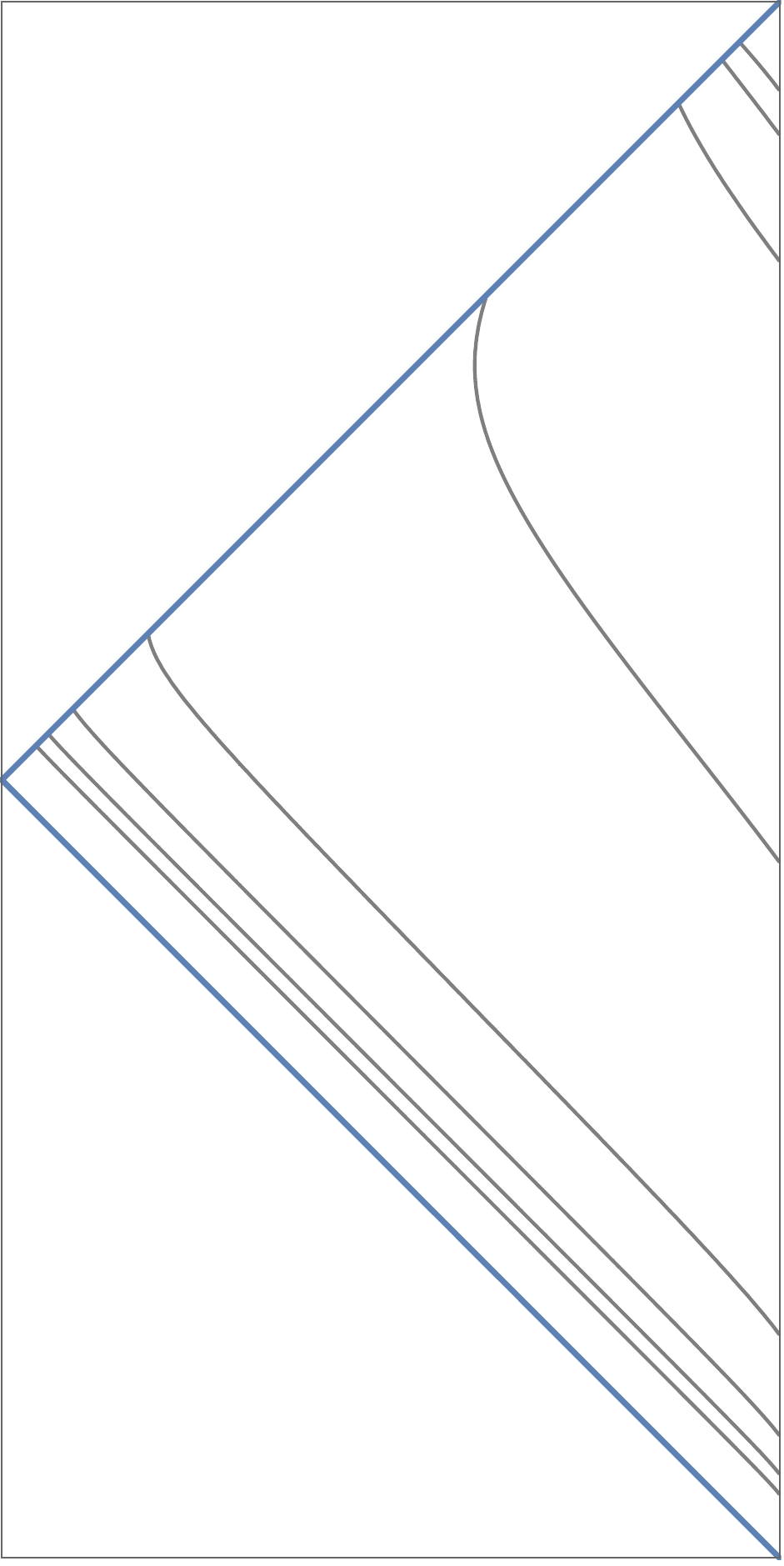}}
\caption{Taxonomy of incoming, timelike NHEK orbits shown here in the conformal diagram of global NHEK. The vertical axis ranges from $\tau =-\pi$ to $\tau = +\pi$ and the horizontal axis from $\text{arctan}(y) =-\frac{\pi}{2}$ to $+\frac{\pi}{2}$, see Section \ref{app:NHEK} for definitions. The blue lines at $45^\circ$ and $135^\circ$ indicate the Poincar\'e horizon. In particular, the blue line at $45^\circ$ is the event horizon. We have used $M=1$ and, for $\ell > \ell_*$ orbits, we have taken $\ell=2\ell_*$.}
\label{fig:taxonomyNHEK}
\end{figure}

\subsection{Timelike equatorial near-NHEK orbits}

After analysis, we obtain the following orbits: 
\bea
\begin{array}{llll}
\kappa > 0 &\ell=\ell_* & e < 0 & \text{Does not exist} \\
\kappa > 0 &\ell=\ell_* & e = 0 & \text{Plunging$_*$}(e=0) \\
\kappa > 0 & \ell=\ell_* &  0 < e < \kappa \ell_* & \text{Plunging$_*$}(e) \\
\kappa > 0 & \ell=\ell_* & e = \kappa \ell_* & \text{Plunging$_*$}(e=\kappa \ell_*) \\
\kappa > 0 &\ell=\ell_* & e > \kappa \ell_* & \text{Plunging$_*$}(e) \\
\kappa > 0 & \ell>\ell_* & e < -\frac{\kappa}{2} \sqrt{3(\ell^2-\ell_*^2)} &\text{Osculating}(e,\ell) \\
\kappa > 0 &\ell > \ell_* & e =  -\frac{\kappa}{2} \sqrt{3(\ell^2-\ell_*^2)} & \text{Circular}(\ell) \\
\kappa > 0 &\ell > \ell_* & e> -\frac{\kappa}{2} \sqrt{3(\ell^2-\ell_*^2)}   & \text{Plunging}(e,\ell) \\
\end{array}
\eea
In addition for the specific initial condition $t_0 \rightarrow -\infty$ there is another family of circular orbits specified by
\bea\begin{array}{llll}
\kappa > 0 & \ell>\ell_* & e = -\kappa \ell &\text{Special Circular}(\ell). \\
\end{array}
\eea
All orbits are detailed in Appendix \ref{formulae_nearNHEK}.   The $\text{Plunging$_*$}(e=0)$ orbit was studied in \cite{Hadar:2014dpa}. The $\text{Plunging$_*$}(e)$ orbits are also known as the near-NHEK fast plunges \cite{Hadar:2015xpa}.  The $\text{Osculating}(e,\ell)$ and $\text{Plunging}(e,\ell)$ orbits were described in \cite{Hadar:2016vmk}. 

\begin{figure}\vspace{-2.6cm}
\centering
\subfigure[\ Plunging$_*$$(e=0)$]{
\includegraphics[width=.28\textwidth]{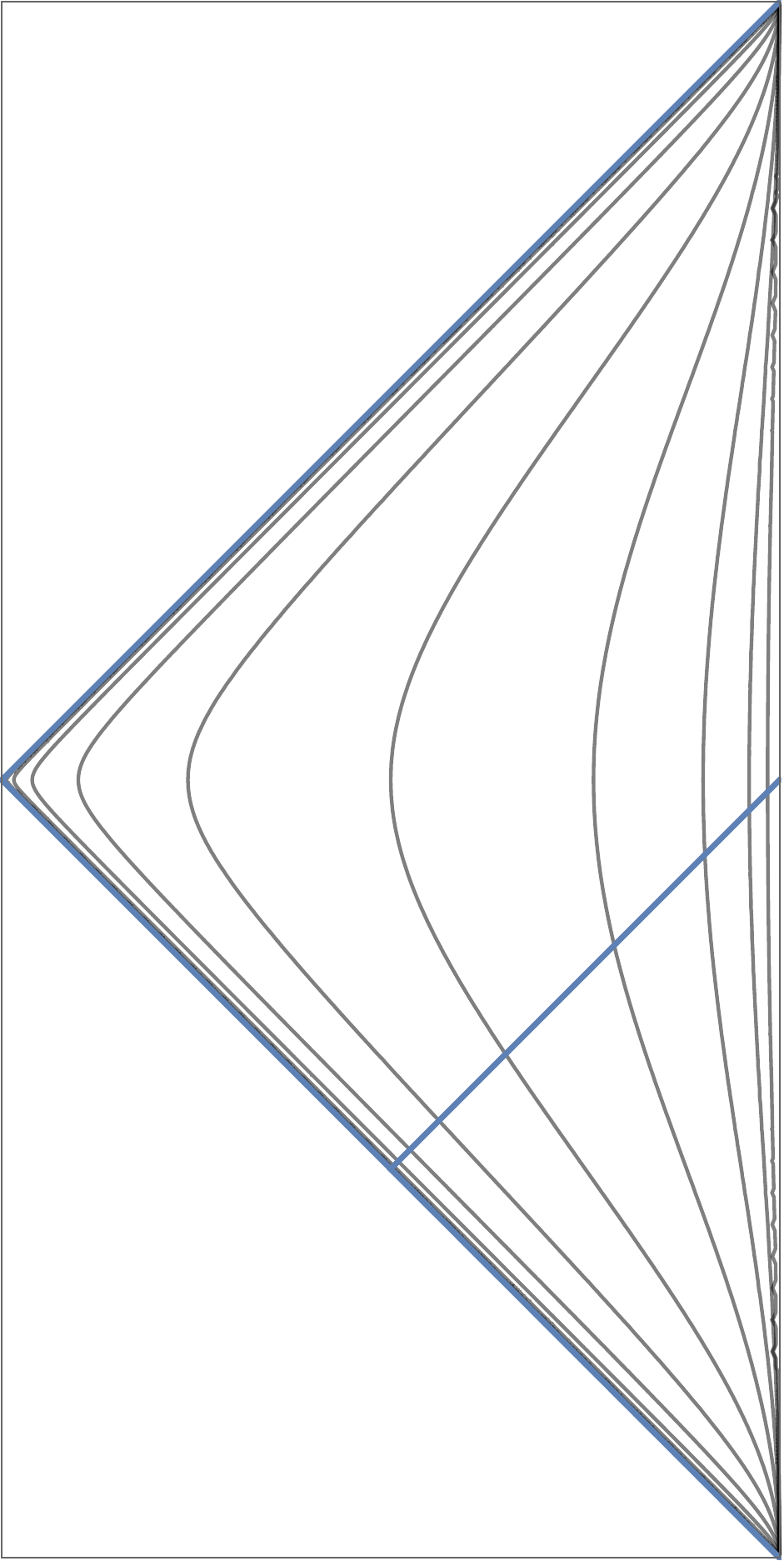}}
\qquad
\subfigure[\ Plunging$_*$$(0 < e < \kappa \ell_*)$]{
\includegraphics[width=.28\textwidth]{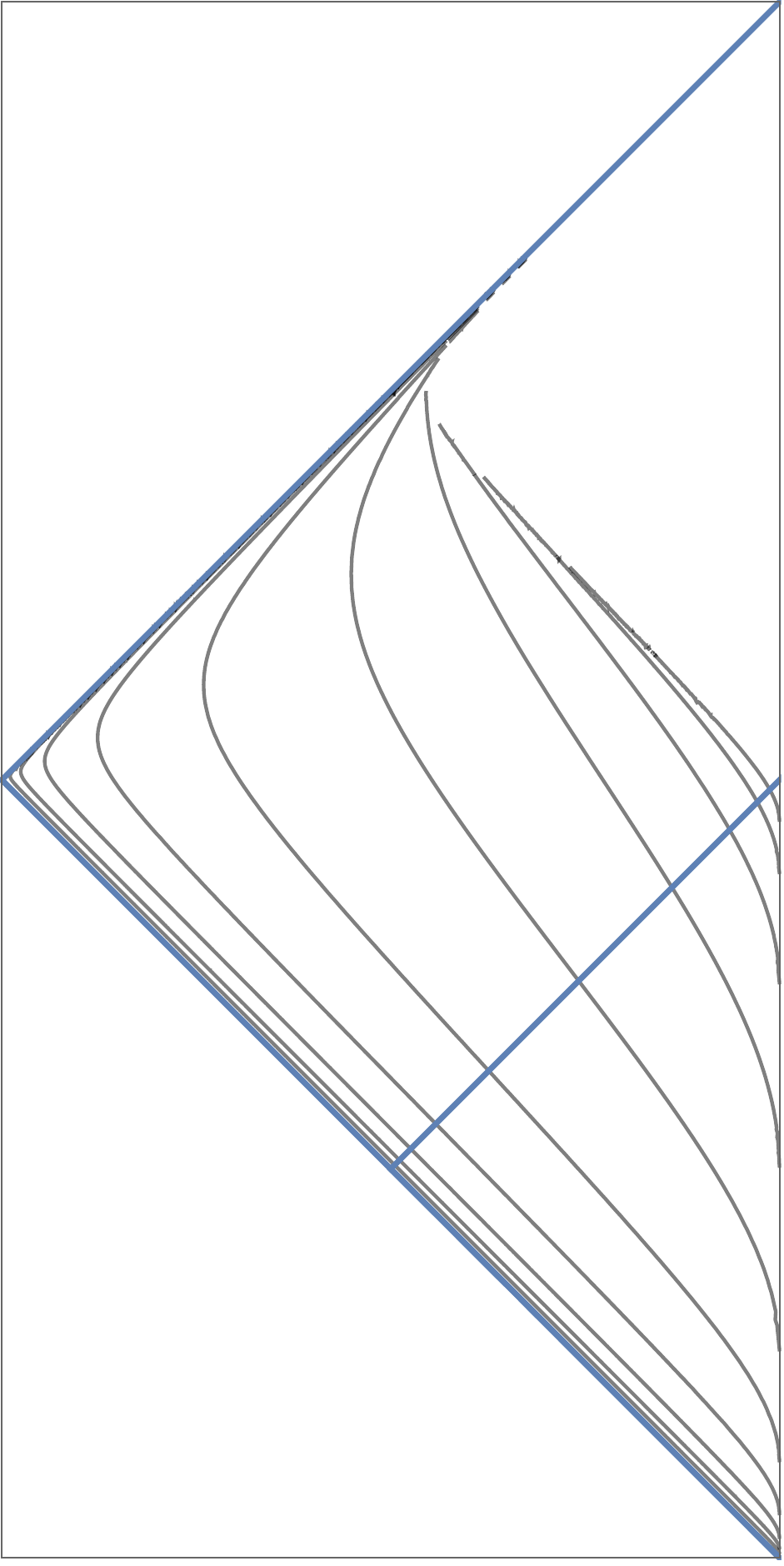}}
\qquad
\subfigure[\ Plunging$_*$($e=\kappa \ell_*$)]{
\includegraphics[width=.28\textwidth]{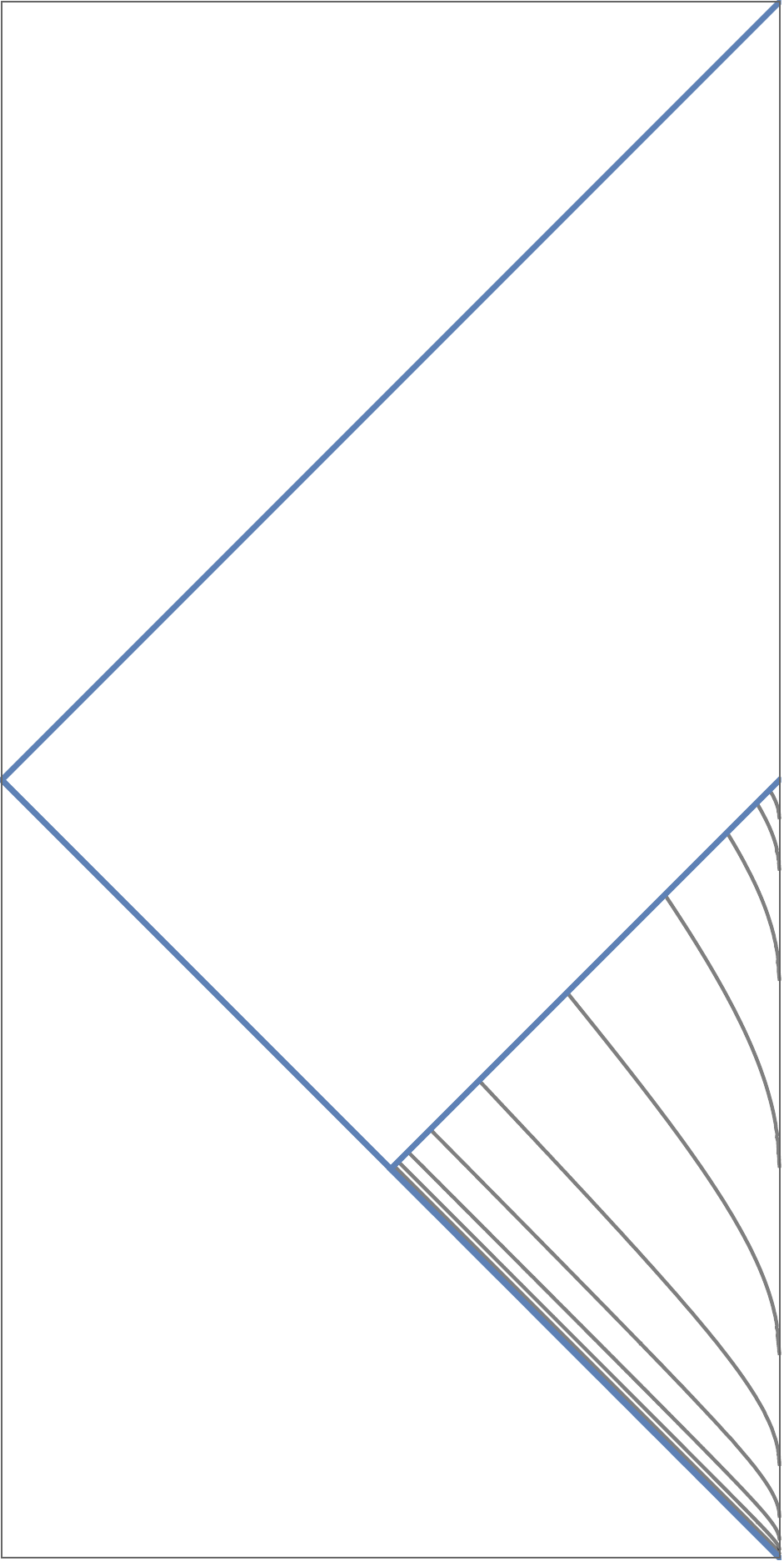}}
\qquad
\subfigure[\ Plunging$_*$$(e > \kappa \ell_*)$]{
\includegraphics[width=.28\textwidth]{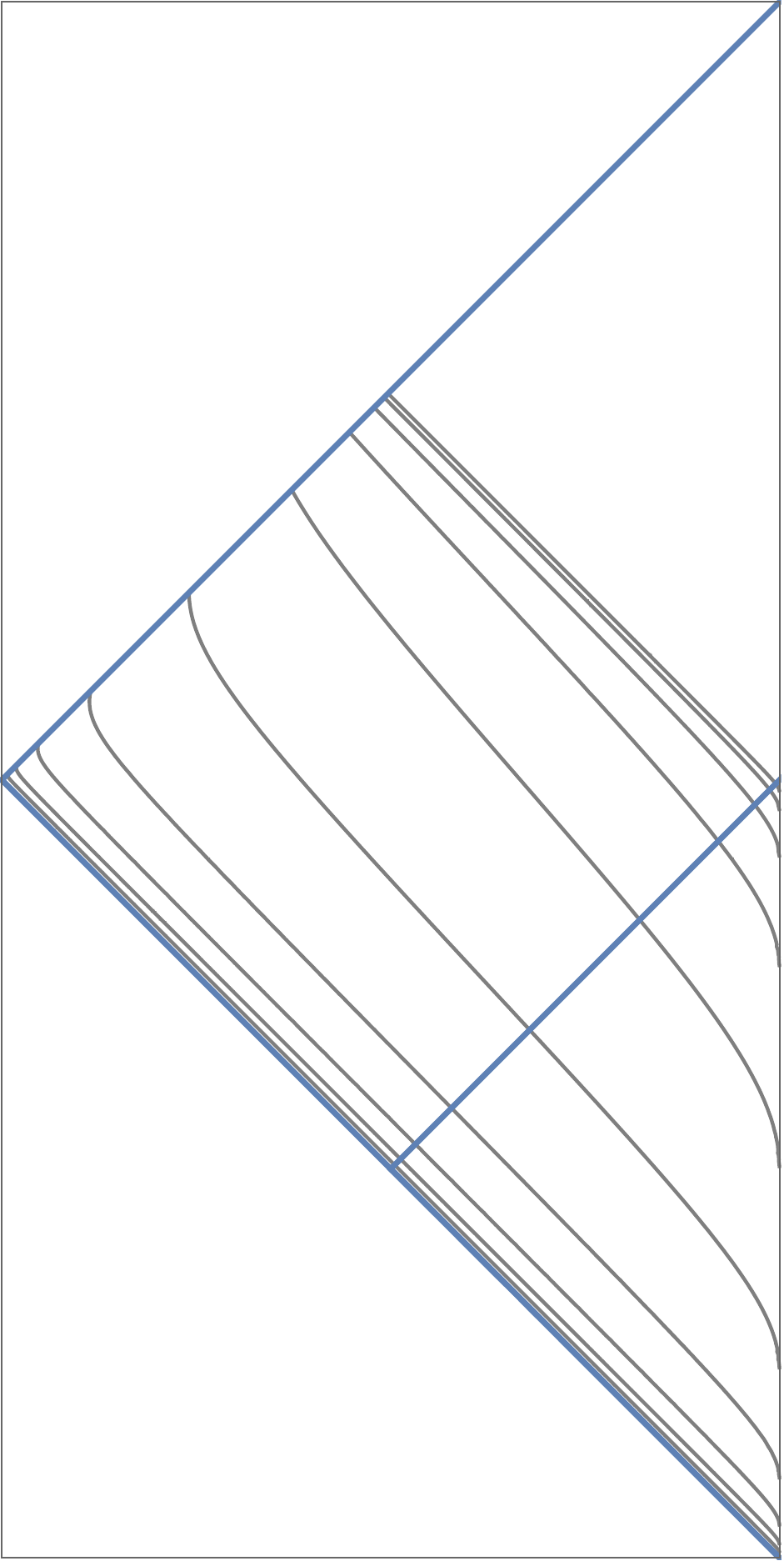}}
\qquad \vspace{-0.3cm}
\caption{Taxonomy of incoming timelike near-NHEK orbits (critical orbits). The axis and choice of parameters are defined as in Figure \ref{fig:taxonomyNHEK}. The outermost blue lines at $45^\circ$ and $135^\circ$ indicate the Poincar\'e horizon $R=0$ of the embedding Poincar\'e NHEK patch. The innermost blue lines at $45^\circ$ and $135^\circ$ (degenerate with the Poincar\'e line) indicate the NHEK black hole horizon $r=0$. In particular, the innermost blue line at $45^\circ$ is the event horizon. We draw the orbits past the event horizon in order to visualize the real conformal maps of orbits (the complex ones do not allow such a visualization).}
\label{fig:taxonomynNHEK}
\end{figure}

\begin{figure}\vspace{-2cm}
\centering
\subfigure[\ Osculating$(e,\ell)$ ($ e < -\frac{\kappa}{2} \sqrt{3(\ell^2-\ell_*^2)}$)]{
\includegraphics[width=.28\textwidth]{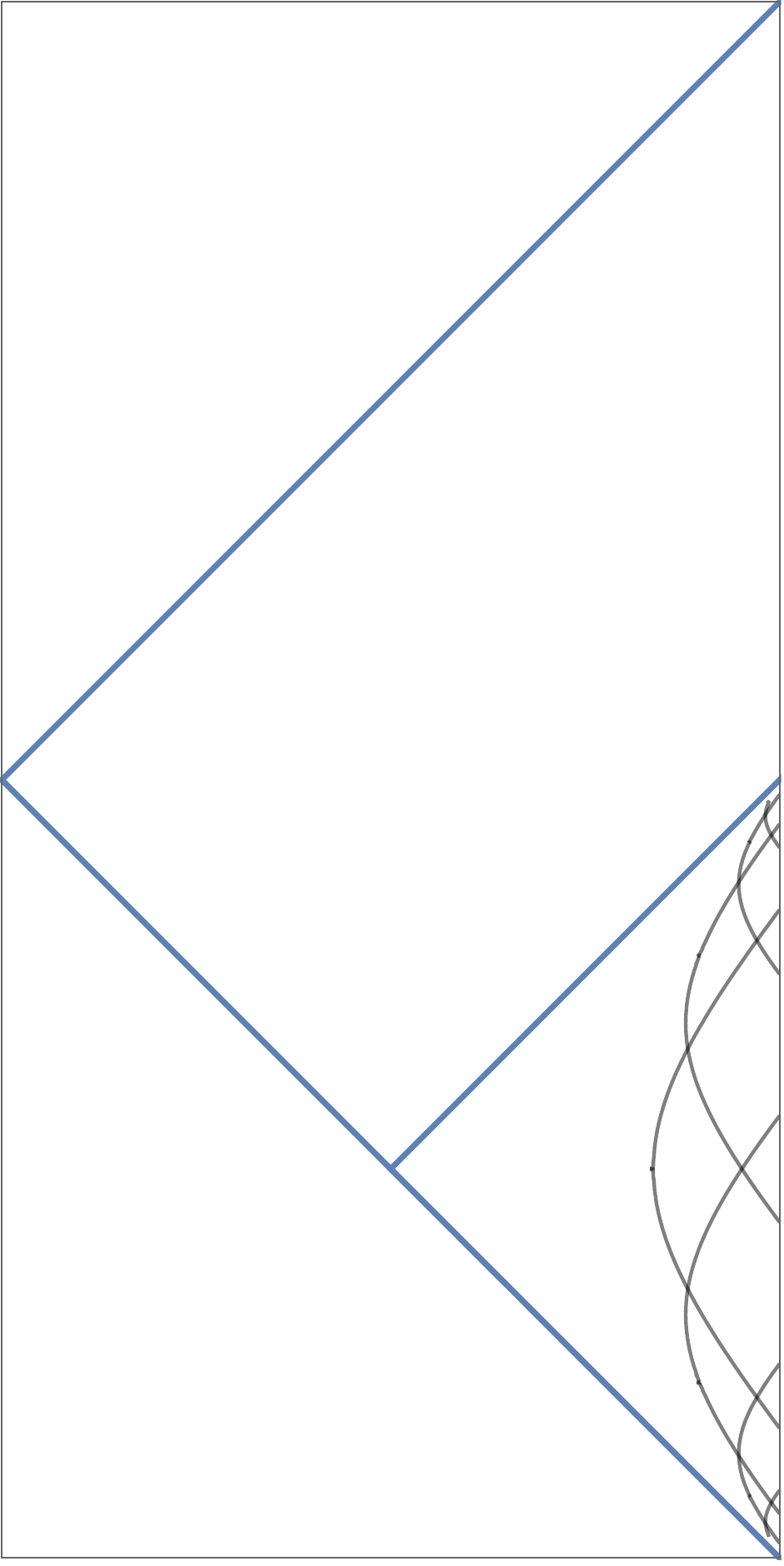}}
\qquad
\subfigure[\ Circular$(\ell)$ ($e= -\frac{\kappa}{2} \sqrt{3(\ell^2-\ell_*^2)}$) or Second Circular$(\ell)$ ($e=-\kappa \ell$)]{
\includegraphics[width=.28\textwidth]{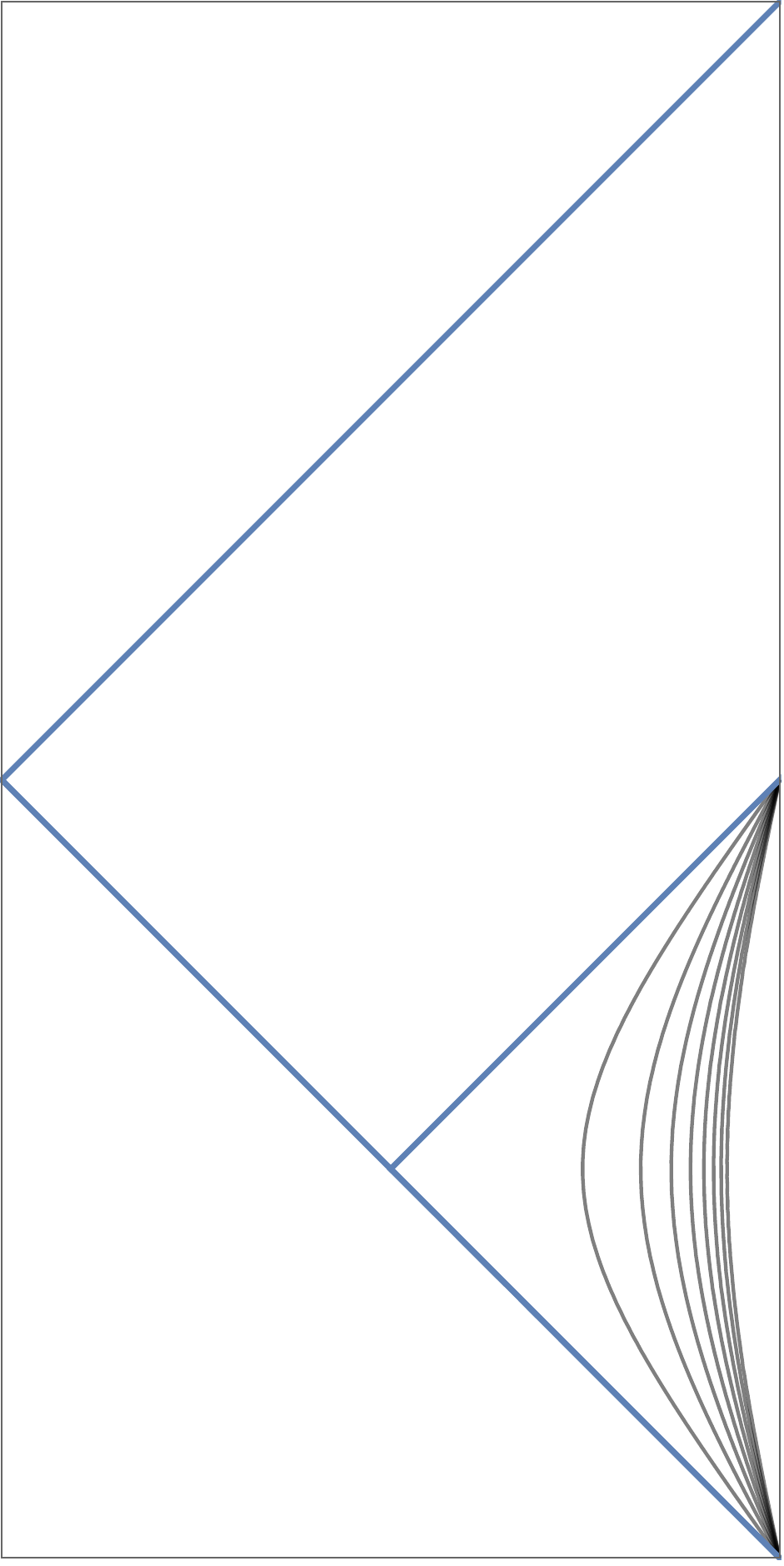}}
\qquad
\subfigure[\ Plunging$(e,\ell)$ ($-\frac{\kappa}{2} \sqrt{3(\ell^2-\ell_*^2)} < e < 0 $)]{
\includegraphics[width=.28\textwidth]{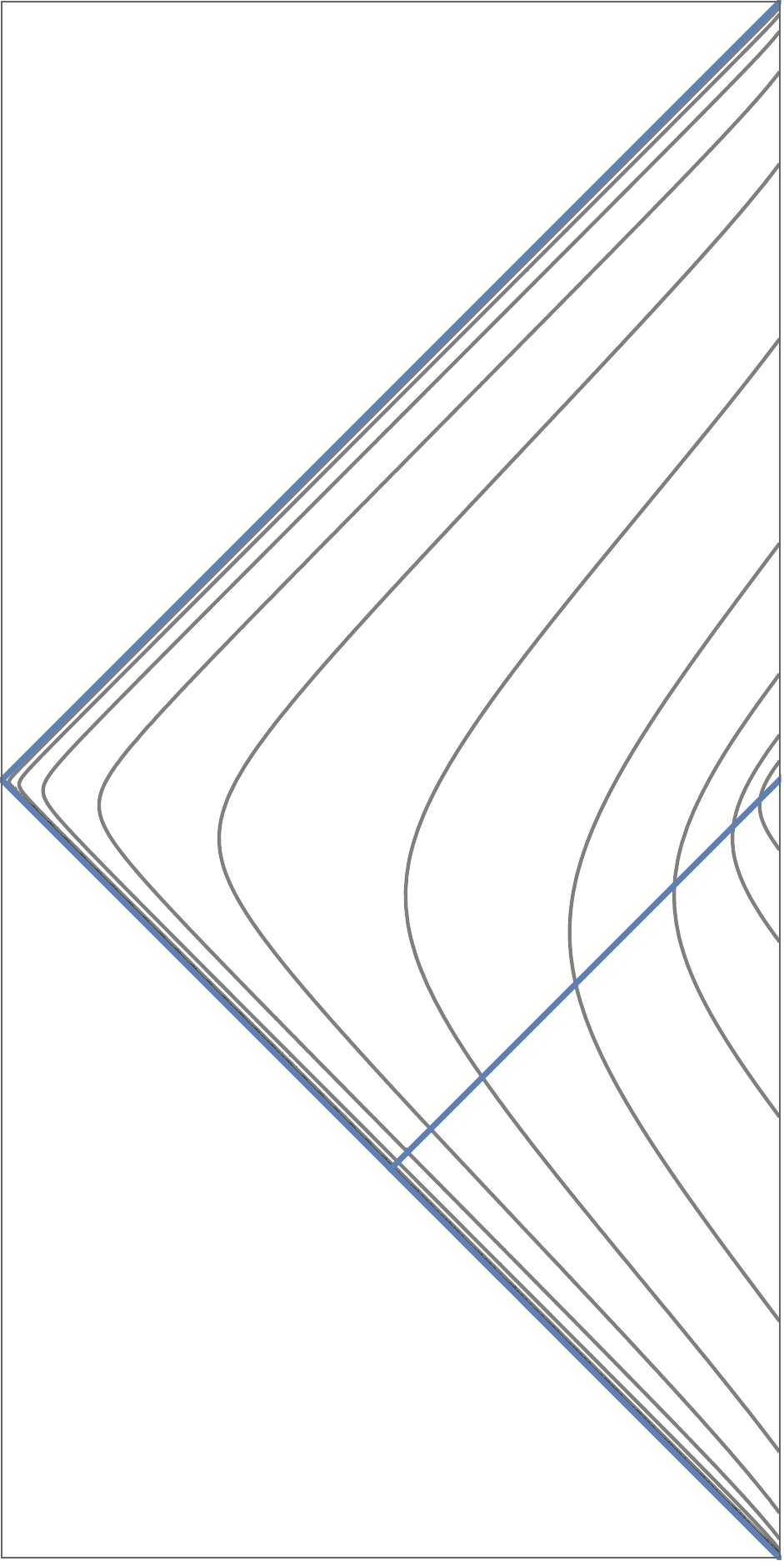}}
\qquad
\subfigure[\ Plunging$(e,\ell)$ ($e=0$)]{
\includegraphics[width=.28\textwidth]{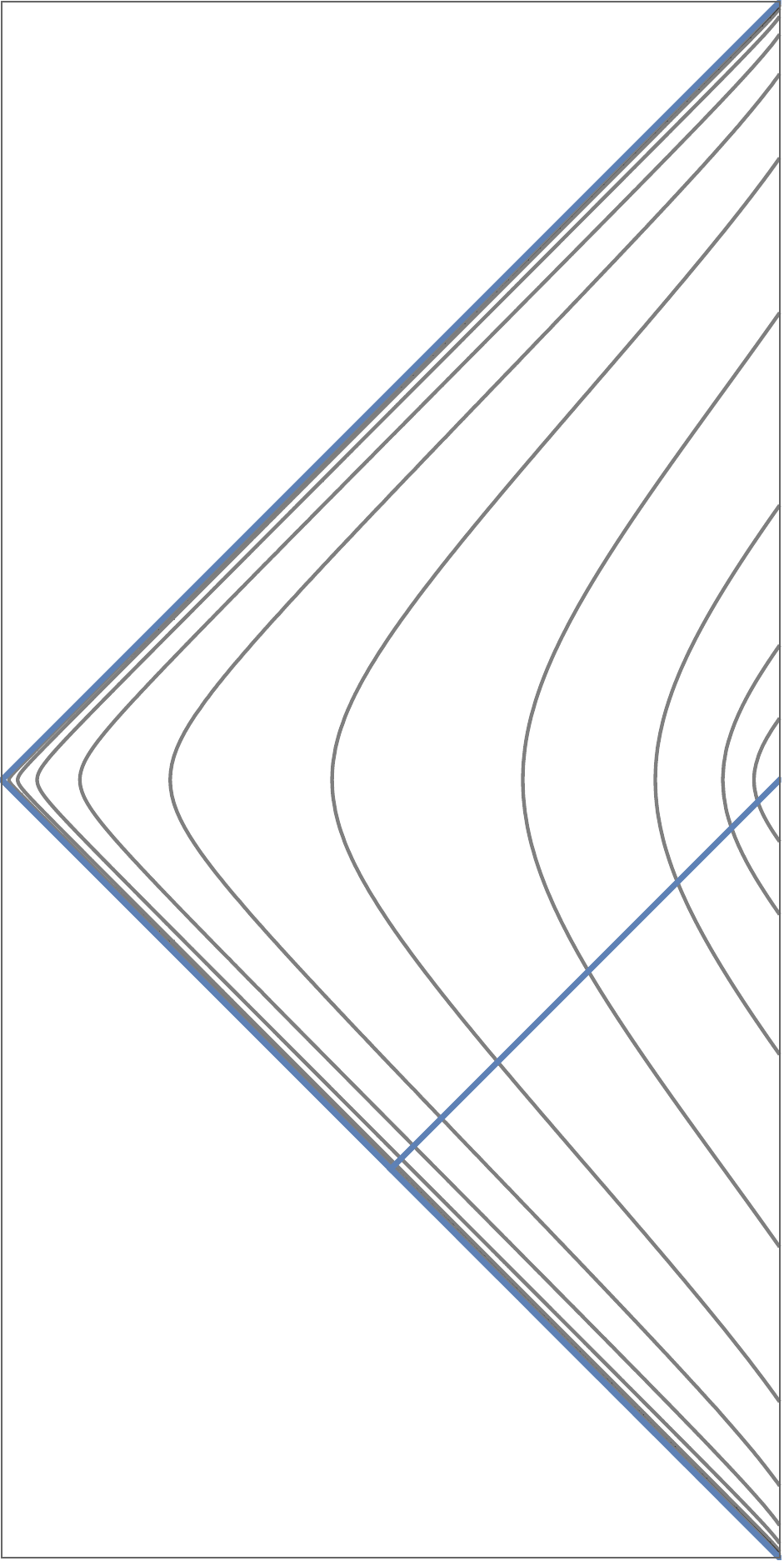}}
\qquad
\subfigure[\ Plunging$(e,\ell)$ with $0 < e < \kappa \ell_* $]{
\includegraphics[width=.28\textwidth]{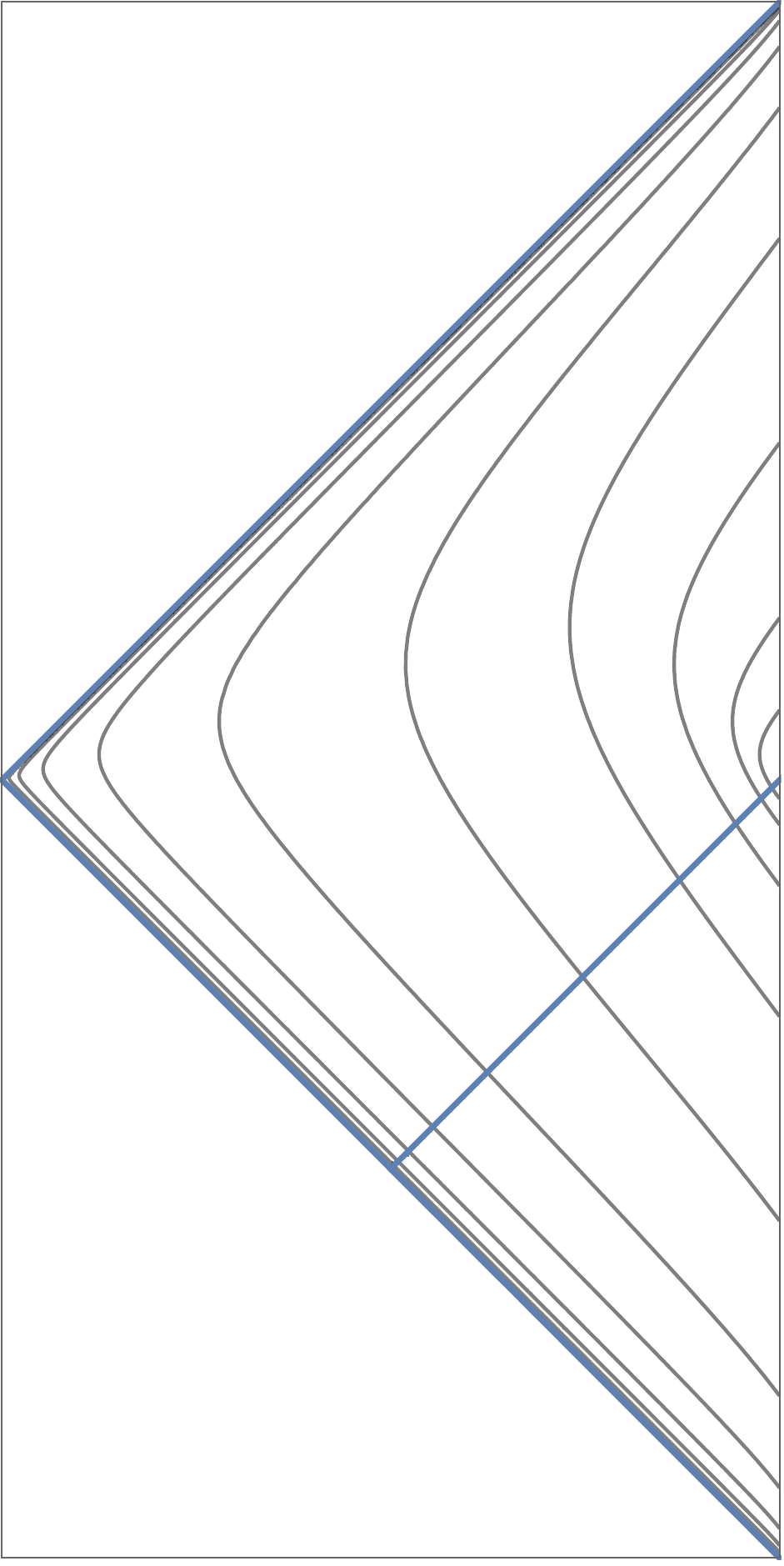}}
\qquad 
\subfigure[\ Plunging$(e,\ell)$ with $e \geq \kappa \ell_*$]{
\includegraphics[width=.28\textwidth]{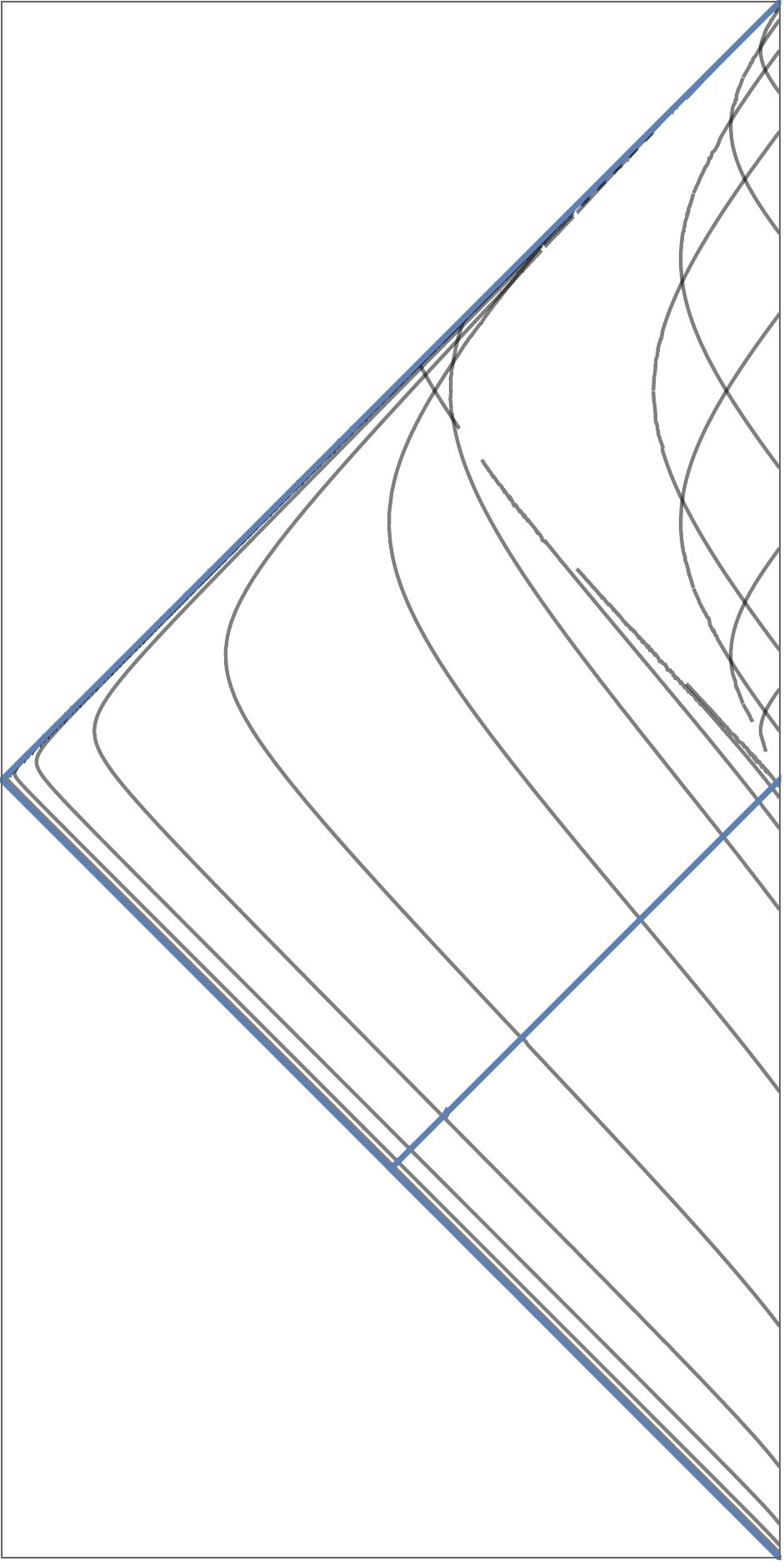}}
\caption{Taxonomy of incoming timelike near-NHEK orbits (supercritical orbits). (The broken lines in (f) are unphysical numerical rendering effects)}\
\label{fig:taxonomynNHEK2}
\end{figure}

We illustrate these orbits in Figures \ref{fig:taxonomynNHEK} and \ref{fig:taxonomynNHEK2} where we show the conformal diagram of the global NHEK geometry from $\tau =-\frac{\pi}{2}$ to $\tau = +\frac{\pi}{2}$, with the Poincar\'e and near-NHEK horizons indicated in blue. We used $M=\kappa=1$ and when depicting $\ell > \ell_*$ orbits, we chose $\ell=2\ell_*$. The $\text{Plunging$_*$}(e)$ orbits with $e$ smaller or bigger than $\kappa \ell_*$ are distinguished by their behavior inside the near-NHEK horizon.

\subsection{Conformal transformations}

Let us now characterize how conformal symmetry or more precisely $SL(2,\mathbb R) \times U(1)$ symmetry can be used to relate various equatorial orbits.  Since $U(1)$ symmetry commutes with all symmetries, the angular momentum is invariant under symmetry transformations.  One can therefore only attempt to map the $\ell=\ell_*$ orbit to another $\ell=\ell_*$ orbit, and a $\ell> \ell_*$ orbit to another $\ell> \ell_*$ orbit. 
 
After analysis, it turns out that all equatorial orbits can be related to either a NHEK circular orbit or near-NHEK circular orbit by a \emph{complex} $SL(2,\mathbb C) \times U(1)$ transformation which could be combined with a PT symmetry flip (time and axial angle). There exist therefore exactly two conjugacy classes of orbits under $SL(2,\mathbb C) \times U(1) \times PT$ symmetry. These are
\begin{itemize}
\item  Circular$_*$ (ISCO) $\Leftrightarrow$ Plunging$_*(E)$  $\Leftrightarrow$ Plunging$_*(e=0)$ $\Leftrightarrow$ Plunging$_*(e)$
\item Circular$(\ell)$ $\Leftrightarrow$ Marginal$(\ell)$ $\Leftrightarrow$ Osculating$(E,\ell)$ $\Leftrightarrow$  Plunging$(E,\ell)$ $\Leftrightarrow$ Osculating$(e,\ell)$ $\Leftrightarrow$  Plunging$(e,\ell)$ 
\end{itemize}
All conformal transformations relating the orbits in each class are detailed in Appendix \ref{formulae_ct}. The main usefulness of these conformal transformations lies in relating all complicated equatorial orbits to the simple NHEK or near-NHEK circular orbits.
This remarkable property will enable us below to obtain analytically the gravitational wave emission of generic corotating equatorial orbits.

\section{Extremal perturbation theory}
\label{sec:Perturbation}

Linear perturbations around the Kerr black hole are governed essentially by one separable partial differential equation, the Teukolsky master equation. A brief review of gravitational perturbations of Kerr as well as the conventions with respect to the Newman-Penrose formalism, in which this approach is formulated, can be found in Appendix \ref{Teuk}. The upshot of this is that we are interested in obtaining the Newman-Penrose scalar $\Psi_{(-2)}$ in the decomposition \eqref{eqn:KerrDecomposition}. 

\subsection{Gravitational perturbations of Kerr}

The near-extremality condition \eqref{lambda} guarantees the existence of a (near-)NHEK region of Kerr. This is equivalent to the condition that the reduced Hawking temperature 
\bea
\tau_H = \frac{\hat r_+-\hat r_-}{\hat r_+} \ll 1. \label{nearext}
\eea
We are interested in gravitational wave emission from moving probes within the (near-)NHEK region. Within this region, waves have a frequency $\hat \omega$ and angular momentum $m$ close to the superradiant bound,
\bea
M |\hat{\omega} - m \Omega_H| \ll 1\label{nears} .
\eea
Since we only consider such corotating modes, we restrict to corotating waves $m \neq 0$ from now on. However, we allow for modes to be slightly below or slightly above the superradiant bound. In the terminology of the near-horizon region, we allow for positive or negative near-horizon energy.

The existence of these small parameters allows the use of the method of matched asymptotic expansions. In this, the near-horizon region is defined in terms of the adimensional Boyer-Lindquist radius $\hat x$ as 
\bea
\hat x \equiv \frac{\hat r - \hat r_+}{\hat r_+} \ll 1. \label{NEAR}
\eea
The radial coordinate which resolves the near-horizon region is taken to be either the NHEK radius $R$, defined as $\hat x = \lambda^{2/3}R$, or the near-NHEK radius $r$, defined as $\hat x = \frac{\lambda}{\kappa} r$ (where $\kappa > 0$ is any fixed constant) as described in Section \ref{app:NHEK}.

The asymptotic or `far' region is defined as
\bea
\hat x \gg \frac{\hat r_+ - \hat r_-}{\hat r_+},\qquad \hat x \gg M |\hat{\omega} - m \Omega_H |, \label{FAR}
\eea
and the intermediate or matching region is defined as 
\bea
 \mbox{max}\left( \frac{\hat r_+ - \hat r_-}{\hat r_+}, M |\hat{\omega} - m \Omega_H | \right) \ll \hat x \ll 1. \label{MATCH}
\eea

We impose the boundary condition that the solution of the Teukolsky equation is outgoing at asymptotic null infinity and ingoing at the horizon. The stress-tensor source is taken to be a non-spinning point particle of rest mass $m_0$ following a geodesic $x_*^\alpha(\tau)$ inside the near-horizon region, 
\bea
T^{\mu\nu} = m_0 \int \frac{d\tau}{\sqrt{-g}} \frac{dx^\mu}{d\tau} \frac{dx^\nu}{d\tau} \delta^{(4)}(x^\alpha - x_*^\alpha(\tau)). \label{Tmunu}
\eea
We now describe the general solution, up to an undetermined coefficient $B$ defined below, which is dictated by the near-horizon physics. The latter requires an explicit computation that depends on the specific geodesic and is described in Section \ref{Emission}.

\paragraph{Asymptotic region:} At zeroth order approximation in the near-extremal \eqref{nearext} and near-superadiant limit \eqref{nears}, the only relevant spheroidal harmonics \eqref{eqn:AngularTeukolsy} are those with 
\bea
\hat \omega =m \Omega_{ext} = \frac{m}{2M}\label{hato}
\eea
thereby leading to the extremal spheroidal harmonics defined in \eqref{Slm}-\eqref{eqn:extAngularTeukolsy}.
In the far region \eqref{FAR}, the radial equation \eqref{eqn:RadialTeukolsky} can be approximated by 
\be
\hat x^{-2s} \frac{\d}{\d \hat x}(\hat x^{2s+2}\frac{\d \hat R^{\text{far}}_{lm \hat{\omega}}}{\d \hat x}) + [\frac{m^2}{4}(\hat x+2)^2+ism \hat x+\frac{3}{4}m^2+s(s+1)- \cE_{lm} ] \hat R^{\text{far}}_{lm \hat{\omega}} = 0
\ee
where $\hat R^{\text{far}}_{lm \hat{\omega}}(\hat x) = \hat R_{lm \hat{\omega}}(\hat r_+ (1+\hat x))$, which agrees with \cite{Porfyriadis:2014fja}. The solutions can be expressed in terms of confluent hypergeometric functions  
\bea
\hat R^{\text{far}}_{lm \hat{\omega}}(\hat x) &=& P \,\hat x^{h-1-s}e^{-im \hat x/2} {}_1F_1(h+im-s,2h,im \hat x) + Q \; (h \mapsto 1-h)
%\nn \\ &&+ Q \, \hat x^{-h-s}e^{-im \hat x/2} {}_1F_1(1-h+im-s,2(1-h),im \hat x)
\label{eqn:Kerrfar}
\eea
where $(h \mapsto 1-h)$ indicates that we should replace $h$ by $1-h$ in the first term and we defined the weight
\be
h = \frac{1}{2} \pm \frac{1}{2}\eta_{lm} ,\qquad \eta_{lm} \equiv \sqrt{1-7m^2 + 4 \cE_{lm}}.  \label{defh}
\ee
The sign is fixed such that $h > \frac{1}{2}$ when $\eta_{lm}^2 >0$ and $\text{Im}(h) < 0 $ when $\eta_{lm}^2 < 0$\footnote{This matches with the convention of  \cite{Yang:2013uba} (after identifying their $\delta$ as our $-i(h-\frac{1}{2})$) but differs from \cite{Porfyriadis:2014fja} if $\Im{(h)} \neq 0$. Our definition of $\eta_{lm}$ coincides with $\eta$ in \cite{Dias:2009ex}.}. The axisymmetric modes ($m=0$) have $\eta_{l0}=1+2l$  and real weight $h=1+l$. For $|s|=2$, all modes with $l=2,3,4$ have a real weight except the modes at the extremities of the range: $|m| = l$ where it is complex. For high $l$ the modes with approximately $|m| < 0.75 l$ are real and those with $|m| > 0.75l$ are complex, see e.g. Figures 1 and 2 of \cite{Dias:2009ex}. 

The outgoing boundary condition at infinity implies
\begin{equation}
P = -Q(-im)^{2h-1} \frac{\Gamma(2-2h)}{\Gamma(2h)} \frac{\Gamma(h-im+s)}{\Gamma(1-h-im+s)}\label{outg}
\end{equation}
such that
\begin{eqnarray}
	\hat R^{\text{far}}_{lm \hat{\omega}}(\hat{x} \to \infty) &=& Q \frac{\Gamma(2-2h)}{\Gamma(1-h+im-s)} (im)^{h-1+im-s}\nn \\ 
	&\times& \left[1-\frac{(-im)^{2h-1}}{(im)^{2h-1}}  \frac{\sin{\pi(h+im)}}{\sin{\pi(h-im)}} \right] \hat x^{-1+im-2s}e^{im \hat x/2} ,\label{eqfar} \\
	\hat R^{\text{far}}_{lm \hat{\omega}}(\hat{x} \to 0) &=&P \hat x^{h-1-s}(1+O(\hat x^1)) + Q \hat x^{-h-s} (1+O(\hat x^1)) .\label{eqn:matchingfar}
\end{eqnarray}
In terms of the radial  tortoise coordinate $\hat r^*$ defined in \eqref{rstar} we have $2^{-i m}e^{i m/2}\hat x^{im}e^{im \hat x/2}=e^{i \hat \omega \hat r_*}+O(\hat x^{-1})$ after using \eqref{hato} and $\Psi_{(s)} \sim \hat x^{-1-2s} e^{i \hat \omega (- \hat t + \hat r_*)+i m \hat \phi}$ is indeed outgoing. The remaining constant $Q$ is determined by matching onto the near-horizon solution in the matching region.

\paragraph{Near-horizon region (NHEK):} The Newman-Penrose scalar $\Psi_{(-2)}$ \eqref{eqn:KerrDecomposition} is expressed in terms of extremal spheroidal harmonics defined in \eqref{eqn:extAngularTeukolsy}. In order to solve for the radial behavior, let us first rewrite the wave perturbation in terms of variables adapted to the NHEK region. Following \eqref{NHEKlimit} we have $e^{-i \hat \omega \hat t + i m \hat \phi} = e^{-i \Omega T + i m \Phi}$ where
\bea
\Omega \equiv \frac{2M}{\lambda^{2/3}}(\hat \omega - \frac{m}{2M})\label{Omega}
\eea
is the near-horizon frequency. From the point of view of a NHEK observer, a finite energy perturbation in the asymptotically flat region has diverging energy in the extremal limit. Substituting $\hat \omega$ from \eqref{Omega} and using the change of coordinates \eqref{NHEKlimit}, the radial equation \eqref{eqn:RadialTeukolsky} exactly reduces in the limit $\lambda \rightarrow 0$ to the NHEK radial equation \eqref{eqn:NHEKradial} for $R_{lm\Omega}$, with 
\bea
R_{lm\Omega}(R) = \hat R_{lm\hat \omega}(M+M \lambda^{2/3}R),\qquad T_{lm\Omega}(R) = \hat T_{lm\hat \omega}(M+M \lambda^{2/3}R).
\eea 
The homogeneous solution to the NHEK radial equation is given in terms of Whittaker functions as a linear combination of
\begin{eqnarray}
\cM^{\text{D}}_{lm\Omega}(R) &=& R^{-s} M_{im+s,h-1/2}(\frac{-2i \Omega}{R}),\\
\cW^{\text{in}}_{lm\Omega}(R) &=& R^{-s}W_{im+s,h-1/2}(\frac{-2i \Omega}{R}),
\end{eqnarray}
where the parameter $h$ was already defined in the asymptotic region in \eqref{defh}. These functions have a special behavior at infinity and at the horizon, respectively. The function $\cW^{\text{in}}_{lm\Omega} \rightarrow (-2 i \Omega)^{im+s}R^{-i m -2s}e^{i \Omega/R}$ is purely ingoing at the horizon $R \rightarrow 0$ while  for $R \rightarrow \infty$
\bea
\cM^{\text{D}}_{lm\Omega} &=& (-2 i \Omega)^h R^{-h-s}(1+O(R^{-1})),\nn \\
\cW^{\text{in}}_{lm\Omega} &=& c_h^{\Omega,m} R^{-h-s}(1+O(R^{-1}))+c_{1-h}^{\Omega,m} R^{h-1-s}(1+O(R^{-1})),\label{ccoef}
\eea
with $c^{\Omega,m}_h = (-2i \Omega)^{h} \frac{\Gamma(1-2h)}{\Gamma(1-h-i m -s)}$. The function $\cM^{\text{D}}$ therefore obeys Dirichlet boundary conditions. 

Away from the source, for radii $R > R_*(\tau)$ where $R_*(\tau)$ describes the source geodesic, the radial function is a solution to the homogenous equation and can be written in terms of two coefficients $A$ and $B$ as 
\bea
R_{lm\Omega}(R) |_{R>R_*(\tau)} =  A\, \cW^{\text{in}}_{lm\Omega}(R) +B \,  \cM^{\text{D}}_{lm\Omega}(R) .\label{Rexo}
\eea
The ``near-horizon coefficients'' $A$ and $B$ can be matched onto the asymptotic coefficients $P$, $Q$ through the matching conditions 
\bea
\delta\psi_4 |_{Kerr} = M^2 \lambda^{4/3}\times  \delta \psi_4|_{NHEK}
\eea
where the prefactor originates from the type III rotation from the Kerr tetrad frame to the NHEK tetrad \eqref{tpsi4}, which induces a scaling of the Newman-Penrose scalar \eqref{tpsi4}. After performing the substitution \eqref{eqn:KerrDecomposition} and \eqref{eqn:NHEKDecomposition}, the matching condition can be rewritten in terms of the radial functions as 
\bea  
\text{lim}_{\hat x \rightarrow 0} \hat R^{far}_{lm\hat \omega}(\hat x) =M^4 \times \frac{2M}{\lambda^{2/3}} \times M^2 \lambda^{4/3}\times  \text{lim}_{R \rightarrow \infty} R_{lm\Omega}(R) |_{R>R_*(\tau)} 
\eea
where the extra $M^4$ factor comes from the difference between $\rho$ \eqref{Kerrspins} and $\eta$ \eqref{defeta} and the $2M \lambda^{-2/3}$ factor from the difference in frequency integration \eqref{Omega}. Using \eqref{eqn:matchingfar} and $\hat x = \lambda^{2/3}R+o(\lambda^{2/3})$ from \eqref{NHEKlimit} we find
\bea
A &=& -{{\frac{Q \lambda^{\frac{2}{3}}}{2 M^7}}} (-im)^{2h-1} (-2i\Omega \lambda^{2/3})^{h-1}  \frac{\Gamma(2-2h)}{\Gamma(2h-1) \Gamma(2h)} \frac{\Gamma(h-im+s) \Gamma(h-im-s)}{\Gamma(1-h-im+s)} ,\nn\\
Q &=&2 M^7 \lambda^{-\frac{2}{3}} B(-2 i \lambda^{2/3} \Omega)^h  \times \nn\\
&&\hspace{-1.5cm} \left( 1 -(-i\lambda^{2/3}\Omega)^{2h-1} (-2i m)^{2h-1} \frac{\Gamma(1-2h)^2}{\Gamma(2h-1)^2} \frac{\Gamma(h-im+s)}{\Gamma(1-h-im+s)} \frac{\Gamma(h-im-s)}{ \Gamma(1-h-im-s)}\right)^{-1}.\label{matchsol}
\eea
The only remaining unknown is $B$, which will be fixed after we find the particular solution for the relevant source. 

In summary, in the Kinnersley tetrad adapted to the Kerr geometry, the asymptotic behavior of the Newman-Penrose scalar $\delta\psi_4 = \rho^4\Psi_{(-2)}$ (where $\rho$ is given in \eqref{Kerrspins} and $\Psi_{(-2)}$ in \eqref{eqn:KerrDecomposition}) reads
\be
\delta\psi_4(\hat{r} \to \infty) =  \frac{M^2}{\sqrt{2 \pi}} \int_{-\infty}^{\infty} \d \Omega\, \sum_{lm} B  \cK  S_{lm}(\theta)e^{im\hat{\phi}}e^{-\frac{i}{2M}(m+\lambda^{2/3}\Omega)\hat{u}} \hat{x}^{-1}.\label{apsi4}
\ee
Here we have used \eqref{eqfar}-\eqref{matchsol}, and we have defined the asymptotic retarded time $\hat{u} = \hat t - \hat r^*$. In this expression
\bea
\cK &\equiv& \frac{(-2 i \lambda^{2/3} \Omega)^{h} k_1}{1 -(-i\lambda^{2/3}\Omega)^{2h-1}k_2 } ,\label{cK}\\
k_1 &\equiv&  \frac{2^{im}e^{-im/2}\Gamma(2-2h)}{\Gamma(1-h+im-s)} (im)^{h-1+im-s} \left[1-\frac{(-im)^{2h-1}}{(im)^{2h-1}} \frac{\sin{\pi(h+im)}}{\sin{\pi(h-im)}} \right], \\
k_2 &\equiv& (-2i  m)^{2h-1} \frac{\Gamma(1-2h)^2}{\Gamma(2h-1)^2} \frac{\Gamma(h-im+s)}{\Gamma(1-h-im+s)} \frac{\Gamma(h-im-s) }{\Gamma(1-h-im-s)},
\eea
and $s=-2$ is understood. 

Note that the existence of a matching region \eqref{MATCH} requires $\Omega \lambda^{2/3} \ll 1$. Therefore care must be taken when evaluating the Fourier integral \eqref{apsi4} in the small $\lambda$ limit. We return to this point below. Note also that all modes with $\text{Re}(h) \gg 1$ are highly suppressed. The leading contributions to \eqref{apsi4} will come from the complex modes $h=\frac{1}{2}+i h_I$.

\paragraph{Near-horizon region (near-NHEK):} Alternatively, if the source is located within the near-NHEK region, we need to match the asymptotically flat solution to the near-NHEK solution. Due to \eqref{chgtnear}, we have $e^{-i \hat{\omega} \hat{t} + im\hat{\phi}} = e^{-i \omega t + i m \phi}$ with
\be
\omega = \frac{2 M \kappa}{\lambda}(\hat{\omega}-\frac{m}{2M}).
\label{omega}
\ee

Using this expression for $\hat{\omega}$ and the change of coordinates  \eqref{chgtnear}, the radial equation \eqref{eqn:RadialTeukolsky} reduces to the near-NHEK radial equation \eqref{eqn:nearNHEKRadial} in the limit $\lambda \to 0$ upon identifying
\bea
R_{lm\omega}(r) = \hat R_{lm\hat \omega}(M+M\frac{\lambda}{\kappa}(r+\kappa)),\qquad T_{lm\omega}(r) = \hat T_{lm\hat \omega}(M+M\frac{\lambda}{\kappa}(r+\kappa)).
\eea 
The homogeneous solutions to the radial near-NHEK equation \eqref{eqn:nearNHEKRadial} are spanned by
\be
	\cR^{\text{in}}(r) = r^{-in/2-s}(\frac{r}{2\kappa}+1)^{i(\frac{n}{2}-m)-s} {}_2F_1(h-im-s,1-h-im-s,1-in-s,-\frac{r}{2 \kappa}) ,
\ee
\be
	\cR^{\text{D}}(r) = r^{-h-s}(\frac{2\kappa}{r}+1)^{i(\frac{n}{2}-m)-s} {}_2F_1(h-im-s,h-im+in,2h,-\frac{2 \kappa}{r})
\ee
where we defined 
\bea
n = \frac{\omega}{\kappa} +m.\label{defn}
\eea
The boundary conditions satisfied by these solutions are again respectively ingoing and Dirichlet. More precisely, 
\bea
\cR^{\text{D}}(r) &=& r^{-h-s}(1+O(r^{-1})), \nn\\
\cR^{\text{in}}(r) &=& d_h^{m,n} r^{-h-s}(1+O(r^{-1}))  + d_{1-h}^{n,m} r^{h-1-s}(1+O(r^{-1})) \label{dcoef}
\eea
with $d_h^{n,m}=(2\kappa)^{h-\frac{in}{2}} \frac{\Gamma(1-2h)\Gamma(1-in -s)}{\Gamma(1-h+i(m-n))\Gamma(1-h-i m -s)}$.

 Outside the source, which is described by $r_{*}(\tau)$, the radial function is again a solution to the homogeneous equation and it can be written as 

\be
	R_{lm\omega}(r)|_{r>r_{*}(\tau)} =  A \cR^{\text{in}}(r) + B \cR^{\text{D}}(r)  .
	\label{eqn:nearNHEKmatching}
\ee
Matching this solution to the far solution using \eqref{eqn:matchingfar} and  $\hat x = \frac{\lambda}{\kappa} r$ from \eqref{chgtnear} we find 
\bea
	A &=& -\frac{Q}{2 M^7} 2^{h+in/2-1}(-im)^{2h-1} \lambda^{h} \kappa^{in/2-1}  \nn \\ &\times& \frac{\Gamma(2-2h)}{\Gamma(2h-1) \Gamma(2h)} \frac{\Gamma(h-im+s) \Gamma(h-im-s) \Gamma(h-i(n-m)) }{\Gamma(1-h-im+s) \Gamma(1-in-s)} , \nn \\ 
	Q &=& 2 M^7 \kappa^{1-h}\lambda^{-1+h}  B  \\ 
&&\hspace{-1cm}\times \left(1 -(-2im \lambda)^{2h-1} \frac{\Gamma(1-2h)^2}{\Gamma(2h-1)^2} \frac{\Gamma(h-im+s)}{\Gamma(1-h-im+s)} \frac{\Gamma(h-im-s) \Gamma(h-in+im)}{\Gamma(1-h-im-s) \Gamma(1-h-in+im)} \right)^{-1} \nn 
\eea
where $s=-2$ is understood. Therefore, 
\be
\delta \psi_4(\hat{r} \to \infty) =  \frac{M^2}{\sqrt{2 \pi}} \int_{-\infty}^{\infty} \d \omega \,  \sum_{lm} B  \cK_{\kappa}  S_{lm}(\theta)e^{im\hat{\phi}}e^{-i\hat{\omega} \hat u}  \hat{x}^{-1}
\label{eqn:nearpsi4}
\ee
with
\be
\cK_{\kappa} \equiv   \frac{  \lambda^{h} \kappa^{-h} k_1}{ 1 - \lambda^{2h-1} k_2 \frac{\Gamma(h-in+im)}{\Gamma(1-h-in+im)}}.\label{Kk}
\ee

\subsection{Quasi-normal mode approximation}

So far, we have given frequency-based waveforms  \eqref{apsi4} and \eqref{eqn:nearpsi4} for the curvature perturbation. Experiment however requires time-based waveforms. In what follows we ignore the short time transients, which are primarily associated with the motion of the source, and the tail at very late times due to the further scattering of gravitational waves. Instead we focus on the contributions from the quasi-normal modes (QNM), which provide a good approximation to the waveform at late (but not very late) times. To select the QNM contribution we deform the frequency integrals over the real axis in \eqref{apsi4} and \eqref{eqn:nearpsi4} in the lower complex plane and rewrite this as a sum of three components: the branch cut corresponding to the tail at very late times \cite{PhysRevD.34.384}, the half lower circle corresponding to the short time transients and finally the sum over quasi-normal modes that accounts for the late-time behavior of the waveform.

The spectrum of Kerr quasi-normal modes bifurcates in the near-extremal limit into ``zero-damped'' and ``damped''  quasi-normal modes \cite{Yang:2013uba}. The damped QNMs decouple from the near-horizon region and can therefore be ignored. The zero-damped QNMs was observed by Hod \cite{2008PhRvD..78h4035H} to match the analytic formulae 
\bea
\text{Re}(\hat \omega_{Nlm}) &=& \frac{m}{2M} + o(\lambda^0), \nn \\
\text{Im}(\hat \omega_{Nlm}) &=& -i \frac{2\pi T_H}{2M \Omega_H} (N+\frac{1}{2}) +o(\lambda)= - i \frac{\lambda}{2M} (N+\frac{1}{2})+o(\lambda)\label{HodQNM}
\eea
where $N=0,1,2,\dots $ is the overtone number, up to small corrections and can be written more precisely at first order in the near-extremal limit as \cite{Yang:2013uba}\footnote{Our convention for $h$ \eqref{defh} is crucial for the validity of this formula. This expression was also obtained in \cite{Hadar:2014dpa} for modes with $\eta^2>0$. Note that in \cite{Yang:2013uba}, it was discussed that this formula is not a good approximation for small but finite $\lambda$ for specific modes $l,m$.} (see also \cite{Cook:2014cta})
\bea
\hat \omega_{Nlm} = \frac{1}{2M} (m - i \lambda (N+h))+o(\lambda)\label{scaling}
\eea
where $h$ is given by \eqref{defh}. The QNMs with $\eta_{lm}^2 > 0$ are usually called the normal modes, and the QNMs with $\eta_{lm}^2 < 0$ are the travelling waves. The real part of $\hat \omega$ is lower than $\frac{m}{2M}$ for all travelling waves. 

Given the scaling in $\lambda$ of \eqref{scaling} and the absence of other QNMs with an intermediate $\lambda^p$, $0< p < 1$ scaling (see however \cite{Hod:2015swa,Zimmerman:2015rua}), we conclude there is no QNM in the NHEK limit. However there are the damped QNMs in the near-NHEK limit. We are therefore allowed to use the QNM approximation when the source is in the near-NHEK region, where the formula \eqref{eqn:nearpsi4} applies. Using \eqref{omega} the near-NHEK quasi-normal frequencies are
\bea
\omega_{Nlm}= -i \kappa (N+h). \label{onlm} 
\eea
The QNM \eqref{onlm} originate from poles of $\Gamma(h-in+im)$. It is important to note that the coefficient $B$ is proportional to $\Gamma(h-in+im)$. Indeed, $B$ is given by the Green function constructed from the homogeneous solutions as
\be
B = \frac{1}{\tilde{W}}\int^{\infty}_{0} \d r' \cR^{\text{in}}(r')\frac{T_{lm\omega}(r')}{r'(r'+2\kappa)}
\ee
where the Wronskian is 
\be
\tilde{W} = -\frac{\Gamma(2h) \Gamma(1-i n-s)}{\Gamma(h-i m-s) \Gamma(h-i n+im)}(2 \kappa)^{1-h-i n/2}. 
\ee
The residue of the QNM can be obtained by expanding $B \mathcal K_\kappa$ \eqref{Kk} around $\omega =\omega_{Nlm}$, 
\bea\label{Be}
B \mathcal K_\kappa = k_1 (\frac{\lambda}{\kappa})^h \frac{B}{\Gamma(h-in+im)}  \frac{i \kappa}{ \omega- \omega_{Nlm}} \frac{(-1)^N}{N!} . 
\eea
For normal modes this is immediate. For travelling waves one has to note that $k_2 \ll 1$ (with some exceptions such as the $(l,m)=(9,7)$ or $(13,10)$ modes where $|k_2| \approx 10^{-4}$, see also \cite{Yang:2013uba}). In those cases, the formula \eqref{Be} is therefore an approximation. 
 
Considering the contribution of these modes only, the residue theorem yields 
\bea
\delta \psi_4(\hat{r} \to \infty) &=&  \sqrt{2 \pi} M^2 \sum_{lm}  \lambda^h \kappa^{1-h} k_1 S_{lm}(\theta)e^{im\hat{\phi}}  \hat{x}^{-1}\nn\\
&& \times  \sum_N \frac{(-1)^Ne^{-i\hat{\omega}_{Nlm}\hat u}}{N!}\left. \frac{B}{\Gamma(h-in+im)} \right|_{\omega= \omega_{Nlm}}  .
\label{eqn:genQNM}
\eea

When the source $T_{lm\omega}(r)$ can be considered as independent of the overtone number $N$, i.e. $T_{lm\omega}(r)= T_{lm}(r)$, we can perform the overtone sum exactly. Gathering, in this approximation, the $N$-dependent part of \eqref{eqn:genQNM} but neglecting the potential overtone dependence in the source results in the following overtone sum
\be
\sum^{\infty}_{N=0} \frac{(-1)^Ne^{-\frac{\lambda N \hat u}{2M}} (1+\frac{2 \kappa}{r'})^{\frac{N}{2}}}{N!} \frac{{}_2F_1(h-im-s,1-h-im-s,1-h-im-s-N;-\frac{r'}{2 \kappa})}{\Gamma(1-h-im-s-N)} 
\label{eqn:Greenovertone} 
\ee
which is computed in Appendix in \eqref{eqn:sum3} with
\bea
x &=& e^{-\frac{\lambda \hat{u}}{2M}  } (1+\frac{2 \kappa}{r'})^{\frac{1}{2}}, \qquad
z = \frac{r'}{r'+2\kappa} ,\\
c_+ &=& h+im+s ,\qquad c_- = h-im-s. 
\eea
It leads to 
\be
\delta \psi_4(\hat{r} \to \infty) =    \sqrt{2 \pi} M^2 \sum_{lm} k_1 \frac{\lambda^{h}}{\kappa^{h}} S_{lm}(\theta)e^{im\hat{\phi}}e^{-i\frac{m-i\lambda h}{2 M}\hat{u}} \hat{x}^{-1} \int^{\infty}_0 \d r' \frac{T_{l m}(r')}{r'(r'+2\kappa)} \cG(\hat u,r') \label{generalsourceQNMresponse}
\ee	
where the Green function, which connects the near-NHEK physics with the asymptotic observer, is given by 
\bea
\cG(\hat u,r')   &\equiv&  - \frac{1}{2} (2\kappa)^{\frac{3h}{2}+\frac{im}{2}}\frac{\Gamma(h-im-s)}{\Gamma(2h)\Gamma(1-h-i m -s) } r'^{-\frac{h}{2}-\frac{im}{2}-s}  (1+\frac{r'}{2\kappa})^{-\frac{h}{2}+\frac{im}{2}} \nn \\ &\times& (1 - (1+\frac{2\kappa}{r'})^{-\frac{1}{2}} e^{-\frac{\lambda \hat u}{2M}})^{-h+im+s} (1 - (1+\frac{2\kappa}{r'})^{+\frac{1}{2}} e^{-\frac{\lambda \hat u}{2M}})^{-h-im-s}.
\eea	

This general expression for the emission illustrates the possible phenomenology of general trajectories in near-NHEK spacetime. For retarded times in the range $0<\frac{\lambda\hat u}{2M}<1$ of the asymptotic observer, we can approximate $1 - e^{-\frac{\lambda \hat u}{2M}} \approx \frac{\lambda \hat u}{2M}$ and we observe that $\delta\psi_4$ exhibits a fall-off in between $\hat u^{-h}$ and $\hat u^{-2h}$, depending on the details of the integral of the Green function. The dominant modes are the travelling waves with $\text{Re}(h)=\frac{1}{2}$. The behavior $\hat u^{-1}$ for such modes in the limit of a source in the region $r' \to \infty$ has been predicted before \cite{Yang:2013uba} and it was observed numerically \cite{Burko:2016sfi}. By contrast, the intermediate behavior $\hat u^{-\chi}$, with $1/2 \leq \chi \leq 1$, remains to be found numerically although the full expression for the (near-horizon) QNM approximation to the Green function has also been derived before \cite{Gralla:2016sxp} \footnote{Our result matches theirs taking into account that a different tetrad was used and that the overtone dependence of the source was neglected.}. In Section \ref{nHnH} we discuss explicit trajectories where this behavior is found. In light of \eqref{generalsourceQNMresponse}, this implies a significant contribution to the gravitational wave signal from deep inside the near-NHEK region.

\section{Emission from circular near-NHEK orbits}
\label{Emission}

The spectrum of emission of a body moving on a circular geodesic in NHEK spacetime at first order in the asymptotically matched expansion was obtained in \cite{Porfyriadis:2014fja,Gralla:2015rpa} by the Green's function method. We repeat this computation in our notation in Appendix \ref{app:circ1} and find complete agreement, up to a global sign. This global sign difference originates from the sign of the stress-tensor in Teukolsky equations \eqref{Master}. This global sign difference leads to phase shift which does not modify the amplitude or energy fluxes of \cite{Porfyriadis:2014fja}, which have been confirmed numerically independently  \cite{Gralla:2015rpa}. 

In this section, we extend the analysis to the case of (unstable) circular orbits in near-NHEK spacetime. The procedure applied to circular orbits in near-NHEK is essentially the same as in NHEK and we invite the interested reader to first review the computation in Appendix \ref{app:circ1}. In what follows we need both $\delta\psi_0$ and $\delta\psi_4$ so we calculate both quantities.

Separating the perturbation equation as in \eqref{eqn:nearNHEKDecomposition} and specializing to circular orbits \eqref{circ1} we have
\bea
\delta\psi_4 &=& \frac{1}{(1-i\cos{\theta})^4}  \sum_{lm} R^{(s=-2)}_{lm \tilde{\omega}}(r) S^{(s=-2)}_{lm}(\theta) e^{im(\phi-\tilde{\omega} t)} 
\label{eqn:nearcircDecomposition} ,\\ 
\delta\psi_0 &=& \sum_{lm} R^{(s=2)}_{lm \tilde{\omega}}(r) S^{(s=2)}_{lm}(\theta) e^{im(\phi-\tilde{\omega} t)}\label{radout0},
\eea
with 
\be
\tilde{\omega} = -\frac{3r_0}{4}(1+\kappa_0),\qquad \kappa_0 \equiv \frac{\kappa}{r_0} = \left( \frac{2\ell}{\sqrt{3(\ell^2-\ell^2_*)}}-1\right)^{-1}
\ee
where we used \eqref{circ1}. Note the bound $0 < \kappa_0< \frac{\sqrt{3}}{2-\sqrt{3}}$. It reflects that the timelike circular orbits in the near-NHEK region lie in between the ISCO $r_0 \to \infty$ (or $\ell \to \ell_*$) and the photon circular orbit $r_0 \to \kappa (\frac{2}{\sqrt{3}}-1)$ (or $\ell \to \infty$). The spheroidal harmonics are independent of the sign of the spin $s$ so we drop their superscript from now on. 

The source term for $s=-2$ is given by 
\begin{eqnarray*}
	\mathcal T_4 &=& \frac{m_0 r^{3}_0}{64 M^7\sqrt{3(1+2\kappa_0)-\kappa_0^2}} \\ &\times & \lbrace 144\left[1+4\kappa_0+\frac{61}{18}\kappa_0^2-\frac{11}{9}\kappa_0^3+\frac{1}{9}\kappa_0^4 \right]  \delta(r-r_0)\delta(\theta-\frac{\pi}{2})\delta(\phi-\tilde{\omega} t) \\&+& 16 \left[ 1+5\kappa_0+7\kappa_0^2+\kappa_0^3-2\kappa_0^4\right] r_0 \delta'(r-r_0)\delta(\theta-\frac{\pi}{2})\delta(\phi-\tilde{\omega} t)\\ &-& 48 i \left[ 1+4\kappa_0+\frac{10}{3}\kappa_0^2-\frac{4}{3}\kappa_0^3 \right]\delta(r-r_0)\delta'(\theta-\frac{\pi}{2} ) \delta(\phi-\tilde{\omega} t) \\ &-& 21\left[1+4\kappa_0+\frac{68}{21}\kappa_0^2-\frac{32}{21}\kappa_0^3+\frac{1}{7}\kappa_0^4 \right]\delta(r-r_0)\delta(\theta-\frac{\pi}{2})\delta'(\phi-\tilde{\omega} t)\\ &-&  8i(1+\kappa_0)(1+2\kappa_0)^2 r_0 \delta'(r-r_0)\delta'(\theta-\frac{\pi}{2})\delta(\phi-\tilde{\omega} t)\\&-& 3 \left[ 1+5\kappa_0+\frac{23}{3} \kappa_0^2 +3\kappa_0^3-\frac{2}{3} \kappa_0^4 \right]r_0 \delta'(r-r_0)\delta(\theta-\frac{\pi}{2})\delta'(\phi-\tilde{\omega} t)\\ &+&6 i \left[1+4\kappa_0+\frac{11}{3}\kappa_0^2-\frac{2}{3} \kappa_0^3 \right]\delta(r-r_0)\delta'(\theta-\frac{\pi}{2})\delta'(\phi-\tilde{\omega} t) \\ &+&2(1+\kappa_0)^2(1+2\kappa_0)^2 r^2_0 \delta''(r-r_0)\delta(\theta-\frac{\pi}{2})\delta(\phi-\tilde{\omega} t) \\ &-& 8(1+2\kappa_0)^2\delta(r-r_0)\delta''(\theta-\frac{\pi}{2})\delta(\phi-\tilde{\omega} t) \\ &+& \frac{9}{8} (1+2\kappa_0-\frac{1}{3}\kappa_0^2)^2\delta(r-r_0)\delta(\theta-\frac{\pi}{2})\delta''(\phi-\tilde{\omega} t) \rbrace ,
	\label{eqn:sourcenearNHEK}
\end{eqnarray*}
and for $s=2$ it is given by 
\begin{eqnarray*}
\mathcal T_0&=&\frac{m_0}{128M^3r_0\sqrt{3(1+2\kappa_0)-\kappa_0^2}} [192\delta(r-r_0)\delta(\theta-\frac{\pi}{2})\delta(\phi-\tilde{\omega} t)\nn\\
&&\hspace{-1.2cm}+\frac{8 (9 + \kappa_0 (36 + 36 \kappa_0 - \kappa_0^3)}{(1 + 2 \kappa_0)^2}\delta(r-r_0)\delta(\theta-\frac{\pi}{2})\delta'(\phi-\tilde{\omega} t)+\nn\\
&&\hspace{-1.2cm}\frac{(-3 + (-6 + \kappa_0) \kappa_0)^2}{(1 + 2 \kappa_0)^2}\delta(r-r_0)\delta(\theta-\frac{\pi}{2})\delta''(\phi-\tilde{\omega} t)-128i\delta(r-r_0)\delta'(\theta-\frac{\pi}{2})\delta(\phi-\tilde{\omega} t)+\nn\\
&&\hspace{-1.2cm}\frac{16 i (-3 + (-6 + \kappa_0) \kappa_0)}{1 + 2 \kappa_0}\delta(r-r_0)\delta'(\theta-\frac{\pi}{2})\delta'(\phi-\tilde{\omega} t)-64\delta(r-r_0)\delta''(\theta-\frac{\pi}{2})\delta(\phi-\tilde{\omega} t)\nn\\
&&\hspace{-1.2cm}-\frac{8 r_0 (1 + \kappa_0) (-3 + (-6 + \kappa_0) \kappa_0)}{1 + 2 \kappa_0}\delta'(r-r_0)\delta(\theta-\frac{\pi}{2})\delta'(\phi-\tilde{\omega} t)\nn\\
&&\hspace{-1.2cm}-64 i r_0 (1 + \kappa_0)\delta'(r-r_0)\delta'(\theta-\frac{\pi}{2})\delta(\phi-\tilde{\omega} t)+16r_0^2(1+\kappa_0)^2\delta''(r-r_0)\delta(\theta-\frac{\pi}{2})\delta(\phi-\tilde{\omega} t)].
\end{eqnarray*}

It can be remarked that $\mathcal T_4$ and $\mathcal T_0$ are related as follows
\be
\mathcal T_4 = \frac{r^2(r+2\kappa)^2}{4 M^4 (1+\cos^2{\theta})}  (\frac{1+i\cos{\theta}}{1-i\cos{\theta}})^2 \left. \mathcal T_0 \right|_{(t,\phi) \to (-t,-\phi)}
\ee
due to the invariance of $T_{\mu \nu}$ under $(t,\phi) \to (-t,-\phi)$. A formula analogous to \eqref{eqn:genseparatedsource} applies to near-NHEK upon adapting the notation such that
\be
T^{(s=-2)}_{lm \tilde{\omega}}(r) = \tilde{a}_0 \delta(r-r_0) + \tilde{a}_1 r_0 \delta'(r-r_0) + \tilde{a}_2 r^2_0 \delta''(r-r_0)
\ee
with
\begin{eqnarray}
\tilde{a_0} &=& -\frac{m_0 r^{3}_0}{16 M^5 \sqrt{-\kappa_0^2+3(1+2 \kappa_0)}} ( S_{lm}(\frac{\pi}{2}) (-\frac{m^2}{8}(-3-6\kappa_0+\kappa_0^2)^2 \\  &+& 8(4+16\kappa_0+21\kappa_0^2+10\kappa_0^3+2\kappa_0^4)+8(1+2\kappa_0)^2   - i m(-3-12\kappa_0-20\kappa_0^2-16\kappa_0^3+3\kappa_0^4)) \nn \\ &-& 2(1+2\kappa_0)(m(-3-6\kappa_0+\kappa_0^2)+8 i (1+2\kappa_0+2\kappa_0^2)) S_{lm}'(\frac{\pi}{2}) - 8(1+2\kappa_0)^2 S_{lm}''(\frac{\pi}{2})) ,\nn \\
\tilde{a_1} &=& -\frac{m_0 r^{3}_0}{16 M^5 \sqrt{-\kappa_0^2+3(1+2 \kappa_0)}} (1+2\kappa_0)(1+\kappa_0)  \\ &\times &((-16-3m i +2\kappa_0(-16-3mi)+\kappa_0^2(-16+i m))S_{lm}(\frac{\pi}{2}) + 8 i (1+2\kappa_0)S_{lm}'(\frac{\pi}{2})), \nn \\
\tilde{a_2} &=&-\frac{m_0 r^{3}_0}{8 M^5} \frac{(1+\kappa_0)^2 (1+2\kappa_0)^2}{\sqrt{-\kappa_0^2+3(1+2 \kappa_0)}}S_{lm}(\frac{\pi}{2}).
\end{eqnarray}
For the $s=2$ case, we can define
\be
T^{(s=2)}_{lm\tilde{\omega}}(r)=-4M^2 \int_0^{2\pi} d\phi e^{-im\phi}\int_0^{\pi}\sin\theta d\theta S_{lm}(\theta)(1+\cos^2\theta)\mathcal T_0.
\ee
Then we find 
\be
T^{(s=2)}_{lm\tilde{\omega}}(r)=\tilde{\alpha}_0\delta(r-r_0)+\tilde{\alpha}_1r_0\delta'(r-r_0)+\tilde{\alpha}_2r_0^2\delta''(r-r_0)
\ee
with 
\bea
\tilde{\alpha}_0&=&(\frac{m_0}{32Mr_0(1+2\kappa_0)^2\sqrt{-\kappa_0^2+3(1+2 \kappa_0)}})[(-128 (1 + 2 \kappa_0)^2 + m^2 (-3 + (-6 + \kappa_0) \kappa_0)^2 \nn\\&&+ 
  8 i m (-9 + \kappa_0 (-36 - 36 \kappa_0 + \kappa_0^3)))S_{lm}(\frac{\pi}{2})+16 (1 + 2 \kappa_0) \nn\\&&\times (-(8 i (1 + 2 \kappa_0) + m (-3 + (-6 + \kappa_0) \kappa_0)) S'_{lm}(\frac{\pi}{2})+4(1+2\kappa_0)S''_{lm}(\frac{\pi}{2}))],\nn\\
  \tilde{\alpha}_1&=&\frac{i m_0 (1 + \kappa_0)  [m (-3 + (-6 + \kappa_0) \kappa_0) S_{lm}(\frac{\pi}{2}) -   8 (1 + 2 \kappa_0) S'_{lm}(\frac{\pi}{2})]}{4Mr_0\sqrt{-\kappa_0^2+3(1+2 \kappa_0)}(1+2\kappa_0)}
,\nn\\
  \tilde{\alpha}_2&=&-\frac{m_0}{2Mr_0\sqrt{-\kappa_0^2+3(1+2 \kappa_0)}}(1+\kappa_0)^2S_{lm}(\frac{\pi}{2}).
\eea
Finally, the solution obeying Dirichlet boundary conditions and ingoing boundary conditions at the horizon reads 
\begin{equation}
R_{lm \tilde{\omega}}(r) = \frac{(r_0(r_0+2 \kappa))^{s}}{\tilde{W}}(\tilde{\cX} \Theta(r_0-r) \cR^{\text{in}}(r) + \tilde{\cZ} \Theta(r-r_0) \cR^{\text{D}}(r)) + \frac{\tilde{a}_2}{1+2\kappa_0} \delta(r-r_0)
\label{eqn:nearRadialSolution}
\end{equation}
with 
\begin{eqnarray*}
	\tilde{W} &=& \frac{(1-2h)\Gamma(2h-1) \Gamma(1-i n-s)}{\Gamma(h-i m-s) \Gamma(h-i (n-m))}(2 \kappa)^{1-h-i n/2} ,\\
	\tilde{\cX} &=& r_0 \cR^{\text{D}}{}'(r_0)(-\tilde{a}_1-2(-s+1)\tilde{a}_2 \frac{1+\kappa_0}{1+2\kappa_0})\\ &+& \cR^{\text{D}}(r_0) (\tilde{a}_0-2s\tilde{a}_1 	\frac{1+\kappa_0}{1+2\kappa_0}+2s\frac{\tilde{a}_2}{1+2\kappa_0} + 4s(s-1) \frac{(1+\kappa_0)^2}{(1+2\kappa_0)^2} \tilde{a}_2+\frac{\tilde{a}_2}{1+2\kappa_0} V(r_0)),\\
	\tilde{\cZ} &=&  r_0 \cR^{\text{in}}{}'(r_0)(-\tilde{a}_1-2(-s+1)\tilde{a}_2 \frac{1+\kappa_0}{1+2\kappa_0}) \\ &+& \cR^{\text{in}}(r_0) (\tilde{a}_0-2s \tilde{a}_1 \frac{1+\kappa_0}{1+2\kappa_0}+2s \frac{\tilde{a}_2}{1+2\kappa_0} + 4s(s-1)\frac{(1+\kappa_0)^2}{(1+2\kappa_0)^2}\tilde{a}_2 + \frac{\tilde{a}_2}{1+2 \kappa_0} V(r_0))
\end{eqnarray*}
where $V(r)$ is given in \eqref{Vr} and 
\bea
n= m+\frac{m\tilde{\omega}}{\kappa} = \frac{m}{4}(1-\frac{3}{\kappa_0}).
\eea 

As it turns out, we will also need the $s=2$ solution obeying Dirichlet boundary conditions and outgoing boundary conditions at the horizon (cf. Section \ref{sec:toNHEKo}). This solution reads 
\be
R^{(s=2)}_{lm\tilde{\omega}}(r)=\frac{(r_0(r_0+2\kappa))^s}{\tilde{W}'}( \tilde{\cX}' \Theta(r_0-r)\cR^{\text{out}}(r)+\tilde{\cZ}'\Theta(r-r_0)\cR^{\text{D}}(r) ) +\frac{\tilde{\alpha}_2}{1+2\kappa_0}\delta(r-r_0)\label{radout}
\ee
where the outgoing solution basis is chosen to be 
\be
\cR^{\text{out}}(r)=r^{\frac{in}{2}}(1+\frac{r}{2\kappa})^{i(m-\frac{n}{2})}{}_2F_1(h + i m + s,1 - h + i m + s , 1 + i n + s, -\frac{r}{2\kappa})
\ee
where $s=2$ is understood  and finally, 
\bea
\tilde{W}'&=&\frac{\Gamma(2 h)\Gamma(1 + i n + s)}{\Gamma(h - i m + i n) \Gamma(h + i m + s)}(2\kappa)^{1 - h +\frac{i n}{2} + s},\nn\\
\tilde{\cX}'&=&r_0 \cR^{\text{D}\prime}(r_0)(\tilde{\alpha}_1-2\tilde{ \alpha }_2(-1 + s)\frac{1+\kappa_0}{1+2\kappa_0})+\cR^\text{D}(r_0) \label{Zt}\\
&\times&(-\tilde{\alpha}_0+2s\tilde{\alpha}_1\frac{1+\kappa_0}{1+2\kappa_0}+\tilde{\alpha}_2(-2s\frac{1}{1+2\kappa_0}+4s(1-s)\frac{(1+\kappa_0)^2}{(1+2\kappa_0)^2})-\frac{\tilde{\alpha}_2}{1+2\kappa_0}{V}(r_0)),\nn\\
\tilde{\cZ}'&=&r_0 \cR^{\text{out} \prime }(r_0)(\tilde{\alpha}_1-2\tilde{ \alpha }_2(-1 + s)\frac{1+\kappa_0}{1+2\kappa_0})+\cR^{\text{out}}(r_0)\nn\\
&\times&(-\tilde{\alpha}_0+2s\tilde{\alpha}_1\frac{1+\kappa_0}{1+2\kappa_0}+\tilde{\alpha}_2(-2s\frac{1}{1+2\kappa_0}+4s(1-s)\frac{(1+\kappa_0)^2}{(1+2\kappa_0)^2})-\frac{\tilde{\alpha}_2}{1+2\kappa_0}{V}(r_0)).\nn
\eea
Finally note the relationship between the outgoing and ingoing solutions: 
\bea
\cR^{\text{out}, (s=2),(m)}(r) = \frac{4\kappa^2}{r^2(r+2\kappa)^2}\cR^{\text{in},(s=-2),(-m)}(r). \label{map1}
\eea

\section{Emission from generic orbits from conformal transformations}
\label{Generic}

We now derive the gravitational wave emission from all other equatorial (corotating) orbits in (near-)NHEK. To do so we apply the conformal transformations described in Section \ref{taxo} to relate the waveforms associated with generic equatorial orbits in (near-)NHEK to one of the solutions in the two sets of circular ``seed orbits'' given in Section \ref{Emission} above. 

In \cite{Hadar:2014dpa,Hadar:2015xpa,Hadar:2016vmk} this procedure was employed for particular orbits, and in \cite{Hadar:2015xpa,Hadar:2016vmk} the analysis was limited to spin 0 probes. Here we generalize these considerations to gravitational wave emission from all equatorial orbits. An important technical subtlety, which we discuss in detail below, arises from the fact that conformal transformations do not in general conserve the form of the tetrad. Instead, the transformations must be accompanied by a Type III frame rotation that transforms the Weyl scalar as in \eqref{tpsi4}. Remarkably, however, we find that the conformal maps do preserve the nature of the boundary conditions at infinity and at the horizon: Dirichlet solutions remain Dirichlet, and ingoing solutions remain ingoing. This reduces the problem of finding the gravitational wave emission in the frequency domain of a plunging or osculating orbit to solving a particular integral over the real line, which arises from the Fourier transform of the conformal map at the boundary of (near-)NHEK spacetime. Quite importantly, this Fourier integral can be analytically solved since it reduces for each orbit to an integral representation of an hypergeometric function or a simpler function. This fact directly originates from conformal symmetry. 

There are four categories of conformal transformations mapping either (near-)NHEK into (near-)NHEK. Below we first derive the main formulae in a notation that is adapted to the case of a near-NHEK circular seed mapped to a NHEK orbit. In particular we use the convention that the final near-horizon coordinates of the physical solution are barred while the coordinates of the seed solution are unbarred, as in Appendix \ref{formulae_ct}. This notation will be easily adapted when we treat the remaining three cases in the rest of this Section, by switching lower and upper cases and updating some formulae while keeping all barred quantities barred.

\subsection{Circular near-NHEK orbit to NHEK orbits}
\label{sec:toNHEKo}

Let us first discuss the maps from the near-NHEK circular orbit, whose gravitational wave emission was just computed in Section \ref{Emission}, to NHEK orbits. There are two classes of such conformal maps, as shown in Section \ref{conf:3}.
In order to find the gravitational waveform in the asymptotically flat region we must determine the coefficient $B$ in the asymptotic solution \eqref{apsi4} of the Newman-Penrose scalar $\delta \psi_4$. This coefficient has its origin as the coefficient multiplying the Dirichlet solution of $\delta \psi_4$ in the near-horizon region. The near-NHEK circular seed solution $\delta \psi^{circ}_4(t,r,\theta,\phi)$ is given in \eqref{eqn:nearcircDecomposition}-\eqref{eqn:nearRadialSolution}. We denote the physical plunging/osculating solution in NHEK by $\delta \psi^{physical}_4(\bar T,\bar R,\theta,\bar \Phi)$. The conformal maps relating both orbits reads  as
\bea
\bar R = \bar R(t,r),\qquad \bar T = \bar T(t,r),\qquad \bar \Phi = \phi + \delta\bar\Phi(t,r). 
\eea
In general, this change of coordinates needs to be accompanied by a Type III frame rotation, $l_{physical}^\mu = F^{-1} l^\mu_{seed}$, $n_{physical}^\mu = F n^\mu_{seed}$ ,with $F=F(\bar T,\bar R)$. The Newman-Penrose scalars are then related as
\bea\label{cgh4}
\psi^{physical}_4(\bar T,\bar R,\theta,\bar \Phi) = F^2(\bar T,\bar R) \psi^{circ}_4(t,r,\theta,\phi) .
\eea
In order to have enough freedom to enforce the two boundary conditions on the physical orbits, we allow to supplement the  seed circular solution \eqref{eqn:nearRadialSolution} with an additional homogeneous solution, i.e. we consider 
\bea
R_{lm \tilde{\omega}}(r) &=& \frac{(r_0(r_0+2 \kappa))^{s}}{\tilde{W}}(\tilde{\cX} \Theta(r_0-r) \cR^{\text{in}}(r) + \tilde{\cZ} \Theta(r-r_0) \cR^{\text{D}}(r) + \tilde \cY  \cR^{\text{in}}(r) ) \nn \\
&& + \frac{\tilde{a}_2}{1+2\kappa_0} \delta(r-r_0)
\label{eqn:nearRadialSolutionMOD}
\eea
where $\tilde \cY$ is not fixed.

After a Fourier transformation, we obtain that the seed and physical solutions within the near-horizon region are related as 
\bea
R_{l m \Omega}(\bar R) = \frac{1}{\sqrt{2\pi}}\int_{-\infty}^{\infty} d\bar T e^{i \Omega \bar T - i m \tilde \omega t(\bar T,\bar R)- i m \delta \bar \Phi(\bar T,\bar R)} F^2( \bar T,\bar R)  R_{lm \tilde \omega}(r( \bar T,\bar R) ) .\label{matchR0}
\eea

All conformal transformations written in Appendix \ref{formulae_ct} have the property that $\bar R \rightarrow \infty$ at fixed $\bar T$ is equivalent to $r \rightarrow \infty$ at fixed $t$, and that $\bar R = F(\bar T)r$, and $t=t(\bar T)$ in the $\bar R \rightarrow \infty$ limit for a specific function $F$. Also, $\lim_{r \rightarrow \infty} F(\bar T,\bar R) = F(\bar T)$ for the same function $F$ and $\lim_{r \rightarrow \infty}\delta\bar\Phi (r,t)= 0$. In the matching region we therefore have
\bea
\lim_{\bar R \rightarrow \infty} R_{l m \Omega}(\bar R) = \frac{1}{\sqrt{2\pi}}\int_{-\infty}^{\infty} d\bar T e^{i (\Omega \bar T - m \tilde \omega t(\bar T))} F^2( \bar T) \lim_{r \rightarrow \infty} R_{lm \tilde \omega}(r) .\label{matchR}
\eea
This equation provides in particular the explicit map between homogeneous solutions. It shows that a Dirichlet mode is mapped to a Dirichlet mode and a Neumann mode is mapped to a Neumann mode because the two modes are proportional to each other in the asymptotic region. Considering the full solution, the transfer matrix from \eqref{matchR} is
\bea
\left[ \begin{array}{cc} (-2i \Omega)^{h} & c^{\Omega,m}_h \\ 0 & c^{\Omega,m}_{1-h}\end{array} \right] \left[ \begin{array}{c} B \\ A \end{array}\right] &=& \frac{r_0^s (r_0+2\kappa)^{s}}{\tilde W} \left[ \begin{array}{cc} T_h^{\Omega, m \tilde \omega}    & T_h^{\Omega, m \tilde \omega}  d^{n, m} _{h} \\ 0 & T_{1-h}^{\Omega, m \tilde \omega}  d^{n, m} _{1-h} \end{array}\right] \left[ \begin{array}{c} \tilde{\mathcal Z}\\ \tilde{ \mathcal Y}\end{array}\right] \label{tr1}
\eea
where $c^{\Omega,m}_h$ and $d^{n,m}_h$ are defined respectively in \eqref{ccoef} and \eqref{dcoef} and the time integral can be absorbed into the coefficient
\bea
T_h^{\Omega, m \tilde \omega} =  \frac{1}{\sqrt{2\pi}}\int_{-\infty}^{\infty} d \bar T e^{i (\Omega \bar T - m \tilde \omega t(\bar T))} F^{h+2+s}( \bar T).\label{Tterm}
\eea

Now, we also have that a purely ingoing mode solution is mapped to a purely ingoing mode solution. This follows from the fact that if there is no outgoing mode from the past black hole horizon then there will be no outgoing mode from the Poincar\'e horizon by continuity of the solutions to the wave equations. This argument is independent of the details of the conformal map. It implies that the ratio of Neumann to Dirichlet modes which characterizes an ingoing solution is preserved under the conformal map. In other words, 
\bea
\frac{c^{\Omega,m}_h }{ c^{\Omega,m}_{1-h}} = \frac{T_{h}^{\Omega, m \tilde \omega}  d^{n, m} _{h}}{T_{1-h}^{\Omega, m \tilde \omega}  d^{n, m} _{1-h} }.
\eea
The final solution for the gravitational wave emission is thus given in \eqref{apsi4} with $B$ given by \eqref{tr1}. This yields\footnote{Arguably a simpler way to get the final result is to start from a Dirichlet solution in the matching region, perform the conformal map, and finally add an ingoing solution in order to obey the correct boundary condition with respect to the asymptotically flat region. Since the final step can always be performed and does not change the coefficient $B$, it can be ignored in the computation of $B$. The coefficient $B$ is obtained from the Dirichlet solution in the matching region: it is a function of $\tilde {\mathcal Z}$ only, and not of $\tilde{\mathcal Y}$, leading to \eqref{solB1}.}
\bea
B = (-2 i \Omega)^{-h}\frac{r_0^s (r_0+2\kappa)^{s}}{\tilde W} T_h^{\Omega, m \tilde \omega} \tilde{\mathcal Z}.\label{solB1}
\eea

We now present the explicit emission formulae for the following two classes of orbits: Marginal$(\ell)$ and Plunging$(E,\ell)$/Osculating$(E,\ell)$. 

\paragraph{Marginal$(\ell)$} As described in Appendix \ref{conf:3}, the conformal map in this case consists of the transformation \eqref{nearNHEKtoNHEK} with final barred coordinates $(\bar T,\bar R,\bar \Phi)$, combined with a $PT$ flip $\bar T \mapsto -\bar T$, $\bar \Phi \mapsto - \bar\Phi$. Ignoring first the PT flip, a type III tetrad rotation is required with 
\bea
F(\bar T, \bar R)=\frac{\bar R}{r(\bar T, \bar R)} = -\frac{1}{\kappa \bar T}+O(\bar R^{-1}). 
\eea
where the asymptotic limit of the map is ($\bar R \to \infty$ with $\bar T$, $\bar \Phi$ fixed)
\bea
r &=&- \kappa \bar R \bar T(1+O(\bar R^{-1})), \\
t &=& -\frac{1}{\kappa}\log{|\bar T|} +O(\bar R^{-1}),  \\
\phi &=& \bar\Phi +O(\bar R^{-1}).
\eea
Now, the PT flip $\bar T \mapsto -\bar T$, $\bar \Phi \mapsto - \bar\Phi$ can be taken into account by flipping the sign of both the angular momentum $m$ and frequency $\Omega$ of the NHEK solution. Since the frequency of the near-NHEK solution is $m \tilde \omega$, it will be automatically flipped as well. The angular equations between the near-NHEK and NHEK solutions however would not match since $S_{l,m}(\theta) \neq S_{l,-m}(\theta)$. Instead, the NHEK solution with $\theta$ angle has to be matched with the near-NHEK solution with $\pi-\theta$ angle. In the equatorial plane, the solutions and physical orbits are simply identified. The radial solutions can then be matched term by term in $l,m$ thanks to the identity $S_{l,m}(\theta) = S_{l,-m}(\pi-\theta)$. No tetrad rotation needs to be performed. One simply formally maps the $\theta$ NHEK solution to the $\pi-\theta$ near-NHEK solution with frequencies and angular momentum flipped. 

The final $B$ coefficient is therefore (here $s=-2$)
\bea
B = (-2 i \Omega)^{-h}\frac{\tilde{\mathcal Z} }{r_0^{2} (r_0+2\kappa)^{2} \tilde W} T_h^{-\Omega, -m \tilde \omega}  \label{solBhere}
\eea
where according to \eqref{Tterm}, 
\bea
T_h^{\Omega, m \tilde \omega}  = \frac{1}{\sqrt{2\pi}} \int_{-\infty}^0 d\bar T e^{i \Omega \bar T} |\bar T|^{\frac{i m \tilde \omega}{\kappa}} (-\kappa \bar T)^{-h}. 
\eea
Here the $\bar T$ integral is cut at $\bar T = 0$ which is the endpoint of the trajectory corresponding to $t = -\infty$. Equivalently, in the notation adapted to the result of the PT flip, 
\bea
T_h^{-\Omega, -m \tilde \omega}  &=&\frac{1}{\sqrt{2\pi}} \int_0^\infty d \bar T e^{i \Omega \bar T} \bar T^{-\frac{i m \tilde \omega}{\kappa}-h}\kappa^{-h} \\
&=& \frac{\kappa^{-h}}{\sqrt{2\pi}}  (-i \Omega)^{h-1+\frac{im \tilde{\omega}}{\kappa}}\Gamma(1-h-\frac{im \tilde{\omega}}{\kappa}) 
\eea
where the last integral is strictly valid for modes $h=\frac{1}{2}+i h_I$, $h_I < 0$, which are the dominant modes as shown below. 

As a cross-check, we can also understand the conformal map followed by the PT flip as a single map from the circular near-NHEK trajectory with outgoing boundary conditions at the horizon to the NHEK Marginal$(\ell)$ orbit with the same $\theta$ angle and same angular momentum $m$. The transformation relating the orbits is given by
\bea
r &=& \kappa \bar R \bar T(1+O(\bar R^{-1})), \\
t &=& -\frac{1}{\kappa}\log{\bar T} +O(\bar R^{-1}),  \\
\phi &=& -\bar\Phi +O(\bar R^{-1}).
\eea
Such a transformation transforms the tetrad frame as
\be
l^{\mu}\to \frac{2M^2(1+\cos^2\theta)}{-r R}N^{\mu},\ n^{\mu}\to \frac{-r R}{2M^2(1+\cos^2\theta)}L^{\mu},\ m^{\mu}\to\frac{1-i \cos\theta}{1+i\cos\theta}\bar{M}^{\mu}. 
\ee
After some algebra we find that the $s=-2$ waveform in NHEK is related to the $s=+2$ waveform in near-NHEK as 
\be
\psi_{(-2)}|_{NHEK}(\bar R)=\frac{r^2\bar{R}^2}{4M^4}\psi_{(2)}|_{near-NHEK}(r).
\ee
This solution with outgoing boundary conditions at the horizon was computed in \eqref{radout0}-\eqref{radout}. We can now use the identity $S^{(2)}_{l,m}(\theta)=S^{(-2)}_{l,-m}(\theta)$ to relate the angular part of the $s=+2$ and $s=-2$ solutions. The radial NHEK solution is therefore obtained from the near-NHEK radial solution with $m \mapsto -m$. After some algebra, we find that $B$ can be written as 
\bea
B = \frac{1}{4M^4}(-2i\Omega)^{-h} \left. \frac{(r_0(r_0+2\kappa))^2 \tilde \cZ'}{\tilde{W}'} \right|_{m \mapsto -m} T_h^{-\Omega, -m \tilde \omega}.\label{solBhere2}
\eea
where $\tilde{W}'$ and $\tilde{\cZ}'$ are defined in \eqref{Zt}. We checked that \eqref{solBhere} and \eqref{solBhere2} identically agree after using the property \eqref{map1}. This provides a non-trivial cross-check of our formul\ae.

The final gravitational wave flux \eqref{apsi4} is 
\bea
\delta \psi_4(\hat{r} \to \infty) &=&  \frac{M^2}{2 \pi} \int_{-\infty}^{\infty} \d \Omega \sum_{lm} \frac{\tilde{\cZ}(2 \kappa)^{-h}}{\tilde{W} (r_0(r_0 + 2\kappa))^{2}} \cK  (-i \Omega)^{-1+\frac{im \tilde{\omega}}{\kappa}} \nn \\ &\times& \Gamma(1-h-\frac{im \tilde{\omega}}{\kappa}) S_{lm}(\theta)e^{im\hat{\phi}}e^{-i\frac{m + \lambda^{2/3} \Omega}{2M}\hat{u}} \hat{x}^{-1} .
\label{eqn:psi4_marginall}
\eea
Effectively, the integral should be cut above $\Omega \sim \lambda^{-2/3}$ in order to remain in the domain of validity of the asymptotically matched expansion scheme. In order to find the large $\hat u$ behavior of the waveform, we require that $\lambda^{2/3}\frac{\hat{u}}{2M}$ is finite  in the limit $\lambda \rightarrow 0$, $\hat u \rightarrow \infty$. Indeed, the NHEK orbit is only valid for a finite range of retarded time $\Delta \hat u$ which we expect by scaling to be $\Delta \hat u \sim M \lambda^{-2/3}$. Defining the new integration variable
\be\label{fix}
\Phi \equiv \hat \omega \hat u  - \frac{m}{2M} \hat u= \frac{\lambda^{2/3} \hat{u}}{2M}\Omega
\ee
and using \eqref{cK}, the large $\hat u$ behavior of the waveform becomes
\be
	\delta \psi_4(\hat{r} \to \infty) = \sum_{lm}\frac{M^2 \tilde{\cZ}\kappa^{-h}}{\tilde{W} (r_0(r_0 + 2\kappa))^{2}} S_{lm}(\theta)e^{im(\hat{\phi}-\frac{\hat u}{2M})}\hat{x}^{-1} \lambda^{-\frac{2im \tilde{\omega}}{3\kappa}}k_1(\frac{\hat{u}}{2M})^{-h-\frac{im\tilde{\omega}}{\kappa}}I^{l,m,\tilde \omega}	\label{eqn:psi_marginall}
\ee
where the remaining integral is 
\bea
I^{l,m,\tilde \omega} = \frac{\Gamma(1-h-\frac{im \tilde{\omega}}{\kappa})}{2\pi}  \lim_{\hat u \rightarrow \infty}\int^{\frac{\hat u}{2M}}_{-\frac{\hat u}{2M}} d\Phi\frac{ (-i\Phi)^{h-1+\frac{im\tilde{\omega}}{\kappa}}e^{-i\Phi}}{1-(-i\Phi\frac{2M}{\hat{u}})^{2h-1}k_2}\label{ri}
\eea
To find the large time behavior, we should distinguish two subcases:
\begin{itemize}
	\item $\text{Re}(h)>\frac{1}{2}$. We can then neglect the denominator inside the integrand and recognize the integral as an inverse Laplace transform \cite[3.381, p348]{Gradshteyn}\footnote{The $\eps$ prescription with $\eps >0,\, \eps \rightarrow 0$ is required to make the integral well-defined. This defines our prescription for avoiding the branch cut. We assume that this regulator originates from subleading correction in the $\lambda \rightarrow 0$ limit.} 
	\bea
	\frac{1}{2 \pi i}\int_{-i\infty+\eps} ^{i\infty+\eps} ds\, s^{h-1+\frac{i m \tilde \omega}{\kappa}}e^{s}  = \frac{1}{\Gamma(1-h-\frac{i m \tilde \omega}{\kappa})}.\label{i5}
	\eea
	The waveform is then
	\bea
	\delta \psi_4(\hat{r} \to \infty) \hspace{-3pt}&\hspace{-3pt}=&\hspace{-3pt}\sum_{lm}\frac{M^2 \tilde{\cZ}\kappa^{-h}}{\tilde{W} (r_0(r_0 + 2\kappa))^{2}} S_{lm}(\theta)\frac{e^{im(\hat{\phi}-\frac{\hat u}{2M})}}{\hat x} \frac{k_1}{\lambda^{\frac{2im \tilde{\omega}}{3\kappa}}}(\frac{\hat{u}}{2M})^{-h-\frac{im\tilde{\omega}}{\kappa}}
	\label{eqn:largeu_marginalwaveform}
	\eea
	and $|\delta \psi_4| \propto (\frac{\hat{u}}{2M})^{-\text{Re}{(h)}}$ in this regime.
	\item $\text{Re}(h)=\frac{1}{2}$. The integral factor will oscillate as $\hat u$ increases. Now, for $h=\frac{1}{2}+i h_I$ and $ h_I<0 $
	\bea
	|k_2| =  e^{\pi h_I} \frac{\cosh \pi(m+h_I)}{\cosh \pi(m-h_I)} < 10^{-3}
	\eea
	for all modes $l \geq 2$, $m>0$. In particular, the modes $(l,m)=(9,7)$ and $(l,m)=(13,10)$ have $|k_2| \approx 10^{-4}$.  We numerically checked for those cases that $I^{l,m,\tilde \omega} \approx 1$ up to a small ($\approx 10^{-3}$) variation. We deduce that $|\delta \psi_4| \propto (\frac{\hat{u}}{2M})^{-\text{Re}{(h)}}$. 
\end{itemize} 

\begin{figure}[!hbt]
	\centering
	\includegraphics[width=0.45 \textwidth]{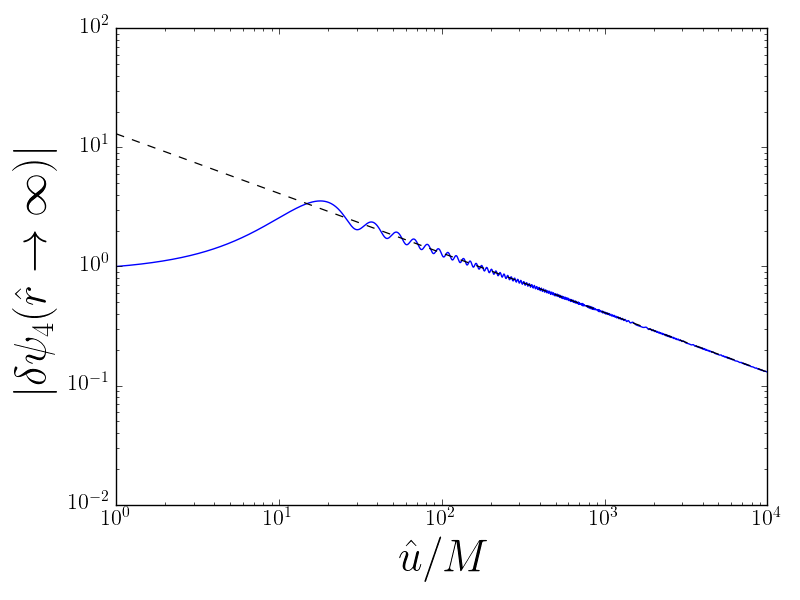}
	\includegraphics[width=0.45 \textwidth]{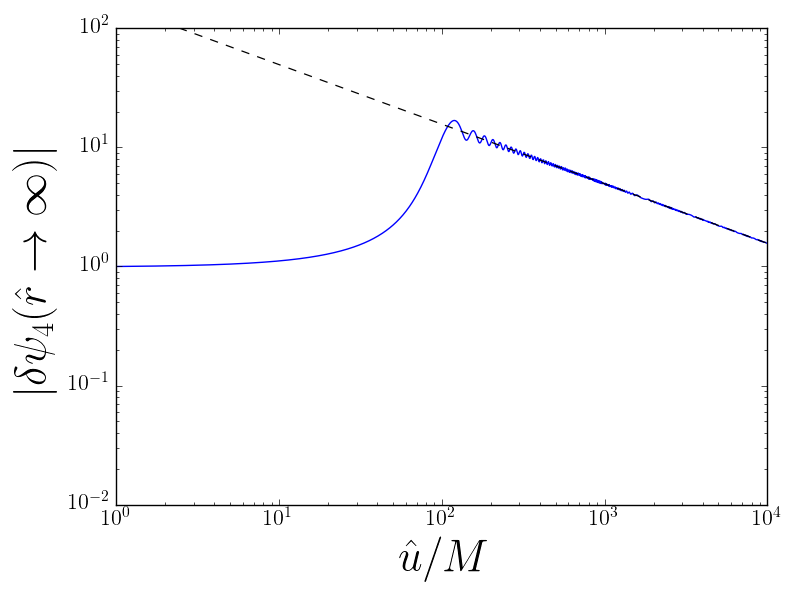}
	\includegraphics[width=0.45 \textwidth]{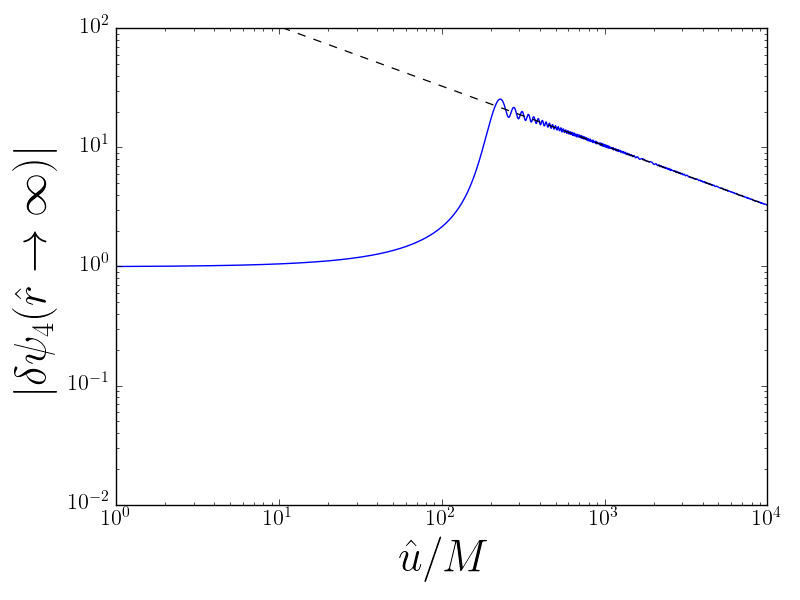}
	\caption{Envelope time-evolution of $\delta \psi_4 (\hat{r} \to \infty)$ (normalized) for the Marginal$(\ell)$ orbits of the modes (top-left) $l=2$, $m=2$; (top-right) $l=50$, $m=38$ and (bottom) $l=50$, $m=50$ with $\lambda =  10^{-3}$ and $\ell = 2 \ell_*$. The late time behavior is given by $(\frac{\hat{u}}{2M})^{-1/2}$ as indicated by the dashed line.}
	\label{fig:marginallevolution}
\end{figure}

Since we assumed $\lambda^{2/3}\frac{\hat{u}}{2M}$ finite we obtain in both cases
\be
|\delta \psi_4| \propto \lambda^0 (\frac{\hat{u}}{2M})^{-\text{Re}{(h)}} \propto \mathcal{O}(\lambda^{\frac{2}{3}\text{Re}{(h)}})
\ee
which is similar to the scaling behavior of the circular NHEK orbit. The modes with the leading amplitude in the near-extremal limit are therefore the modes with $\text{Re}(h)=\frac{1}{2}$ and they fall-off as $1/\sqrt{\hat u}$.\footnote{This result is in tension with a recent numerical simulation \cite{Burko:2016sfi} which claims to obtain a $1/\hat u$ behavior for one example of Marginal$(\ell)$ orbit (in our terminology).} As an illustration, the time evolution of $|\delta \psi_4(\hat{r} \to \infty)|$ (normalized to unity at $\hat u = 1$) is shown for three different modes with $\text{Re}(h)=\frac{1}{2}$  in Figure \ref{fig:marginallevolution}. 

\paragraph{Plunging$(E,\ell)$/Osculating$(E,\ell)$}

 The transformation \eqref{NHEKtonearNHEK} followed by a PT flip $\Phi \mapsto -\Phi$, $T \mapsto -T$ followed by \eqref{barc2}, which connects a Plunging$(E,\ell)$ orbit with a near-NHEK circular orbit requires the following frame transformation
 \bea
&&l^{\mu}\to\frac{-2M^2(1+\cos^2\theta)}{r\bar{R}}\frac{1+\bar{R}\bar{T}+\bar{R}\cot\frac{\zeta}{2}}{-1+\bar{R}\bar{T}+\bar{R}\cot\frac{\zeta}{2}}N^{\mu},\nn\\
&& n^{\mu}\to\frac{r\bar{R}}{-2M^2(1+\cos^2\theta)}\frac{-1+\bar{R}\bar{T}+\bar{R}\cot\frac{\zeta}{2}}{1+\bar{R}\bar{T}+\bar{R}\cot\frac{\zeta}{2}}L^{\mu},\\
&&m^{\mu}\to\frac{1-i\cos\theta}{1+i\cos\theta}\bar{M}^{\mu}\nn %\nn\\&& \bar{m}^{\mu}\to\frac{1+i\cos\theta}{1-i\cos\theta}M^{\mu}
\eea
 where $\zeta$ is related to the orbit parameters as \eqref{EPEl}. It implies that 
 \be
\psi_{(-2)}|_{NHEK}(\bar{R})=\frac{r^2\bar{R}^2}{4M^4}(\frac{-1+\bar{R}\bar{T}+\bar{R}\cot\frac{\zeta}{2}}{1+\bar{R}\bar{T}+\bar{R}\cot\frac{\zeta}{2}})^2\psi_{(2)}|_{near-NHEK}(r).
\ee
The coordinate transformation is given asymptotically by
\bea
r &=& \kappa \bar{R}(\bar{T} \cos{\zeta}-\frac{1}{2}(1-\bar{T}^2)\sin{\zeta})(1+O(\bar{R}^{-1})) ,\\
t &=& -\frac{1}{\kappa} \log{\frac{\bar{T}\cos{\zeta}-\frac{1}{2}(1-\bar{T}^2)\sin{\zeta}}{(\cos{\frac{\zeta}{2}}+\sin{\frac{\zeta}{2}}\bar{T})^2}}(1+O(\bar{R}^{-1})) ,\\
\phi &=& -\bar{\Phi}(1+O(\bar{R}^{-1})). 
\eea
The trajectory ends at $\bar T \rightarrow +\infty$ and starts at $t \rightarrow +\infty$ (because of the PT flip) which is $\bar T = \tan\frac{\zeta}{2}$. 

Repeating the same analysis as in the Marginal$(\ell)$ case, one finds the two equivalent forms
\bea
B_{lm\Omega}&=&\frac{(-2i\kappa\Omega)^{-h}}{\sqrt{2\pi}} \frac{r_0^2(r_0+2\kappa)^2 \tilde{\mathcal Z}'}{4M^4 \tilde W'}|_{m \mapsto -m,s=2} T_h^{\Omega,m \tilde \omega}\\ 
&=& \frac{(-2i\kappa\Omega)^{-h}}{\sqrt{2\pi}} \frac{\tilde{\mathcal Z}}{r_0^2(r_0+2\kappa)^2  \tilde W}|_{s=-2}  T_h^{\Omega,m \tilde \omega}
\eea
with 
\bea
T_h^{\Omega,m \tilde \omega} &=& \int_{\tan \frac{\zeta}{2}}^\infty d\bar{T}e^{i\Omega \bar{T}}(\bar{T}\cos\zeta-\frac{1}{2}(1-\bar{T}^2)\sin\zeta)^{-h-\frac{im\tilde{\omega}}{\kappa}}(\bar{T}\sin\frac{\zeta}{2}+\cos\frac{\zeta}{2})^{\frac{2im\tilde{\omega}}{\kappa}}, \\
&=&  (\cot{\frac{\zeta}{2}})^{-\frac{im\tilde{\omega}}{\kappa}}  \Gamma(1-h-\frac{im\tilde{\omega}}{\kappa})  (-i \Omega)^{h-1}e^{-i \Omega \cot \zeta}  W_{\frac{im\tilde{\omega}}{\kappa},\frac{1}{2}-h}(\frac{-2i \Omega}{\sin\zeta})\nn
\eea
after using \eqref{eqn:integral4} with $p = -i \Omega$ and $y = \frac{im\tilde{\omega}}{\kappa}$. The Weyl scalar in the asymptotically flat spacetime then follows from \eqref{apsi4} and \eqref{cK}
\bea
\delta \psi_4(\hat{r} \to \infty) \hspace{-7pt}&=& \frac{M^2}{2 \pi} \int_{-\infty}^{\infty} \d \Omega \sum_{lm}   \frac{\tilde{\cZ} (2 \kappa)^{-h} }{\tilde{W} (r_0(r_0+2\kappa))^{2}}  S_{lm}(\theta)e^{im\hat{\phi}}e^{-i\frac{m+\lambda^{2/3} \Omega}{2M}\hat{u}}  \nn\\ 
&& \hspace{-15pt} \times ((\cot{\frac{\zeta}{2}})^{-\frac{im\tilde{\omega}}{\kappa}} (-i \Omega)^{-1}  \cK    e^{-i \Omega \cot \zeta}  \Gamma(1-h-\frac{im\tilde{\omega}}{\kappa})  W_{\frac{im\tilde{\omega}}{\kappa},\frac{1}{2}-h}(\frac{-2i \Omega}{\sin\zeta})) \hat{x}^{-1} \nn\\
&=& \sum_{lm}   \frac{M^2 \tilde{\cZ} \kappa^{-h} }{\tilde{W} (r_0(r_0+2\kappa))^{2}}  S_{lm}(\theta)\frac{e^{im\hat{\phi} -i\frac{m}{2M}\hat{u}}}{\hat x} k_1(\frac{\hat{u}}{2M})^{-h}I_{\hat{u},\lambda}^{l,m}.
\label{eqn:psi4_plungingEl}
\eea
The integral can be expressed as an inverse Laplace transform and it can be evaluated explicitly in a similar approximation as in the Marginal$(\ell)$ case using \cite[7.522, p823]{Gradshteyn}
\bea
I_{\hat{u},\lambda}^{l,m}&=& \frac{\Gamma(1-h-\frac{im\tilde{\omega}}{\kappa}) }{2 \pi (\cot{\frac{\zeta}{2}})^{\frac{im\tilde{\omega}}{\kappa}}}\int_{-\infty}^{\infty}d\Phi \frac{(-i\Phi)^{h-1}e^{-i(1+\cot \zeta\frac{2M}{\hat{u}\lambda^{2/3}})\Phi}}{1-(-2iM\Phi/\hat{u})^{2h-1}k_2} W_{\frac{im\tilde{\omega}}{\kappa},\frac{1}{2}-h}(\frac{-4iM\Phi}{\sin\zeta\hat{u}\lambda^{2/3}}) \\
&\approx& \frac{\Gamma(1-h-\frac{im\tilde{\omega}}{\kappa}) }{(\cot{\frac{\zeta}{2}})^{\frac{im\tilde{\omega}}{\kappa}} 2 \pi i} \int_{-i\infty  +\eps}^{i\infty  +\eps} \d s e^{s(1+\cot{\zeta} \frac{2 M}{\hat u \lambda^{2/3}})} s^{h-1} W_{\frac{im\tilde{\omega}}{\kappa},\frac{1}{2}-h}( \frac{4M}{\sin\zeta\hat{u}\lambda^{2/3}} s) \nn\\
&=& (\frac{2 M}{ \cos^2{\frac{\zeta}{2}} \hat u \lambda^{2/3}})^{\frac{i m \tilde{ \omega}}{\kappa}} (1-\frac{2 M}{\hat u \lambda^{2/3}} \tan{\frac{\zeta}{2}})^{-h-\frac{i m\tilde{\omega}}{\kappa}} (1+\frac{\sin{\zeta} \hat u \lambda^{2/3}}{4 M}(1-\frac{2 M}{\hat u \lambda^{2/3}} \tan{\frac{\zeta}{2}}))^{-h+\frac{im\tilde{ \omega}}{\kappa}}.\nn
\eea
The auxiliary parameters $(\zeta,T_0)$  can be traded for the physical NHEK energy $E$ and the initial NHEK time impact parameter $\bar T_0$ via \eqref{EPEl}, which we rewrite for convenience: 
\be
E = \frac{\sqrt{3(\ell^2-\ell_*^2)}}{2}(\sin\zeta+T_0 (\cos\zeta -1)),\qquad
\bar{T}_0= \frac{-\cos\zeta +T_0 \sin\zeta}{\sin\zeta + T_0(\cos\zeta -1)}. 
\ee

\begin{figure}
	\centering
	\includegraphics[width=0.45 \textwidth]{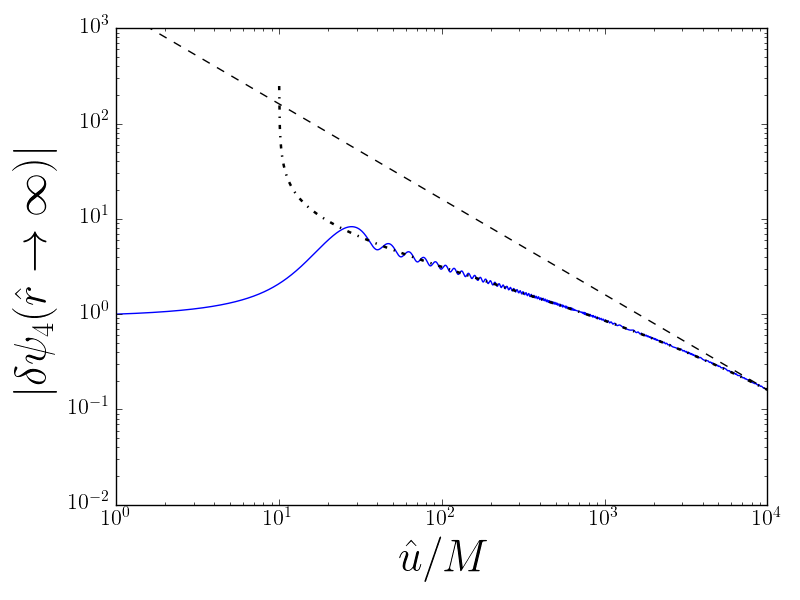}
	\includegraphics[width=0.45 \textwidth]{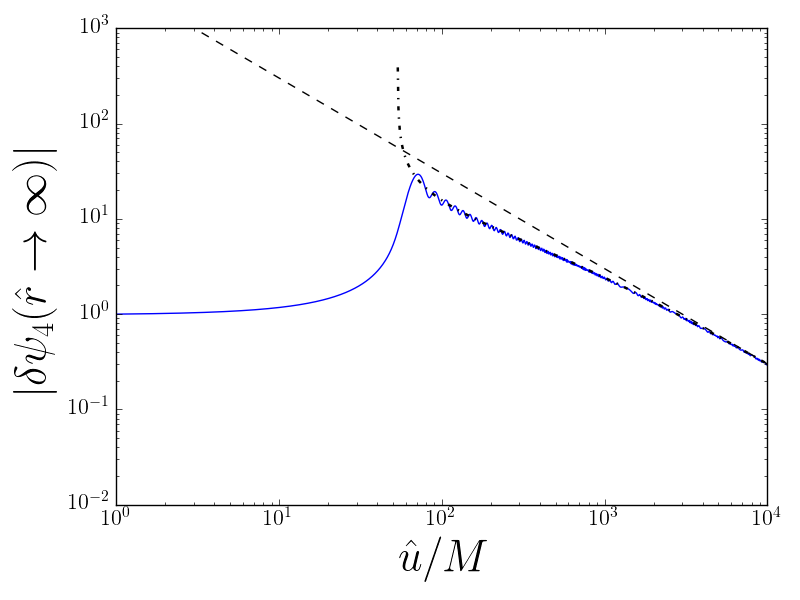}
	\caption{Envelope time-evolution of the (normalized) $l=2$, $m=2$ mode $\delta \psi_4 (\hat{r} \to \infty)$ for the Plunging($E,\ell$) orbits with (left) $E \approx 0.1\frac{3 \ell_*}{2}$ and (right) $E = \frac{3 \ell_*}{4}$, both with $\lambda = 10^{-3}$ and $\ell = 2 \ell_*$. The late time behavior is given by $(\frac{\hat{u}}{2M})^{-1}$ as indicated by the dashed line. The dotted-dashed line gives $\delta \psi_4 (\hat{r} \to \infty)$  using the approximate analytical result for $I_{\hat{u},\lambda}^{l,m}$. }
	\label{fig:plungingElevolution}
\end{figure}

If $\zeta = 0$, or equivalently, $E=0$, $\bar T_0 \rightarrow -\infty$, this matches the result \eqref{eqn:largeu_marginalwaveform} of the Marginal$(\ell)$ case valid for any $\bar T_0$, which is a consistency check. In that case $|\delta \psi_4| \propto (\frac{\hat{u}}{2M})^{-\text{Re}{(h)}}$. For more general energies, the integral contributes an additional factor $(\frac{\hat u \lambda^{2/3}}{2 M})^{-h}$ at large $\hat u \lambda^{2/3}$ such that $|\delta \psi_4| \propto (\frac{\hat{u}}{2M})^{-2\text{Re}{(h)}}$. This is illustrated in Figure \ref{fig:plungingElevolution} for $E \approx 0.1\frac{3 \ell_*}{2}$ and $E = \frac{3 \ell_*}{4}$. The interpolation between the $\hat u^{-h}$ and $\hat u^{-2h}$ polynomial behaviors depending on the source parameters is reminiscent of the near-NHEK signal during the polynomial ringdown phase as expressed in \eqref{generalsourceQNMresponse}. This is also what will be seen explicitly for the Plunging$(e,\ell)$ orbits. The scaling with $\lambda$ is the same as for the Marginal$(\ell)$ orbits
\be
|\delta \psi_4|  \propto \mathcal{O}(\lambda^{\frac{2}{3}\text{Re}{(h)}}).\label{genericNHEK}
\ee

\subsection{Circular NHEK orbit to NHEK orbits}
\label{NtonN}

We are aimed at finding $B$ in \eqref{apsi4} which is determined from the near-horizon physics. The NHEK circular solution is written in \eqref{eqn:circDecomposition}, \eqref{eqn:NHEKsolution}. Since there is only one class of orbits conformally related to the ISCO, namely the Plunging$_*$$(E)$ orbit, we will make all formulae explicit. 

\paragraph{Plunging$_*$$(E)$}

\begin{figure}[!htb]
	\centering
	\includegraphics[width=0.5 \textwidth]{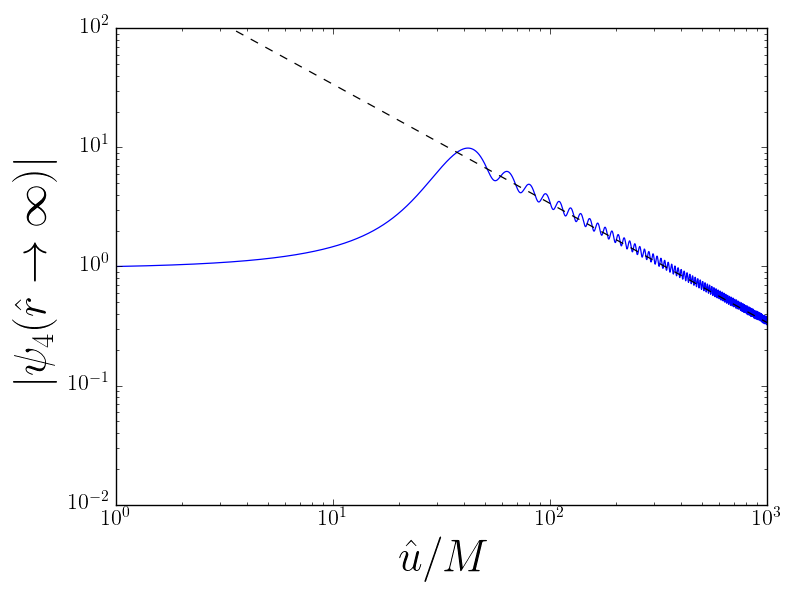}
	\caption{Envelope time-evolution of the (normalized) $l=2$, $m=2$ mode $\delta \psi_4 (\hat{r} \to \infty)$ for the Plunging$_*$($E$) orbits with $\lambda =  10^{-3}$ and $E = 2 \ell_*$. The late time behavior is given by $(\frac{\hat{u}}{2M})^{-1}$ as indicated by the dashed line.}
	\label{fig:plungingEevolution}
\end{figure}

This orbit is related to a circular NHEK orbit by \eqref{NHEKdiff} which requires an accompanying type III tetrad rotation with
\be
F = \frac{\bar{R}^2}{(1+\bar{R} \bar{T})^2} = \frac{1}{\bar T^2}+O(\bar R^{-1}). 
\ee
Asymptotically, the conformal map takes the form
\bea
T &=& - \frac{1}{\bar{T}}, \qquad R= \bar{R} \bar{T}^2 ,\qquad \Phi = \bar{\Phi}	.	
\eea
Even though the final coordinates are barred, we will keep the final frequency $\Omega$ unbarred. The origin of the orbit $\bar T = 0$ corresponds to $T = -\infty$.   Combining these elements with the solution for the circular NHEK orbit results in 
\bea
B &=& \frac{(-2im \tilde{\Omega})^{h}}{(-2i \Omega)^{h}}  \frac{\cZ}{W R_0^{4}} \frac{1}{\sqrt{2\pi}}\int^{\infty}_{0} \d \bar{T} e^{i \Omega \bar{T}} e^{\frac{im \tilde{\Omega}}{\bar{T}}} \bar{T}^{-2h} \nn \\
&=& \frac{2 (-2im\tilde{\Omega})^{\frac{1}{2}}}{\sqrt{2 \pi} (-2 i \Omega)^{\frac{1}{2}}} \frac{\cZ}{W R_0^{4}}   K_{1-2h}(2 \sqrt{-m\Omega \tilde{\Omega}})
\eea
after using \eqref{eqn:integral5} with $p = -i\Omega$ and $y = -im\tilde\Omega$\footnote{We are considering a $+ i \eps$ prescription in both frequencies to be able to use the explicit integral. We then set $\eps=0$.}. The asymptotic curvature perturbation \eqref{apsi4} reads as 
\bea
\delta\psi_4(\hat{r} \to \infty) \hspace{-10pt}&=&  \frac{M^2}{ \pi \hat x} \int_{-\infty}^{\infty} \d \Omega\, \sum_{lm} \frac{ (-2im\tilde{\Omega})^{\frac{1}{2}}}{(-2 i \Omega)^{\frac{1}{2}}} \frac{\cZ \cK}{W R_0^{4}}   K_{1-2h}(2 \sqrt{-m\Omega \tilde{\Omega}})    S_{lm}(\theta)e^{im\hat{\phi}}e^{-i\frac{m+\lambda^{2/3}\Omega}{2 M}\hat{u}}  \nn\\&=& \frac{M^2}{ \hat x}\sum_{lm}\lambda^{-1/3}(\frac{\hat{u}}{2M})^{-1/2-h} (-2im\tilde{\Omega})^{\frac{1}{2}}\frac{\cZ (2)^{h-\frac{1}{2}}k_1}{W R_0^{4}}S_{lm}(\theta)e^{im\hat{\phi}}e^{-i\frac{m}{2 M}\hat{u}} I^{l,m}_{\hat u,\lambda}  \label{eqn:psi_plunging_critical}
\eea
where $\cK$ is defined in \eqref{cK}, $\cZ,W$ in \eqref{defW}, $\tilde\Omega$ in \eqref{deftO}. Using \eqref{fix} and keeping $\hat u \lambda^{2/3}$ fixed we have 
\bea
I^{l,m}_{\hat u,\lambda} & \equiv & \frac{1}{ \pi}\int_{-\infty}^{\infty}d\Phi \frac{(-i\Phi)^{h-1/2}}{1-(-i\frac{2M\Phi}{\hat{u}})^{2h-1}k_2}e^{-i\Phi}K_{1-2h}(2\sqrt{-m\tilde{\Omega}\Phi\frac{2M}{\hat{u}\lambda^{2/3}}})\nn\\&\approx& (-im\tilde{\Omega}\frac{2M}{\hat{u}\lambda^{2/3}})^{h-1/2}e^{im\tilde{\Omega}\frac{2M}{\hat{u}\lambda^{2/3}}},
\eea
where the second step approximates the integral by neglecting the denominator such that one can recognize it as the inverse Laplace transform of \cite[7.629, p836]{Gradshteyn}. It follows that for Plunging$_*$$(E)$ orbits the late time signal has an inverse time behavior. Performing the integral numerically leads to results which are qualitatively very similar to the previous cases. This is illustrated in Figure \ref{fig:plungingEevolution}. In addition, for all NHEK orbits the amplitude of the signal is suppressed by the near-extremality factor $\lambda^{1/3}$.

\subsection{Circular NHEK orbit to near-NHEK orbits}
\label{NtonN2}

%The near-NHEK orbits are expected to become relevant for extremely high spin. %(e.g. $\lambda < 10^{-8}$). 
Let us first consider the near-NHEK orbits related via conformal map to the ISCO in the NHEK spacetime. As previously, we use the barred notation for the final coordinates where the orbit is described but keep $\omega$ unbarred. After mapping a generic near-NHEK orbit of frequency $\omega$ to the ISCO orbit of frequency $\Omega=m \tilde \Omega$ and using \eqref{ccoef}-\eqref{dcoef} and \eqref{cgh4}, we find that the $B$ coefficient in \eqref{eqn:nearpsi4} is given by
\bea
B &=& \frac{\cZ}{R_0^{4}W} (-2i m  \tilde\Omega)^{h} \frac{1}{\sqrt{2 \pi}}  \int_{\bar t_-}^{\bar t_+} \d \bar t  F^h(\bar t) e^{i \omega \bar t - im \tilde\Omega T(\bar t)} .
\eea
where $\bar r = F(\bar t) R$ is the leading order change of coordinate near the boundary $\bar r \rightarrow \infty$ and $\bar t_\pm$ are the initial and final times of the trajectory which are deduced from the change of coordinates from the ISCO. 

There are 2 such classes of orbits: Plunging$_*$$(e=0)$ and Plunging$_*$$(e)$ as described in Section \ref{sec:map3}, which we consider sequentially in what follows.

\paragraph{Plunging$_*$$(e=0)$}

 In the limit $\bar r \to \infty$ with $\bar t$ and $\bar \phi$ fixed the transformation \eqref{NHEKtonearNHEK} relating a circular orbit in NHEK to a Plunging$_*$(e=0) orbit \eqref{Plungingstare0}  becomes
\bea
R &=& \frac{e^{\kappa \bar t}}{\kappa}\bar r(1+O(\bar r^{-1}))  ,\\
T &=& - e^{-\kappa \bar t} +O(\bar r^{-1}), \\
\Phi &=& \bar\phi + O(\bar r^{-1}).
\eea
The trajectory starts at $\bar t \rightarrow -\infty$ and ends at $\bar t \rightarrow +\infty$. The matching of NHEK \eqref{KinNHEK} and near-NHEK \eqref{KinnearNHEK}  tetrads requires a type III rotation with 
\be
F(\bar t)= \frac{\bar r}{R} =\kappa e^{-\kappa \bar t}.
\ee
%Using the map of parameters \eqref{mapp1} 
We find  
\bea
B &=&\frac{\cZ}{R_0^{4}W} (-2i m \tilde \Omega)^{h}  \frac{1}{\sqrt{2 \pi}}  \int_{-\infty}^{\infty} \d \bar t (\kappa e^{-\kappa \bar t})^h e^{i \omega \bar t+im\tilde \Omega e^{-\kappa \bar t}} \\ &=& \frac{1}{\sqrt{2 \pi}} 2^{h-i\frac{\omega}{\kappa}}\kappa^{h-1} \frac{\cZ}{R_0^{4}W} (-2im\tilde \Omega)^{i\frac{\omega}{\kappa}} \Gamma(h-i\frac{\omega}{\kappa})
\eea
after using the integral \eqref{eqn:integral1} in Appendix \ref{formulae_integrals}. This now fixes the Weyl scalar in the asymptotically flat spacetime. 

In order to obtain the late time behavior, we can make the quasi-normal mode approximation and the resulting Weyl scalar is \eqref{eqn:genQNM}. The factor $\Gamma(h-i\frac{\omega}{\kappa})$ exactly cancels $1/\Gamma(h-i(n-m))$. The overtone sum can be performed explicitly using 
\be
\sum_N \frac{(-1)^N}{N! \, 2^N} (-2im\tilde{\Omega})^{N+h} e^{-i\frac{m-i\lambda(N+h)}{2M}\hat{u}} =  e^{-i\frac{m-i\lambda h}{2M}\hat{u}+im \tilde{\Omega} e^{-\frac{\lambda \hat{u}}{2M}}} (-2im\tilde{\Omega})^h .
\ee
The final answer in the quasi-normal mode approximation is
\be
\delta \psi_4 (\hat{r} \to \infty) =  \sum_{l, m}    M^2 \lambda^h  k_1 \frac{\cZ}{R^{4}_0 W} (-2im\tilde{\Omega})^{h}    \\   e^{-i\frac{m-i \lambda h}{2M}\hat{u}+im\tilde{\Omega} e^{-\frac{\lambda \hat{u}}{2M}}}  e^{im\hat{\phi}} S_{lm}(\theta) \hat{x}^{-1}. \nn
\label{eqn:plunging(e=0)QNMpsi4}
\ee
The dominant late time behavior is now 
\bea
|\delta \psi_4(\hat{r} \to \infty)| &\propto& \lambda^{1/2} e^{-\frac{\lambda \hat u}{4M}}.
\eea

\paragraph{Plunging$_*$(e)} Similarly, the Plunging$_*$(e) orbits \eqref{Plungingstare} are in the same equivalence class as the NHEK circular orbits. It is described as \eqref{NHEKtonearNHEK} followed with \eqref{tautr}. The map of parameters is given in \eqref{mapp2}. For general $\zeta$, the transformation is 
\bea
R&=&-\frac{\bar{r}}{\kappa}(\cos\zeta-\sin\zeta\sinh\kappa\bar{t})T^{-1}+o(\bar r),\nn\\
T&=&-\sqrt{\frac{\cosh\kappa\bar{t}-\cos\zeta\sinh\kappa\bar{t}-\sin\zeta}{\cosh\kappa \bar{t}+\cos\zeta\sinh\kappa\bar{t}+\sin\zeta}}+o(\bar r^0) \equiv T(\bar t)+o(\bar r^0) ,\nn\\
\Phi&=&\bar{\phi}+o(\bar r^0)\nn
\eea
We find a type III rotation with 
\bea
F(\bar{r},\bar{t})&=&\frac{\sqrt{\bar{r}(\bar{r}+2\kappa)}}{R}\frac{\sqrt{\bar{r}(\bar{r}+2\kappa)}\sin\zeta \sinh \kappa\bar{t}-(\bar{r}+\kappa)\cos\zeta+\kappa}{(\bar{r}+\kappa)\sin\zeta\sinh\kappa \bar{t}-\sqrt{\bar{r}(\bar{r}+2\kappa)}\cos\zeta+\kappa\sin\zeta\cosh\kappa \bar{t}},\nn\\
&=& \frac{-\kappa\,  T(\bar t)}{\cos\zeta - \sin\zeta \sinh \kappa \bar t} +o(\bar r^0).
\eea
Then
\bea
B&=& \frac{1}{\sqrt{2 \pi}}\kappa^{h}(-2im\tilde{\Omega})^{h}\frac{\cZ}{R^{4}_0 W} I^{l,m}_{\omega}
\eea
where the remaining integral is 
\bea
I^{l,m}_{\omega}&=&\int^{\infty}_{\frac{\log(-\tan\frac{\zeta}{2})}{\kappa}}\d \bar{t} e^{i \omega \bar{t}} e^{im\tilde{\Omega}\sqrt{\frac{\cosh\kappa\bar{t}-\cos\zeta\sinh\kappa\bar{t}-\sin\zeta}{\cosh\kappa \bar{t}+\cos\zeta\sinh\kappa\bar{t}+\sin\zeta}} }\nn \\
&& \times ( \frac{1 }{\cos\zeta-\sin\zeta\sinh\kappa\bar{t}} \sqrt{\frac{\cosh\kappa\bar{t}-\cos\zeta\sinh\kappa\bar{t}-\sin\zeta}{\cosh\kappa \bar{t}+\cos\zeta\sinh\kappa\bar{t}+\sin\zeta}} )^{h}\nn\\
		&=&2^h\kappa^{-1}(-\tan\frac{\zeta}{2})^{\frac{i\omega}{\kappa}}(-\csc\zeta)^h e^{-im\tilde{\Omega}\tan\frac{\zeta}{2}}\Gamma(h-i\frac{\omega}{\kappa})U(h-i\frac{\omega}{\kappa},2h,2im\tilde{\Omega}\csc\zeta)\nn \\ 
		&=&2^{h}\kappa^{-1}(-2 i m \tilde \Omega)^{-h}\Gamma(h-\frac{i \omega}{\kappa})(-\tan\frac{\zeta}{2})^{\frac{i\omega}{\kappa}}e^{im\tilde{\Omega}\cot\zeta}W_{\frac{i\omega}{\kappa},h-\frac{1}{2}}(2im\tilde{\Omega}\csc\zeta).
\eea
In the second line, we performed the change of variables $\bar{t}=\frac{1}{\kappa}\log(-\frac{(1+x)\tan\frac{\zeta}{2}}{x})$, which allowed to recognize the integral representation of the hypergeometric $U$ function. We then used the definition of the Whittaker $W$ function, $W_{a,b}(z)=e^{-\frac{z}{2}}z^{b+\frac{1}{2}}U(b-a+\frac{1}{2},1+2b,z)$. By using the QNM approximation \eqref{eqn:genQNM}, we find the waveform to be 
\bea
\delta \psi_4(\hat{r} \to \infty) &=&  \sum_{lm} (-im \tilde{\Omega})^{h} M^2 \lambda^h k_1 \frac{\cZ}{R^{4}_0 W}S_{lm}(\theta)e^{im\hat{\phi}}e^{-i\frac{m-i \lambda h}{2 M} \hat{u}}e^{-im\tilde{\Omega}\tan\frac{\zeta}{2}}\hat{x}^{-1}\nn\\&&\hspace{-10pt}\times(\tan\frac{\zeta}{2}\csc\zeta)^h(\frac{1+\tan\frac{\zeta}{2}e^{-\lambda\hat{u}/(2M)}}{2})^{-2h}e^{2im\tilde{\Omega}\csc\zeta \frac{\tan\frac{\zeta}{2}e^{-\lambda\hat{u}/(2M)}}{1+\tan\frac{\zeta}{2}e^{-\lambda\hat{u}/(2M)}}}\label{gen77}
\eea
after resumming the QNM using \eqref{oneW}. The dictionary with physical orbit parameters is given in \eqref{mapp2}.

In particular, in the case where $\zeta=-\frac{\pi}{2}$, we have more simply 
\bea
B &=& \frac{1}{\sqrt{2 \pi}}(\frac{\kappa}{2})^{h}(-2im\Omega)^{h}\frac{\cZ}{R^{4}_0 W} \int^{\infty}_{0} \d \bar t e^{i \omega \bar t} (\sinh{\frac{\kappa \bar t}{2}})^{-2h}e^{im\Omega \coth{\frac{\kappa \bar t}{2}}}\nn\\
&=& \frac{2^{h}}{\sqrt{2 \pi}}\kappa^{h-1}\frac{\cZ}{R^{4}_0 W} \Gamma(h-\frac{i \omega}{\kappa})W_{\frac{i\omega}{\kappa},-h{{+}}\frac{1}{2}}(-2im\tilde{\Omega})\label{Bql4}
\eea
after using \eqref{eqn:integral2} in Appendix \ref{formulae_integrals} with $p=-i\omega/\kappa$, $z = \kappa \bar t$ and $y = -im \tilde{\Omega}$. We can then compute the overtone sum explicitly using \eqref{oneW} to obtain
\be
\delta \psi_4(\hat{r} \to \infty) =   \sum_{lm} (-im \tilde{\Omega})^{h} k_1 \frac{\lambda^h  M^2 \cZ}{R^{4}_0 W}  S_{lm}(\theta)e^{im\hat{\phi}}  \hat{x}^{-1}  (\frac{1-e^{-\frac{\lambda \hat{u}}{2M}}}{2})^{-2h} e^{-i\frac{m-i \lambda h}{2 M} \hat{u}} e^{im\tilde{\Omega} \coth{\frac{\lambda \hat{u}}{4 M}}} 
\label{eqn:plunging(e)QNMpsi4}
\ee
consistently with \eqref{gen77} when $\zeta=-\frac{\pi}{2}$. The dominant late time behavior is now
\bea
|\delta \psi_4(\hat{r} \to \infty)| &\propto& \frac{\lambda^{1/2}}{\sinh \frac{\lambda \hat{u}}{4 M}} .
\eea
%Consistently with the dictionary \eqref{mapp2}, distinct $\zeta$'s just amounts to shifting $\hat u$. % {\geo Why $e$ does not matter?}

\subsection{Circular near-NHEK orbit to near-NHEK orbits}
\label{nHnH}

There is only one set of orbits to consider. 

\paragraph{Osculating$(e,\ell)$/Plunging$(e,\ell)$} The transformation \eqref{eq:1} which maps a Plunging$(e,\ell)$ orbit to a near-NHEK circular orbit behaves asymptotically as

\bea
r &=& \bar{r}(\sinh{\kappa \bar t} + \chi (\cosh{\kappa \bar t}-1))(1+O(\bar{r}^{-1})), \\
t &=& -\frac{1}{\kappa} \log{(\coth{\frac{\kappa \bar{t}}{2}}+\chi)}(1+O(\bar{r}^{-1})), \\
\phi &=& \bar{\phi}(1+O(\bar{r}^{-1}))
\eea
where $\chi$ is related to the orbit parameters as \eqref{defechi}. To match to the asymptotically flat space in the Kinnersley tetrad, this transformation needs to be accompanied by type III rotation characterized by 
	\be
	F=\frac{\sqrt{\bar{r}(\bar{r}+2\kappa)}}{-\chi \sqrt{\bar{r}(\bar{r}+2\kappa)}+(\kappa+(\kappa+\bar{r})\chi)\cosh\kappa\bar{t}+(\kappa+\bar{r}+\kappa\chi)\sinh\kappa\bar{t}}. 
	\ee
Together with  the near-NHEK circular solution \eqref{eqn:nearRadialSolution} and \eqref{eqn:nearNHEKmatching} this leads to 
\be
B = \frac{1}{\sqrt{2 \pi}}\frac{\tilde{\cZ}}{(r_0(r_0+2\kappa))^2\tilde{W}} \int^{\infty}_{0} \d \bar{t} e^{i \omega \bar{t}} (\coth{\frac{(\kappa \bar{t})}{2}}+\chi)^{\frac{im\tilde{\omega}}{k}}(\sinh{(\kappa \bar{t})}+\chi(\cosh{(\kappa \bar{t})-1}))^{-h}.
\ee
For $\chi > -1$ it becomes, using \eqref{eqn:integral3} from Appendix \ref{formulae_integrals} with $p=-i\omega/\kappa$, $z = \kappa \bar{t}$ and $y = -im \tilde{\omega}/\kappa$,
\bea
B &=& \frac{1}{\sqrt{2 \pi}}\frac{\tilde{\cZ}}{(r_0(r_0+2\kappa))^2\tilde{W}} \kappa^{-1}(1+\chi)^{\frac{im\tilde{\omega}}{\kappa}-h}2^{h} \\ &\times& B(h-\frac{i \omega}{\kappa}, 1-h-\frac{im \tilde{\omega}}{\kappa}) {}_2F_1(h-\frac{im\tilde{\omega}}{\kappa}, h-\frac{i \omega}{\kappa},1-\frac{im\tilde{\omega}}{\kappa}-\frac{i \omega}{\kappa}, -\frac{1-\chi}{1+\chi} ) \nn
\eea
where $B(x,y)=\Gamma(x)\Gamma(y)/\Gamma(x+y)$ is the beta function. 

In the QNM mode approximation, the overtone sum 
\bea
&&\hspace{-1cm}\sum^{\infty}_{N=0} \frac{(-1)^Ne^{-i\frac{m-i\lambda(N+h)}{2M}\hat{u}}}{N!} \frac{\Gamma(1-h-\frac{im \tilde{\omega}}{\kappa})}{\Gamma(1-h-\frac{im \tilde{\omega}}{\kappa}-N)}  {}_2F_1(h-\frac{im\tilde{\omega}}{\kappa}, -N,1-\frac{im\tilde{\omega}}{\kappa}-N-h, -\frac{1-\chi}{1+\chi}) \nn \\  &=& e^{-i\frac{m-i\lambda h}{2M}\hat{u}}(1+\frac{1-\chi}{1+\chi} e^{-\frac{\lambda\hat{u}}{2M}})^{-h+\frac{im \tilde{\omega}}{\kappa}}(1-e^{-\frac{\lambda \hat{u}}{2M}})^{-h-\frac{im \tilde{\omega}}{\kappa}}	
\eea
is computed by \eqref{eqn:sum3first} with
\bea
c_+ &=& h+\frac{im\tilde{\omega}}{\kappa} , \qquad c_- = h-\frac{im \tilde{\omega}}{\kappa} , \qquad z = -\frac{1-\chi}{1+\chi} ,\qquad x = e^{-\frac{\lambda\hat{u}}{2M}}.
\eea
resulting in
\bea
\delta \psi_4(\hat{r} \to \infty) &=&    \sum_{lm}  M^2 2^{h} \lambda^h \kappa^{-h} k_1 \frac{\tilde{\cZ}}{(r_0(r_0+2\kappa))^2\tilde{W}}  S_{lm}(\theta)e^{im\hat{\phi}} \hat{x}^{-1} e^{-i\frac{m-i\lambda h}{2M}\hat{u}} \nn \\ &&\times   (1+\chi+(1-\chi) e^{-\frac{\lambda \hat{u}}{2M}})^{-h+\frac{im \tilde{\omega}}{\kappa}}(1-e^{-\frac{\lambda\hat{u}}{2M}})^{-h-\frac{im \tilde{\omega}}{\kappa}}.
\label{eqn:plunging(e,l)QNMpsi4}
\eea
The metric perturbation can also be obtained easily in the limit $\lambda \rightarrow 0$ by double integration over $\hat u$. Indeed, the QNM frequency $\hat \omega \approx \frac{m}{2M}$ so from \eqref{eqn:asymptoticmetricperturbations} and because the evolution timescale becomes exceedingly long with respect to the oscillation timescale we can just multiply the expression for each mode $m$ by $-2 (2M)^2/m^2$ to obtain
\bea
(h_+ - i h_\times)|_{\hat{r} \to \infty} &=&  -  \sum_{lm}  \frac{ 2^{h+3}}{m^2 \kappa^h} k_1 \frac{M^4 \lambda^{h}\tilde{\cZ}}{(r_0(r_0+2\kappa))^2\tilde{W}}  S_{lm}(\theta)e^{im\hat{\phi}} \hat{x}^{-1} e^{-i\frac{m-i\lambda h}{2M}\hat{u}} \nn \\ &&\times   (1+\chi+(1-\chi) e^{-\frac{\lambda \hat{u}}{2M}})^{-h+\frac{im \tilde{\omega}}{\kappa}}(1-e^{-\frac{\lambda\hat{u}}{2M}})^{-h-\frac{im \tilde{\omega}}{\kappa}}.
\label{eqn:plunging(e,l)QNMh}
\eea
where we fixed the two integration constants (proportional to $\hat u^0$ and $\hat u^1$) so that $(h_+ - i h_\times)|_{\hat{r} \to \infty} \rightarrow 0$ at $\hat u \rightarrow \infty$. 

The dominant late time behavior is now
\bea
|\delta \psi_4(\hat{r} \to \infty)| &\propto& \frac{\lambda^{1/2}}{(\sinh \frac{\lambda \hat{u}}{4 M})^{1/2}(\cosh \frac{\lambda \hat{u}}{4 M}+\chi \sinh \frac{\lambda \hat{u}}{4 M})^{1/2}} .
\eea
 The perturbation is better expressed in terms of the physical quantities: $\hat E$, the energy of the plunging body in the asymptotic frame and $\ell$ its angular momentum. Using \eqref{defechi} and \eqref{defE}, we get 
\bea
\chi = \frac{2(2 M \hat E - \ell)}{\lambda \sqrt{3(\ell^2 - \ell^2_*)}},\qquad \frac{1}{\kappa_0}\equiv \frac{r_0}{\kappa}= \frac{2\ell}{\sqrt{3(\ell^2-\ell^2_*)}}-1. \label{parchi}
\eea
When expressed as a function of $\kappa_0$, the perturbation \eqref{eqn:plunging(e,l)QNMh} is independent of $r_0$, which cancels out. 

\subsection{Summary and comparison}
\label{intsum}

In this section we have computed the coefficient $B$ that enters in the expressions \eqref{apsi4} and \eqref{eqn:nearpsi4} for the asymptotic curvature perturbation sourced by a massive probe object on a generic corotating equatorial orbit in NHEK or near-NHEK. This solves the problem of gravitational wave emission from such sources to leading order in the extremality asymptotic matched expansion scheme, and in the frequency domain, for which the matching condition \eqref{MATCH} holds. From this we also have computed the corresponding GW signal in the time-domain, for sufficiently late times, for sources in both NHEK and near-NHEK.

\begin{figure}[!hbt]
	%\centering
	\vspace{-1cm}
	\includegraphics[width=.94 \textwidth]{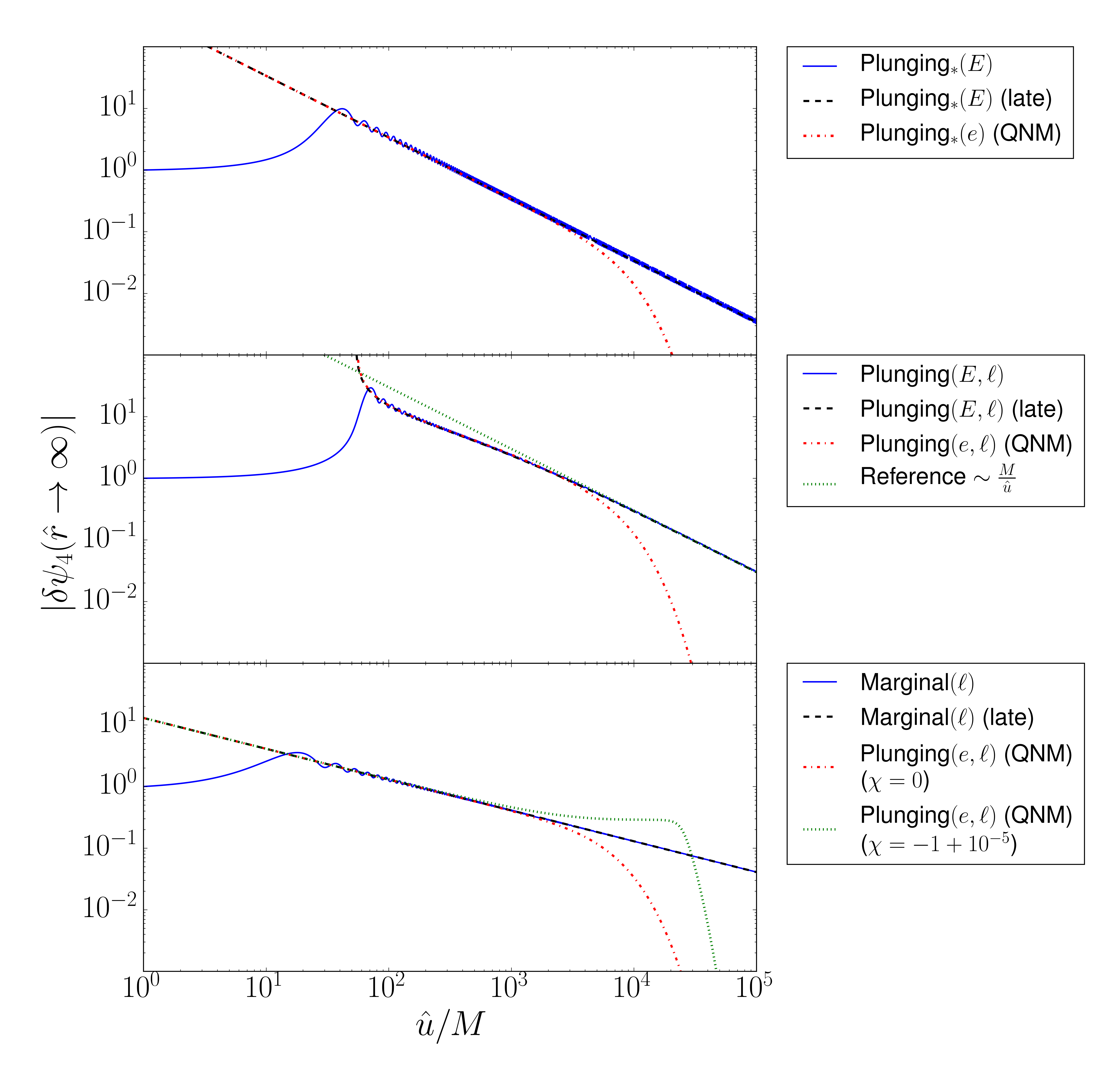}\hspace{-3cm}\vspace{-0.2cm}
	\caption{Envelope time evolution of the $l=2$, $m=2$ asymptotic curvature perturbation $|\delta \psi_4(\hat r \to \infty)|$ (artificially normalized to $|\delta \psi_4(\hat r \to \infty)|_{|\hat{u}=M}$=1), sourced by probe objects moving on a number of different geodesics in the near-horizon region of a nearly extremal black hole. We have taken $\lambda = 10^{-3}$ in all panels. The solid (blue) lines show the time evolution corresponding to (from top to bottom) the NHEK orbits \eqref{eqn:psi_plunging_critical} ($E = 2 \ell_*$), \eqref{eqn:psi4_plungingEl} ($E = \frac{3 \ell_*}{4}$, $\ell = 2\ell_*$) and \eqref{eqn:psi_marginall} ($\ell = 2 \ell_* $). The dashed (black) lines are late time approximations to these expressions. The dashed-dotted (red) lines represent the curvature perturbations due to analogous near-NHEK orbits in the QNM approximations (\eqref{eqn:plunging(e)QNMpsi4} and \eqref{eqn:plunging(e,l)QNMpsi4}). The dotted (green) line finally indicates the $\hat u^{-1}$ behavior (central panel) to which the signal from generic Plunging$(E,\ell)$ orbits asymptotes, and (bottom panel) the (QNM) response to a different Plunging$(E,\ell)$ orbit with a $\hat{u}^{-1/2}$ fall-off. The near-NHEK orbits transition from polynomial to exponential decay at times of order $\hat u \sim M\lambda^{-1}$.\vspace{-0.8cm}}
	\label{fig:summary_time_evolution}
\end{figure}

In Figure \ref{fig:summary_time_evolution} we combine the different features which together govern the qualitative time evolution of individual modes of the gravitational perturbation $|\delta \psi(\hat r \to \infty)|$ in the asymptotically flat domain, while ignoring the overall normalization constant which will be discussed in Section \ref{Seccrit}. Shown in Figure \ref{fig:summary_time_evolution} is the normalized time evolution of the $l=2$, $m=2$ mode for a number of representative examples of orbits in NHEK and near-NHEK. One sees that once the peak amplitude is reached, the overall evolution is almost immediately captured by the late times approximate waveforms we discussed, except for an additional amplitude modulation, which we could not integrate analytically. 

One typical signature is the polynomial decay of the signal. The signals from the Marginal($\ell$) and Plunging$_*(E)$ NHEK orbits decay respectively as $\hat{u}^{-1/2}$ and $\hat{u}^{-1}$. The signal from more general Plunging($E,\ell$) orbits generically interpolates between an initial $\hat{u}^{-1/2}$ behavior and a final $\hat{u}^{-1}$ decay, where the $\hat{u}^{-1/2}$ behavior becomes more and more pronounced as $E \to 0$. The polynomial ringdown stages of the near-NHEK counterparts to these NHEK signal are, at least on this level of single mode envelope evolution, completely analogous. For $e \to 0$, the Plunging${(e,\ell)}$ signal decays like $\hat{u}^{-1/2}$ for the entire range $-\frac{\kappa}{2}\sqrt{3(\ell^2-\ell_*^2)} < e <0$ while the Plunging$_*(e)$ orbit behaves as $\hat{u}^{-1}$. More general Plunging${(e,\ell)}$ orbits give rise to polynomial ringdowns which exactly match the functional time dependence of the Plunging($E,\ell$) NHEK orbits, going from $\hat{u}^{-1/2}$ to $\hat{u}^{-1}$. The matching of parameters between both cases is qualitatively given by $\lambda^{1/3} \chi \sim \sin{\zeta}$. 

Figure \ref{fig:summary_time_evolution} shows that the near-NHEK signals transition from polynomial to exponential decay at times of order $\hat u \sim M \lambda^{-1}$. This corresponds to the timescale of the lowest zero-damped QNM given in \eqref{scaling}. At this point, the NHEK and near-NHEK results diverge from each other. This is precisely what one expects because on such timescales, the spacings between the zero-damped QNMs are resolved. Therefore, the corresponding frequency differences make an important contribution to the signal. Since the NHEK limit ignores these differences the NHEK waveforms become unreliable at such late times. 

In the next section we turn to the overall amplitudes of the GW signals in the asymptotically flat domain.

\section{Critical behavior}
\label{Seccrit}

The critical angular momentum per unit probe mass $\ell_* = \frac{2}{\sqrt{3}}M$ has taken on a special role throughout our computations. It is time to explicitly show that critical behavior occurs at and around $\ell = \ell_*$. We have seen that exactly two representative orbits under (complexified) $SL(2,\mathbb R) \times U(1)$ conformal transformations control the physics of probe orbits in the near-horizon region of nearly extremal Kerr: the circular orbit in NHEK which is critical, with $\ell = \ell_*$, and the family of circular orbits in near-NHEK which are supercritical, $\ell > \ell_*$. 

The overall amplitude of the supercritical plunging orbits discussed in Sections \ref{sec:toNHEKo} and \ref{nHnH} depend on the following coefficients
\bea
\mathcal A^{\ell > \ell_*}_{lm} = \frac{8M^5 k_1 \tilde {\mathcal Z}}{m^2 \kappa^h m_0 r_0^2 (r_0+2 \kappa)^2 \tilde {W}}  . 
\eea
However one can verify that $M$, $m_0$ and $r_0$ (but not $\kappa/r_0$) explicitly drop out in this expression, leaving us with coefficients that depend on the physical impact parameters of the orbit only; the energy, angular momentum and initial plunging time. 

Similarly, the overall amplitude of the critical plunging orbits discussed in Sections \ref{NtonN} and \ref{NtonN2} depend upon the overall coefficients 
\bea
\mathcal A^{\ell = \ell_*}_{lm}=  \frac{8 M^5 k_1  {\mathcal Z}}{m^2 m_0 R_0^4 W}(-2 i m  \tilde \Omega/R_0)^{h} , \label{coef2}
\eea
which likewise involve the physical impact parameters of the orbit only\footnote{We remind the reader that in all Figures in the previous section, these overall coefficients were artificially suppressed by a unit normalization of the amplitude.}. We now derive the critical behavior associated with the plunging orbits on a case by case basis, by explicitly evaluating these coefficients. 

The gravitational wave signal emitted by plunging probes strongly depends upon the $\theta$ incidence angle. We consider the two borderline cases: face-on ($\theta=0$) and edge-on ($\theta=\frac{\pi}{2}$). We take for simplicity $\hat \phi = 0$. 

In the face-on case, the leading signal comes from the $m=l=2$ harmonic\footnote{This is because the higher (spin 2) spheroidal harmonics $S_{lm}(0)$ are highly suppressed: the next-to-leading terms are $S_{33}(0)\simeq 10^{-10}$, $S_{44}(0)\simeq 10^{-20}$, \dots}. The generic plunging supercritical near-NHEK orbits denoted as Plunging$(e,\ell)$ leads to the metric perturbation \eqref{eqn:plunging(e,l)QNMh} consisting of a polynomial ringdown followed by an exponential decay. All in all, the dominant behavior for a generic orbit is given by 
\bea
\vert h_+ + i h_\times\vert = \mathcal A_{FO}(\frac{\ell}{\ell_*}) \frac{  m_0}{D}  \frac{\sqrt{\lambda}}{(\sinh \frac{\lambda \hat{u}}{4 M})^{1/2}(\cosh \frac{\lambda \hat{u}}{4 M}+\chi \sinh \frac{\lambda \hat{u}}{4 M})^{1/2}} \label{FO5}
\eea
where $D=M \hat x=r$ is the luminosity distance to the source, $m_0$ is the mass of the probe and $\mathcal A_{FO}(\frac{\ell}{\ell_*}) = 2^{-1/2}| \mathcal A_{22}^{\ell > \ell_*} |S_{22}(0)$ is the residual numerical coefficient which only depends upon the ratio $\ell/\ell_*$.

\begin{figure}[!htb]
	\centering
	\includegraphics[angle=270,width=0.50 \textwidth]{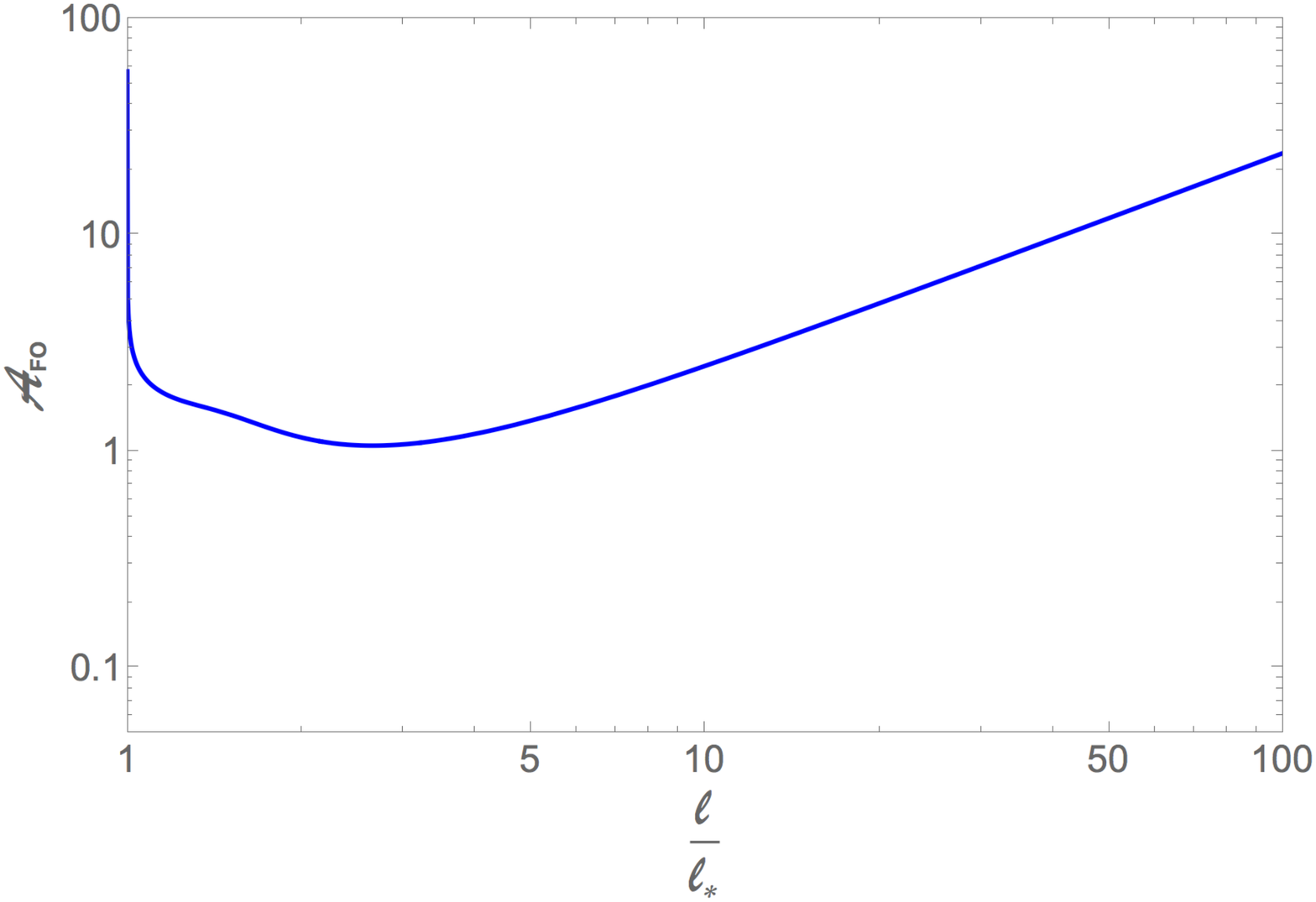}
	\caption{Amplitude coefficient (face-on) exhibits critical behavior in the limit $\ell \rightarrow \ell_*$} \label{fig:AFO}
\end{figure}

We obtain numerically for close to critical and large orbital angular momentum, 
\bea
\mathcal A_{FO}(\frac{\ell}{\ell_*} \rightarrow 1) &=& 1.1 (\frac{\ell}{\ell_*}-1)^{-\frac{1}{4}},\label{critAM}\\
\mathcal A_{FO}(\frac{\ell}{\ell_*} \rightarrow \infty) &=& 0.2 \frac{\ell}{\ell_*} .\label{critAM2}
\eea
The minimum value of the amplitude occurs around $\ell = 2.6 \ell_*$ where $\mathcal A_{FO}=1.1$. The function $\mathcal A_{FO}(\frac{\ell}{\ell_*})$ is plotted in Figure \ref{fig:AFO}. The behavior in the limit $\ell \rightarrow \ell_*$ is the critical behavior that we anticipated above, with critical exponent $-1/4$. 

If the near-NHEK energy $e$ is finite and non-zero around the critical point $\ell \rightarrow \ell_*$, $\chi$ in \eqref{FO5} diverges (see \eqref{defechi}). The second term in the denominator in \eqref{FO5} then dominates and diverges, and exactly cancels the divergence in $\mathcal A_{FO}$. This yields 
\bea \label{FO}
\lim_{\ell \rightarrow \ell_*,\, e \text{ finite}}\vert h_+ + i h_\times\vert = 1.2 \frac{m_0}{D} \sqrt{\frac{\kappa \ell_*}{e}} \frac{\sqrt{\lambda}}{\sinh \frac{\lambda \hat u}{4M}}.
\eea
Again, we find critical behavior in the limit $\ell \rightarrow \ell_*$ and $e \rightarrow 0$. 
For the special case $\hat E = \frac{\ell}{2M}$, or equivalently $\chi = 0$, the near-NHEK energy $e=0$ and \eqref{FO5} gives the critical behavior
\bea \label{FOcrit}
\lim_{\ell \rightarrow \ell_*, e=0}\vert h_+ + i h_\times\vert = 1.6 (\frac{\ell}{\ell_*}-1)^{-\frac{1}{4}} \frac{  m_0}{D}  \frac{\sqrt{\lambda}}{\sqrt{\sinh \frac{\lambda \hat{u}}{2 M} }} .
\eea
As expected from a critical system, the asymptotic behavior close to the critical point depends on how the critical point is approached.

In the NHEK region, the waveform of the Plunging$(E,\ell)$ orbits has the dominant behavior
\be
\vert h_+ + i h_\times\vert = \mathcal A^{NHEK}_{FO}(\frac{\ell}{\ell_*})	
 \frac{m_0}{D} \frac{\lambda^{\frac{1}{3}}}{\sqrt{|(\frac{\hat u\lambda^{2/3}}{4M}\cos\frac{\zeta}{2}-\frac{1}{2}\sin\frac{\zeta}{2})(\cos\frac{\zeta}{2}+\sin\frac{\zeta}{2}\frac{\hat u\lambda^{2/3}}{2M})|}}
\ee
where $\mathcal A^{NHEK}_{FO}(\frac{\ell}{\ell_*})=2^{-1/2}| \mathcal A_{22}^{\ell > \ell_*} |S_{22}(0)$. The overall amplitude is numerically equal to  (\ref{critAM}) and (\ref{critAM2}). We can relate $\zeta$ to impact parameters using \eqref{EPEl}. The amplitude depends upon the NHEK energy $E$ and the initial value $\bar T_0$.

Now, for $\ell = \ell_*$, the plunging orbits belong to a different class: the Plunging$_*(e)$ or Plunging$_*(e=0)$ orbits, whose amplitude is now controlled by the other coefficient \eqref{coef2}. The face-on amplitude of perturbations is easily obtained from \eqref{eqn:plunging(e)QNMpsi4} and \eqref{eqn:plunging(e=0)QNMpsi4} and given by  
\bea
\vert h_+ + i h_\times\vert \Big|_{\ell=\ell_*,e \neq 0} &=&\hat{ \mathcal A}_{FO} \sqrt{\frac{\kappa \ell_*}{e}}\frac{  m_0}{D}  \frac{\sqrt{\lambda}}{\sinh \frac{\lambda \hat{u}}{4 M} } ,\\
\vert h_+ + i h_\times\vert \Big|_{\ell=\ell_*,e=0} &=&\hat{ \mathcal A}_{FO}  \sqrt{\frac{e^{\kappa \bar t_0}}{\kappa}} \frac{  m_0}{D}  \sqrt{2\lambda} e^{-\frac{\lambda \hat{u}}{4 M} } ,
\eea
where we used the map of parameters \eqref{mapp1}-\eqref{mapp2} (with final barred coordinates) and defined $\hat{ \mathcal A}_{FO}  =2^{-1/2} |\mathcal A^{\ell = \ell_*}_{22}|S_{22}(0)$. A special feature of the amplitude of the Plunging$_*(e=0)$ orbit is that it depends on the initial value $\bar t_0$ of the time parameter defined in \eqref{Plungingstare0}. We obtain numerically $\hat{ \mathcal A}_{FO} =1.2$.

As a last example of face-on amplitude, the Plunging$_*(E)$ orbits leads to the overall signal\footnote{Remember that $\hat u \lambda^{2/3}$ is kept fixed in the limit $\lambda \rightarrow 0$, $\hat u \rightarrow \infty$. Also, the overall scaling with $\lambda^{1/3}$ should be interpreted with care in the NHEK (but not near-NHEK) limit, see footnote \ref{footn}.}
\bea
\vert h_+ + i h_\times\vert \Big|_{\ell=\ell_*,E\, \text{finite}} &=&\bar{ \mathcal A}_{FO}  \sqrt{\frac{\ell_*}{E}} \frac{  m_0}{D}  {\lambda^{\frac{1}{3}}} \frac{4M}{\hat u \lambda^{2/3}},
\eea
where $\bar{ \mathcal A}_{FO}  = 2^{-1/2} |\mathcal A^{\ell = \ell_*}_{22}|S_{22}(0)$.

In the edge-on case, the leading contribution to the signal comes from a large number of harmonics which admit conformal weights with real part $1/2$\footnote{The first such harmonics are $(l,m)=(2,2),\, (3,3),\, (4,4),\, (5,4),\, (5,5),\, (6,5),\, (6,6),\, (7,6),\, (7,7) \dots$}. For the generic Plunging$(e,\ell)$ orbit, the dominant behavior of the envelope metric perturbation for a generic orbit is now given by  
\bea
\vert h_+ + i h_\times\vert = \frac{  m_0}{D}  \frac{\sqrt{\lambda}}{(\sinh \frac{\lambda \hat{u}}{4 M})^{1/2}(\cosh \frac{\lambda \hat{u}}{4 M}+\chi \sinh \frac{\lambda \hat{u}}{4 M})^{1/2}} \left| \sum_{m =2}^{\infty} \mathcal A^m_{EO}(\frac{\ell}{\ell_*},\hat E ; \frac{\hat u}{M};\lambda)  \right| \label{EOampli}
\eea
where
\bea
\mathcal A^m_{EO}(\frac{\ell}{\ell_*},\hat E ; \frac{\hat u}{M} ; \lambda) &=& -\sum_{l=m}^{l_{max}(m)} 2^{-h} \lambda^{h-\frac{1}{2}}\mathcal A_{lm}^{\ell > \ell_*} S_{lm}(\frac{\pi}{2})e^{-\frac{i m\hat u}{2M}} \left( \sinh \frac{\lambda \hat u}{4M}\right)^{\frac{1}{2}-h-\frac{i m \tilde \omega}{\kappa}} \nn\\
&& \times \left( \cosh \frac{\lambda \hat u}{4M} + \chi \sinh \frac{\lambda \hat u}{4M} \right)^{\frac{1}{2}-h+\frac{i m \tilde \omega}{\kappa}}. \label{EOsum}
\eea
\begin{figure}[!htb]
	\centering
	\includegraphics[angle=270,width=0.45 \textwidth]{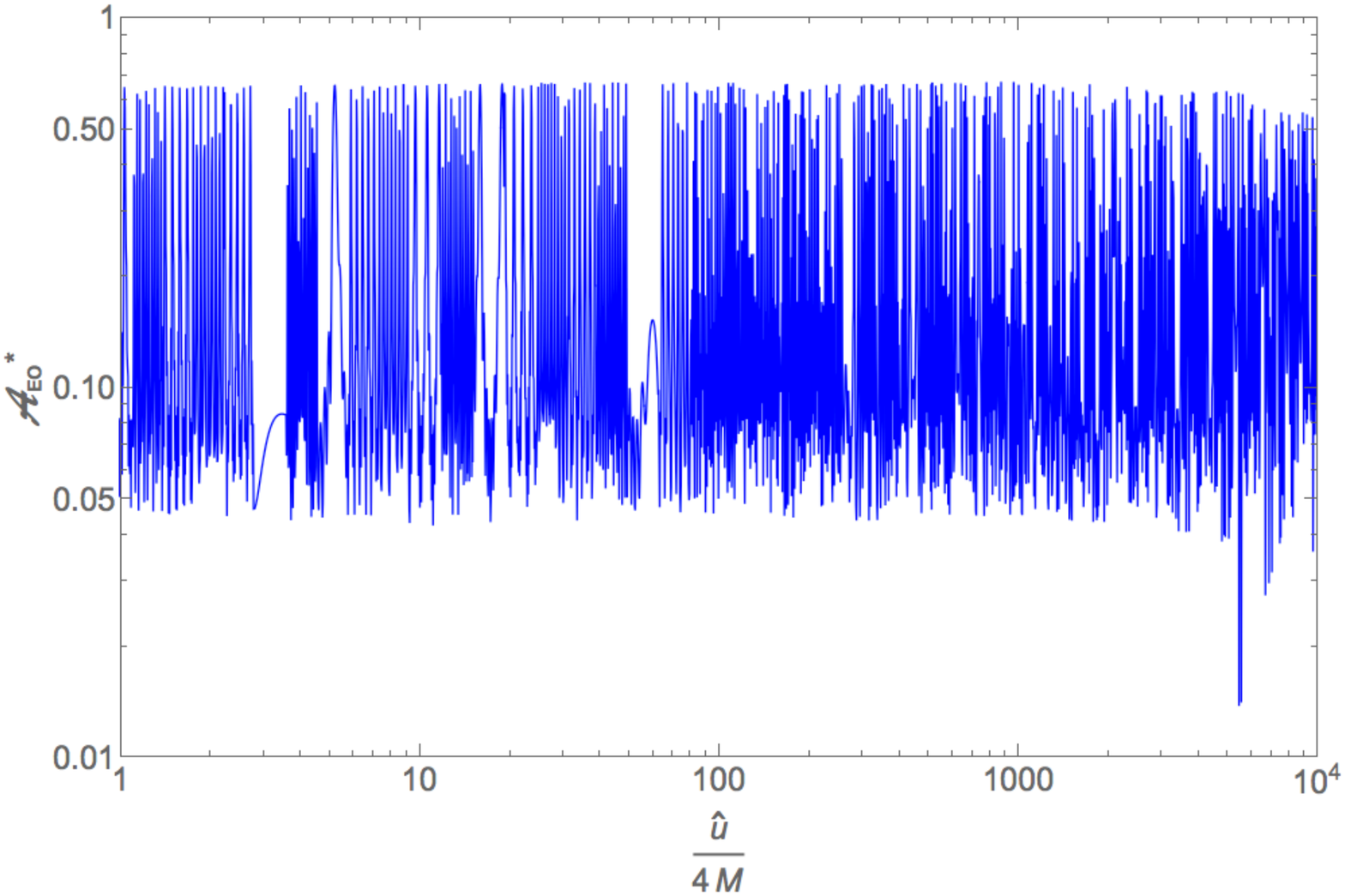}
	\includegraphics[angle=270,width=0.45 \textwidth]{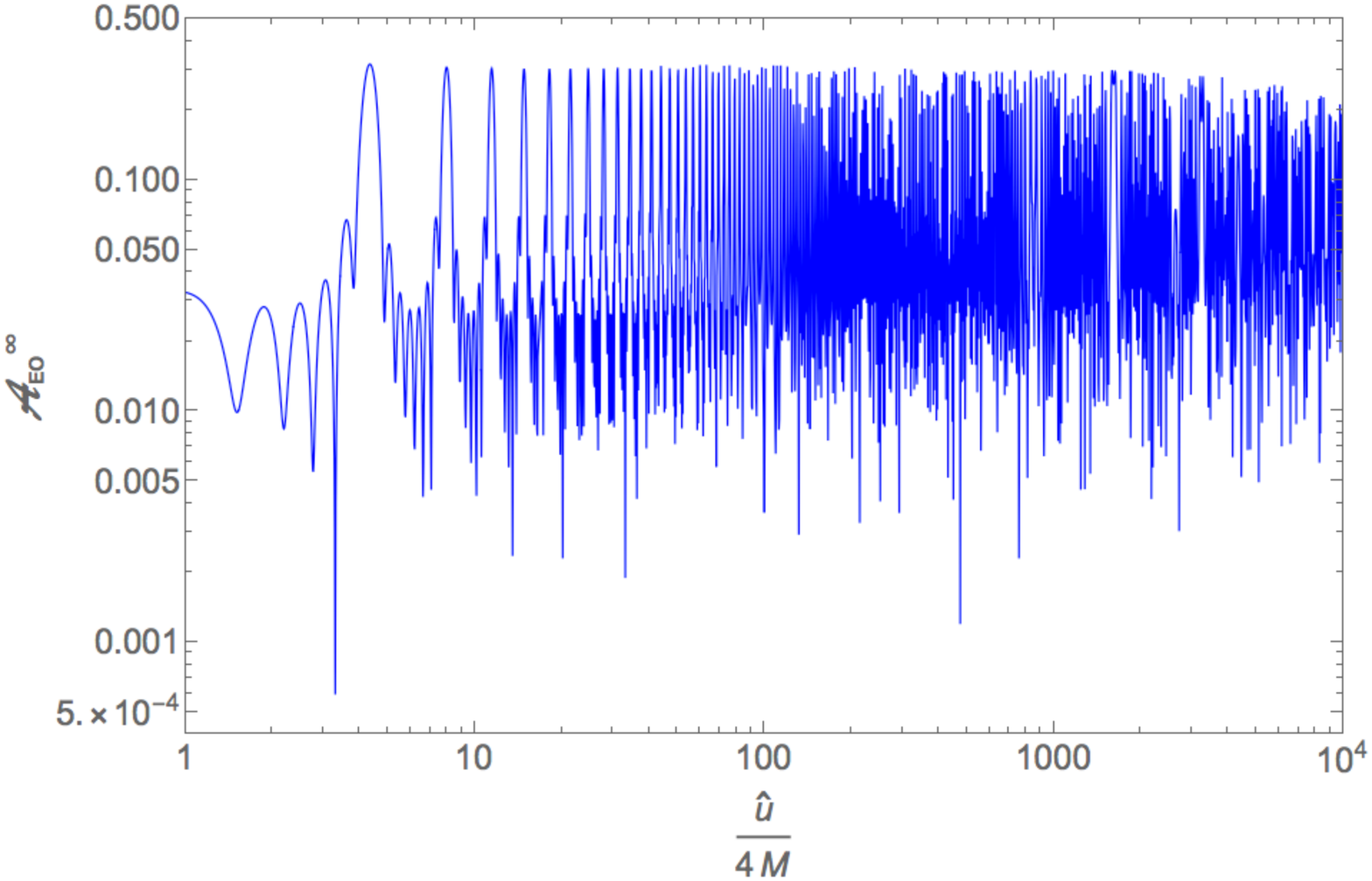}
	\caption{Amplitude coefficients (edge-on) for $\hat E=\frac{\ell_*}{2M}$ and $\lambda=10^{-3}$. Left: near the critical limit $\ell \rightarrow \ell_*$. Right:  for large specific angular momentum $\ell \rightarrow \infty$.} \label{fig:AEOstar1}
\end{figure}
In this expression, the dependence on $M$, $m_0$ and $r_0$ cancels out, but the dependence on $\lambda$ remains. The upper bound $l_{max}(m)$ in the sum in \eqref{EOsum} is the maximum value of $l$ such that $\text{Re}(h)=1/2$. It is given by $\lfloor \frac{m}{0.754}\rfloor$ in the range $2 \leq m \leq 20$. Convergence of the amplitude requires a large $m$. We find numerically a convergence up to $\sim 5\%$ by including all $2 \leq m \leq 20$. The overall amplitude in \eqref{EOampli} as a function of the ratio $\ell \rightarrow \ell_*$ exhibits similar (critical and linear) behavior as the face-on case,
\bea
 \left| \sum_{m =2}^{\infty} \mathcal A^m_{EO}(\frac{\ell}{\ell_*} \rightarrow 1,\hat E; \frac{\hat u}{M} ; \lambda)  \right| &=& \mathcal A_{EO}^*(\hat E ; \frac{\hat u}{M})  (\frac{\ell}{\ell_*}-1)^{-\frac{1}{4}},\label{EOampl}\\
 \left| \sum_{m =2}^{\infty} \mathcal A^m_{EO}(\frac{\ell}{\ell_*}\rightarrow \infty ,\hat E ; \frac{\hat u}{M} ; \lambda)  \right| &=& \mathcal A_{EO}^\infty(\hat E ; \frac{\hat u}{M}) \frac{\ell}{\ell_*} .
\eea
We plot $\mathcal A_{EO}^*(\hat E ; \frac{\hat u}{M})$ and $\mathcal A_{EO}^\infty(\hat E ; \frac{\hat u}{M})$ in Figure \ref{fig:AEOstar1} for values $\hat E=\frac{\ell_*}{2M}$ and $\lambda=10^{-3}$. 

Similar critial behavior appears for generic Plunging$(E,\ell)$ orbits in NHEK region, we just write down the waveform
	\bea
	|h_++ih_{\times}|=\frac{m_0}{D} \frac{\lambda^{1/3}}{\sqrt{|(\frac{\hat u\lambda^{2/3}}{2M}\cos\frac{\zeta}{2}-\sin\frac{\zeta}{2})(\cos\frac{\zeta}{2} +\sin\frac{\zeta}{2}\frac{\hat u\lambda^{2/3}}{2M})|}}\times |\sum_{m=2}^{\infty}\mathcal{A}_{EO,NHEK}^m(\frac{\ell}{\ell_*},\hat{E};\frac{\hat{u}}{M}; \lambda)|,\nn\\
	\eea
	where 
	\bea
	\mathcal{A}_{EO,NHEK}^m(\frac{\ell}{\ell_*},\hat{E};\frac{\hat{u}}{M};\lambda)&=&-\sum_{l=m}^{l_{max}(m)} \mathcal A_{lm}^{\ell > \ell_*} S_{lm}(\frac{\pi}{2})e^{-i\frac{m}{2M}\hat{u}}\lambda^{\frac{2}{3}(h-\frac{1}{2})}\nn\\&&\times (\frac{\hat u\lambda^{2/3}}{2M}\cos\frac{\zeta}{2}-\sin\frac{\zeta}{2})^{\frac{1}{2}-h-i\frac{m\tilde{\omega}}{\kappa}} (\cos\frac{\zeta}{2}+\sin\frac{\zeta}{2}\frac{\hat u\lambda^{2/3}}{2M})^{\frac{1}{2}-h+i\frac{m\tilde{\omega}}{\kappa}}.\nn\\
\eea

\section{Observability by LISA}
\label{sec:LISA}

Let us assume for a moment that nearly extremal supermassive black holes exist in Nature, preferably even at rather low redshift. 
It is then an interesting and timely question whether the gravitational wave signals from the final plunges of EMRIs that we have computed, are potentially observable by LISA. In this section we compute the signal to noise (SNR) ratio of our waveforms for parameters in the LISA band and assuming no prior knowledge about the orbit from the earlier inspiral. Of course this won't be the case in realistic situations, but it serves to estimate the observability of our waveforms as independent signals on their own. In the analysis of \cite{Gralla:2016qfw}, the near-extremality parameter  $\lambda = 10^{-2}$ was assumed, since the near-horizon analytic result models up to $\sim 10\%$ precision a full e-fold of the adiabatically evolved inspiral into Gargantua \cite{Gralla:2015rpa,Gralla:2016qfw}. In our analysis, given the lack of a parallel numerical analysis that would allow to check the precision of the near-horizon results for plunges, we will be more conservative and choose as a reference the near-extremality parameter $\lambda = 10^{-3}$. The existence of such sources is more speculative, but our analytical results are more accurate.

The frequencies of the gravitational wave oscillations as seen from a detector in the asymptotically flat region will be harmonics of the angular frequency of the central, extremal Kerr black hole,
\bea
f_\infty = \frac{\Omega_{ext}}{2\pi} = \frac{1}{4\pi M}= 1.6 \times 10^{-2} Hz \left( \frac{10^6 M_\odot}{M} \right).\label{typf}
\eea
with the lowest harmonics dominating the signal in the regime of interest. The typical LISA source masses $M$ lie in the range $10^5 - 10^7 M_\odot$. In the corresponding range of frequencies $f_\infty$ the sky-averaged strain sensitivity of the LISA observatory expressed as power spectral density $S_h(f)$ is approximately constant and equal to $4 \times 10^{-40}\ \mathrm{Hz}^{-1}$ \cite{Audley:2017drz}.  

The SNR of a monochromatic measured gravitational wave $h(t)$ is given by
\be\label{SNR}
\left(\frac{S}{N}\right)^2 = \frac{2}{S_h(\Omega_{ext}/\pi)} \int_{\hat u_i}^{\hat u_f}  | h(\hat u) |^2 d\hat u
\ee
where $U \equiv \hat u_f-\hat u_i$ is the time over which the signal is measured. 

\begin{figure}[!htb]
	\centering
	\includegraphics[angle=270,width=0.48 \textwidth]{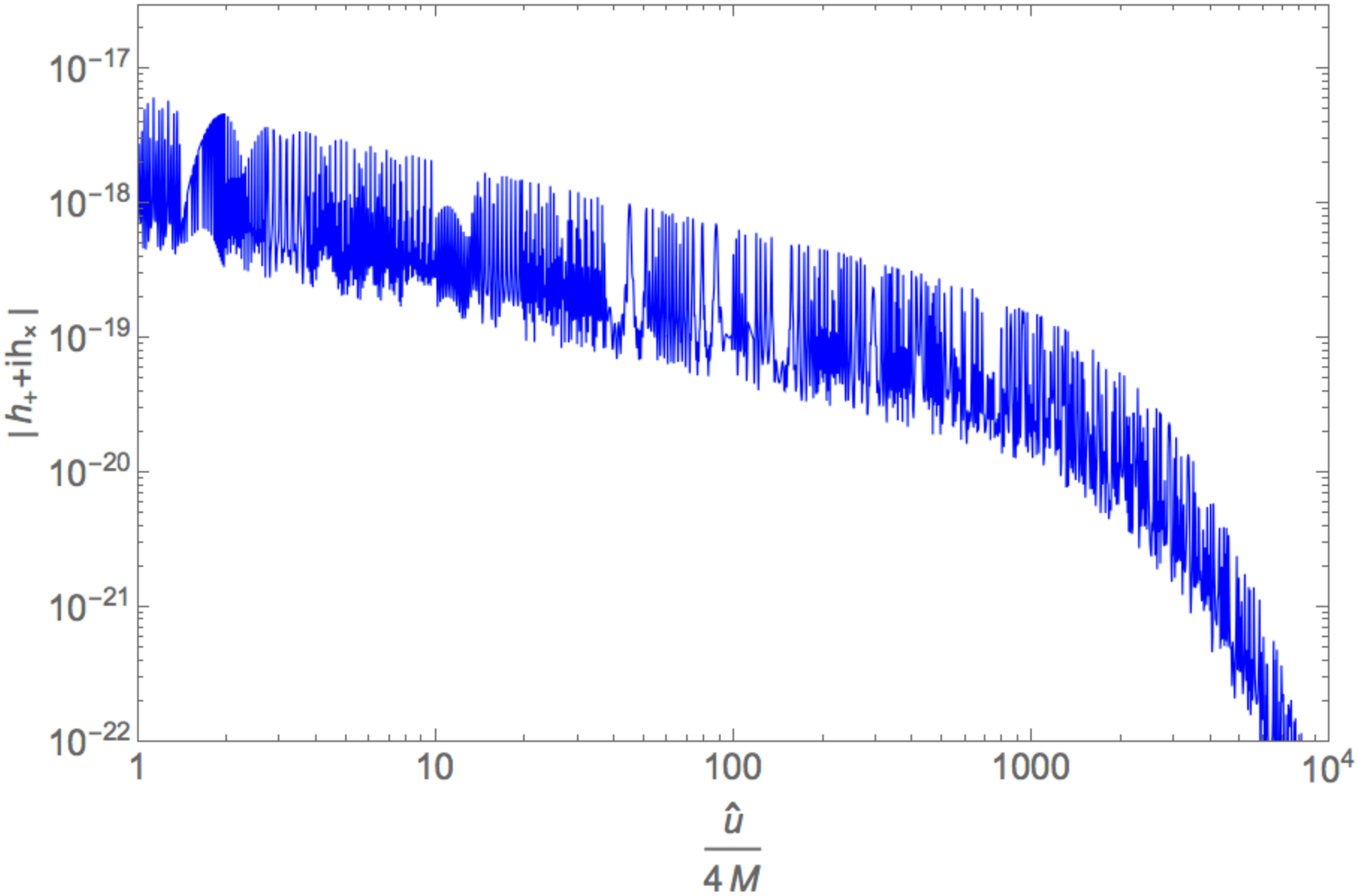}
		\includegraphics[angle=270,width=0.48 \textwidth]{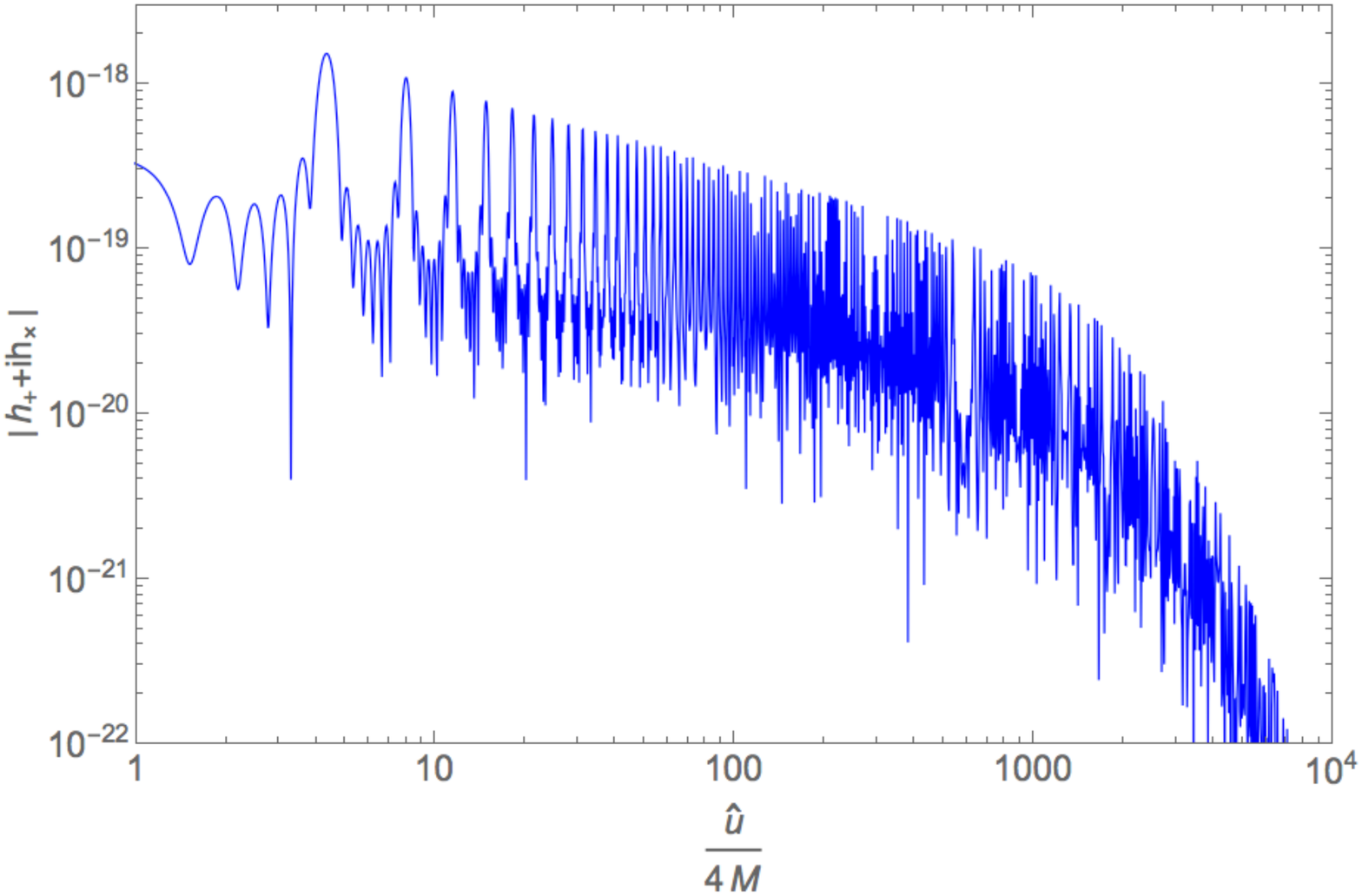}
	\caption{Strain for an edge-on Plunging$(e,\ell)$ orbit with $\hat E=\frac{\ell}{2M}$ and $\lambda=10^{-3}$. Left: a nearly critical orbit with $\ell \rightarrow \ell_*$ and $\frac{m_0}{D} \left( \frac{\ell}{\ell_*}-1\right)^{-1/4}=10^{-17}$. Right: an orbit with large angular momentum $\ell$ and $\frac{m_0}{D} \frac{\ell}{\ell_*}=10^{-17}$.} \label{fig:AEOstar}
\end{figure}

\begin{figure}[!htb]
	\centering
	\includegraphics[angle=270,width=0.48 \textwidth]{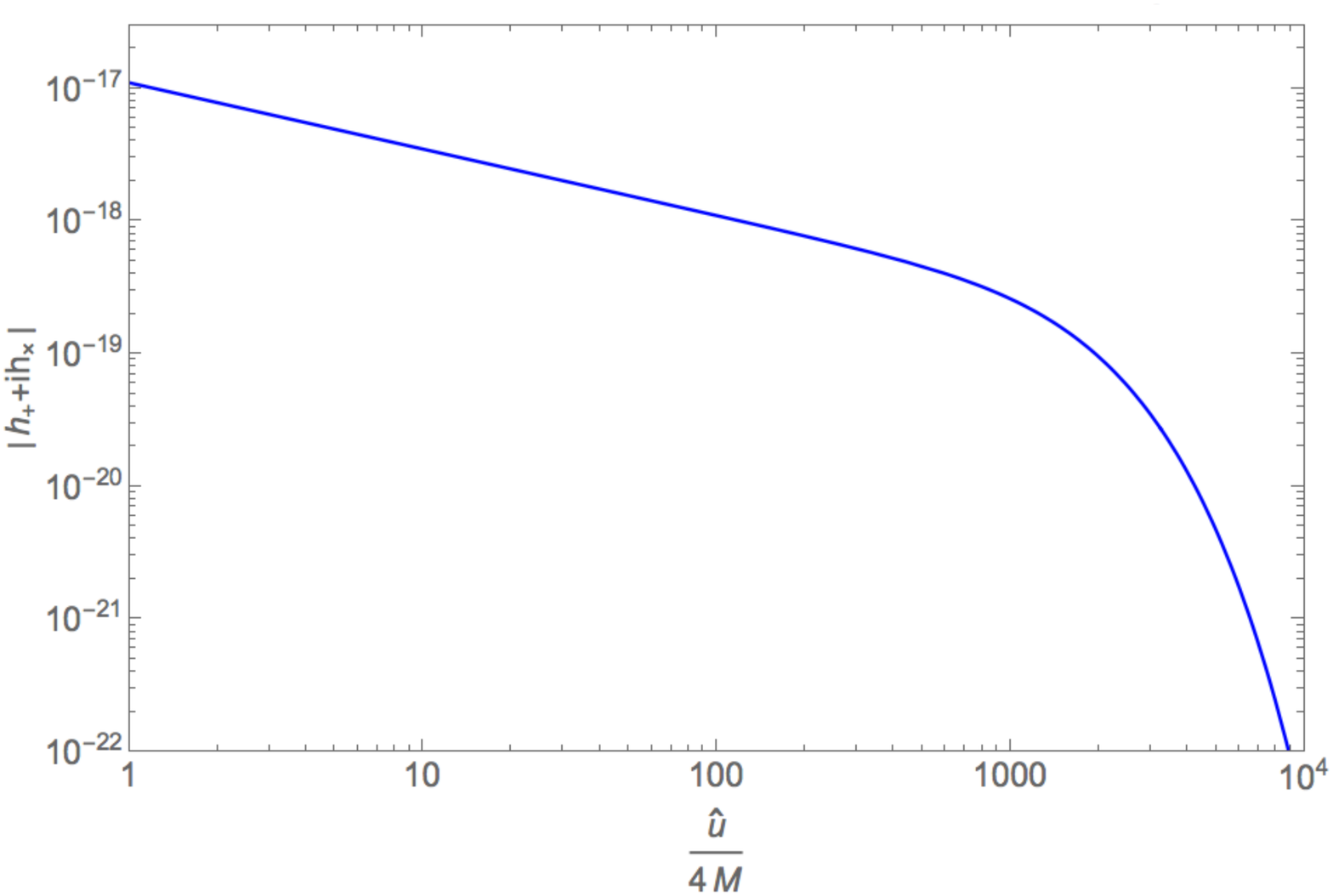}
	\caption{Strain for a face-on Plunging$(e,\ell)$ orbit  with energy $\hat E=\frac{\ell}{2M}$, $\lambda=10^{-3}$, $\ell =2.6 \ell_*$ and with $\frac{m_0}{D} =10^{-17}$. } \label{fig:AFOmin}
\end{figure}

We now use this to estimate the observability of the signal for a number of representative examples of waveforms associated with generic edge-on or face-on Plunging$(e,\ell)$ orbits for which both the polynomial and the onset of the exponential ringdown lie within the LISA range. Some of the waveforms are shown in Figure \ref{fig:AEOstar} and Figure \ref{fig:AFOmin}.

%%%%%%%%%%%%%%%%
%%%%%%%%%%%%%%%%%

We first consider the $e=0$, face-on plunging orbit with near-critical angular momentum $\ell \rightarrow \ell_*$ exhibiting critical behavior. Substituting its waveform \eqref{FOcrit} in \eqref{SNR} yields
\be
\left(\frac{S}{N}\right)^2 = \frac{ m_0^2M}{D^2}  \left(\frac{\ell}{\ell_*}-1\right)^{-\frac{1}{2}} \frac{10.24}{S_h(\Omega_{ext}/\pi)} 
(-\ln (\lambda \hat u_i/4M))
\ee
where we have taken the final time $\hat u_f$ in \eqref{SNR} well beyond the transition from polynomial to exponential decay, which occurs around $\hat u \sim 2M/\lambda$. The initial time corresponds to the onset of the plunging near-NHEK regime. It could be determined by a numerical simulation. A conservative order of magnitude estimate is to take $\hat u_i$ to be a few times $M$. This timescale is certainly consistent with considering the zero-damped modes only, as the damped modes should typically decay on timescale of order $M$. To determine the LISA observable volume of this signal, consider a measurement at threshold SNR $\rho_{th}$, which we assume to be around 15. With $\lambda = 10^{-3}$ and $\hat u_i = 10M$, this yields the maximal luminosity distance
\be
D_{FO,e=0}^{max} \approx 0.08 \ \mathrm{Gpc} \left( \frac{M}{10^7 M_\odot} \right)^{1/2}  \left( \frac{m_0}{10 M_\odot} \right) \left(\frac{\ell}{\ell_*}-1\right)^{-\frac{1}{4}}  \left( \frac{15}{\rho_{th}}\right) .
\ee
The result for face-on orbits in the large angular momentum $\ell \gg \ell_*$ takes a similar form but with the critical enhancement factor above replaced by the linear enhancement in \eqref{critAM2}.

Next consider an $e\neq 0$ face-on plunging orbit, again with near-critical angular momentum $\ell \rightarrow \ell_*$. 
Substituting the waveform \eqref{FO} in \eqref{SNR} and integrating over the time over which the signal is measured yields the following SNR,
\be
\left(\frac{S}{N}\right)^2 = \frac{ m_0^2M}{\lambda D^2}  \left(\frac{\kappa l_*}{e}\right) \left(\frac{M}{\hat u_i}\right)
\frac{32}{S_h(\Omega_{ext}/\pi)} .
\ee
Taking again $\lambda = 10^{-3}$ and $\hat u_i = 10M$, the distance to which we can expect LISA to detect the signal of a plunging orbit of this kind is
\be
D_{FO,e \neq 0}^{max} \approx 0.63 \ \mathrm{Gpc} \left( \frac{M}{10^7 M_\odot} \right)^{1/2}  \left( \frac{m_0}{10 M_\odot} \right)  \sqrt{\frac{\kappa \ell_*}{e}} \left( \frac{15}{\rho_{th}}\right) .
\ee
The observable volume $\sim (D^{max})^3$ of other face-on orbits can be obtained in a similar manner from the explicit formulae for the waveforms in Section \ref{Seccrit}.

%%%%%%%%%%%%%%%%%%
%%%%%%%%%%%%%%%%%

We conclude with an example of a $\chi=0$ edge-on plunging orbit (cf. Figure \ref{fig:AFOmin}), again with near-critical angular momentum $\ell \rightarrow \ell_*$. In this case a large number of modes significantly contribute to the measured signal. The waveform is given in \eqref{EOampli} with amplitude \eqref{EOampl} exhibiting critical behavior. Figure \ref{fig:AEOstar} shows that the amplitude factor $\mathcal{A}_{EO}^*$ rapidly oscillates over the time of observation around a constant value $\sim 0.1$. We note in passing that the amplitude of the signal is strongly $\theta$ dependent, which should allow for a precise sky localization of the signal. An estimate of the observability can be obtained by substituting the waveform in \eqref{SNR}, yielding
\be
\left(\frac{S}{N}\right)^2 = \frac{ m_0^2M}{D^2}  \left(\frac{\ell}{\ell_*}-1\right)^{-\frac{1}{2}} \frac{8(\mathcal{A}_{EO}^*)^2}{S_h(\Omega_{ext}/\pi)} 
(-\ln (\lambda \hat u_i/4M))
\ee
which then gives, for $u_i=10M$,
\be
D_{EO,e=0}^{max} \approx 0.07 \ \mathrm{Gpc} \left( \frac{M}{10^7 M_\odot} \right)^{1/2}  \left( \frac{m_0}{10 M_\odot} \right)   \left(\frac{\ell}{\ell_*}-1\right)^{-\frac{1}{4}}  \left( \frac{15}{\rho_{th}}\right) .
\ee
We conclude that the maximal luminosity distance for edge-on or face-on signals is similar. The observable volume depends significantly on the enhancement factor, whose form is itself determined by the conformal class and the impact parameters of the orbit. Redshifts $z \sim 1$ or $D^{max} \approx 7 \, \text{Gpc}$ require a large enhancement factor associated with critical behavior.

\section{Summary}
\label{sec:ccl}

We have analytically computed the gravitational wave emission from the final stages of a class of EMRIs in which a non-spinning compact object on a generic corotating equatorial orbit enters the near-horizon (either NHEK or near-NHEK) region of Gargantua, i.e. a supermassive nearly extremal Kerr black hole.

To do so we have first enumerated nearly corotating test particle orbits in the equatorial plane of the near-horizon region of a nearly extremal Kerr black hole. We found three classes of plunging orbits in both NHEK and near-NHEK spacetimes, one class of circular orbits as well as one class of osculating orbits. We also found that there is a minimum angular momentum per unit probe mass $\ell_*= \frac{2}{\sqrt{3}}M$, which plays a fundamental role in our later analysis. This specific angular momentum coincides with the one of the ISCO and is therefore physically interesting. Studying these orbits in terms of complex representations of the conformal symmetry group $SL(2,\mathbb R) \times U(1)$, which leave the angular momentum invariant but change the energy, we found that there are only two conjugacy classes under complexified conformal symmetry combined with PT symmetry. Their representatives can be chosen to be the (stable) circular orbit in NHEK with critical angular momentum $\ell = \ell_*$ and the (unstable) circular orbit in near-NHEK with supercritical angular momentum $\ell > \ell_*$. 

We have computed by brute force methods the exact curvature perturbation induced by a probe on a circular orbit in near-NHEK, thereby completing the twin computation of the circular orbit in NHEK \cite{Porfyriadis:2014fja}. Conformal symmetry allows one to obtain the gravitational waveforms for generic equatorial trajectories from the waveforms of these two circular ``seed orbits'' by employing complex transformations, as pioneered for a specific class of orbits and a real transformation in \cite{Hadar:2014dpa}. We have used these waveforms and the calculational power brought about by conformal symmetry to obtain analytic expressions for the leading order time domain gravitational waveforms from plunges on all corotating equatorial orbits into nearly extremal Kerr black holes. Conformal symmetry not only allows one to shorten the computation but also structurally implies that the waveforms involve analytically tractable integral representations of hypergeometric functions (or simpler) and tabulated inverse Laplace transforms. 

The oscillation timescale of any gravitational perturbation originating from the near-horizon region is set by the Lense-Thirring corotation scale of the central nearly extremal Kerr black hole, $\Omega_{ext}^{-1}=2M$. This kinematical feature is a straightforward consequence of the near-horizon limit. A less obvious feature is that the asymptotic amplitude of the gravitational perturbation for all plunging orbits is suppressed as $\lambda^{\frac{2}{3}\text{Re}(h)}$ in the particular NHEK scaling that preserves the ISCO, as does the emission from the ISCO orbit \cite{Porfyriadis:2014fja,Gralla:2016qfw}. This suppression factor becomes universally $\lambda^{\text{Re}(h)}$ in the near-NHEK limit. This means that the modes with conformal weight $h=\frac{1}{2}+i h_I$, $h_I <0$ provide the dominant contribution in the near-horizon region, leading to $|\delta \psi_4|=O(\lambda^{1/3})$ and $O(\lambda^{1/2})$, respectively. 

A new ``smoking gun'' signature of the asymptotic GW signal from plunging orbits in the near-horizon region is the existence of two phases: a polynomial decay with a specific exponent, followed by an exponential decay. The polynomial phase arises from the coherent stacking of zero-damped quasi-normal modes \cite{Yang:2013uba}. We find that the polynomial decay generically interpolates between an initial $\hat u^{-1/2}$ behavior and a final $\hat u^{-1}$ fall-off, where $\hat u$ is the retarded asymptotic time. The details of this interpolation depend on the impact parameters. To the best of our knowledge this behavior has not been seen yet in numerical simulations (see however \cite{Burko:2016sfi}). It should however be present sufficiently close to extremality, say for $\lambda \leq 10^{-3}$. In all near-NHEK orbits the polynomial decay is followed by a universal exponential decay with a characteristic timescale $\Delta \hat u = 2M/\lambda$ governed by the lowest zero-damped QNM\footnote{The NHEK limit does not resolve this exponential phase.}. 

A second unique and remarkable feature concerns the amplitude of the asymptotic curvature perturbation sourced by plunges in the near-horizon region. It critically depends on $\ell$, the conserved angular momentum per unit rest mass of the orbit. In particular the amplitude of the GW signal grows linearly with $\ell$ when $\ell$ is large. By contrast it exhibits critical behavior when $\ell$ tends to its minimum value $\ell_*= \frac{2}{\sqrt{3}}M$, which corresponds to the physically relevant specific angular momentum of the ISCO orbit. This critical behavior comes with specific critical exponents, which we computed.  The metric perturbations $h_+$ and $h_{\times}$ are approximately proportional to the curvature perturbation because the oscillation timescale $2M$ is extremely short compared with the timescale of the overall evolution of the amplitude, $2M/\lambda$. This means that the amplitude of the asymptotic GW signal inherits this critical behavior in the limit $\ell \rightarrow \ell_*$. We intend to study the nature of this critical behavior in more detail in future work. It would be interesting for instance to study whether self-force effects regulate the gravitational wave emission and what would be the critical behavior including self-force effects. More work is also required to see a possible relationship of this near-extremal critical behavior with other critical phenomena in black hole mergers \cite{Pretorius:2007jn,Sperhake:2009jz,Berti:2010ce,Gundlach:2012aj}.

Our results for the gravitational waveforms generated in the final near-horizon stages of corotating equatorial EMRIs into Gargantua complete all previous results for scalar and gravitational probes \cite{Porfyriadis:2014fja,Hadar:2014dpa,Hadar:2015xpa,Gralla:2015rpa,Hadar:2016vmk,Gralla:2016qfw}. They exactly or qualitatively  agree in all cases where comparison is possible. It would be interesting to explore numerically what is the range of values of the extremality parameter $\lambda$ for which our analytic waveforms are accurate. At any rate, our precise analytic results serve as benchmarks for effective EMRI waveform models and numerical simulations. 

Whether observatories like LISA will observe GW signals from plunges into nearly-extremal Kerr black holes will depend not only on the range of $\lambda$ for which these signals are produced, but also on whether supermassive black holes in this range exist at all. The standard geometrically thin disk model does not produce massive black holes in this range in light of the Thorne bound \cite{1974ApJ...191..507T}. However, other disk models might be realized in Nature. In particular systems sustained by magnetic fields can exceed the bound \cite{1980AcA....30...35A,2011A&A...532A..41S}. Alternatively, high spin black holes are producable in black hole collisions \cite{Sperhake:2009jz,Colleoni:2015ena}. Assuming such black holes exist and assuming no prior knowledge about the orbit from the earlier inspiral, we find the LISA observable volume of these GW signals extends out to redshifts $z \sim 0.15$. Evidently this is significantly larger for nearly critical orbits or high orbital angular momentum orbits -- if only we were so lucky\dots

%%%%%%%%%%%%%%%%
\vspace{.2in}
\noindent {\bf Acknowledgments} We thank Niels Warburton for making public his code for computing spheroidal harmonics and our referees for constructive suggestions. This work makes use of the Black Hole Perturbation Toolkit. We thank Sam Gralla, Achilleas Porfyriadis and Andy Strominger for their comments on the manuscript.  This work was supported in part by the European Research Council grant ERC-2013-CoG 616732 HoloQosmos, the ERC Starting Grant 335146 ``HoloBHC", by the C16/16/005 grant of the KU Leuven, and by the FWO grant G092617N. We also acknowledge networking support by the COST Action GWverse CA16104.  G.C. is a Research Associate of the Fonds de la Recherche Scientifique F.R.S.-FNRS (Belgium). We finally thank Adrien Druart and Caroline Jonas for pointing out to some typos in the published version that are corrected in this erratum version.

\appendix

\section{Teukolsky's formalism}
\label{Teuk}

Unless otherwise noted we use the conventions of \cite{Merlin:2016boc}. In particular, we use the $(-+++)$ signature while most of the literature on Teukolsky's formalism uses $(+---)$ signature  \cite{1973ApJ...185..635T,Chandrasekhar:1985kt}. We also use Einstein's equations in the form $G_{\mu\nu}=+8\pi T_{\mu\nu}$ which differ by a sign from \cite{1973ApJ...185..635T}. Our convention for the normalization of spheroidal harmonics however differs from \cite{Merlin:2016boc} and matches with \cite{Porfyriadis:2014fja}.

\subsection{Newman-Penrose tetrad}

A Newman-Penrose tetrad consists of two real null vectors $l^{\mu}$, $n^{\mu}$, and one complex null vector $m^{\mu}$ satisfying $l^{\mu}n_{\mu} = - m^{\mu}\bar{m}_{\mu} = -1$ and such that $g_{\mu\nu}=-l_\mu n_\mu - n_\mu l_\nu + m_\mu \bar m_\mu + \bar m_\mu m_\nu$. The directional derivatives along the tetrads are defined as\footnote{We warn the reader that the notation $\Delta$ is also used for \eqref{defD}, $\delta$ is also used for denoting a variation and $\kappa$ and $\lambda$ are also defined in the (near)-NHEK limit. The meaning of each symbol should be clear in each context.}
\bea
D = l^{\mu}\partial_{\mu} ,\; \;\;\;
\Delta = n^{\mu}\partial_{\mu} ,\; \;\;\;
\delta = m^{\mu}\partial_{\mu} ,\; \;\;\;
\bar{\delta} = \bar{m}^{\mu}\partial_{\mu} .
\eea
The spin coefficients are defined as 
%suggestion: change bars to * to avoid overlap/confusion with barred coordinates
\bea
	\kappa &=& -m^{\mu}l^{\nu}\nabla_{\nu}l_{\mu} \qquad 
	\sigma = -m^{\mu}m^{\nu}\nabla_{\nu}l_{\mu}   \\
	\lambda &=& -n^{\mu}\bar{m}^{\nu}\nabla_{\nu}\bar{m}_{\mu}  \qquad
	\nu = -n^{\mu}n^{\nu}\nabla_{\nu}\bar{m}_{\mu}  \\
	\rho	 &=& -m^{\mu}\bar{m}^{\nu}\nabla_{\nu}l_{\mu}  \qquad
	\mu	 = -n^{\mu}m^{\nu}\nabla_{\nu}\bar{m}_{\mu}   \\
	\tau	 &=& -m^{\mu}n^{\nu}\nabla_{\nu}l_{\mu}  \qquad
	\varpi	 = -n^{\mu}l^{\nu}\nabla_{\nu}\bar{m}_{\mu}  \\
	\epsilon &=& -\frac{1}{2}(n^{\mu}l^{\nu}\nabla_{\nu}l_{\mu} + m^{\mu}l^{\nu}\nabla_{\nu}\bar{m}_{\mu}  ) \\
	\gamma &=& -\frac{1}{2}(n^{\mu}n^{\nu}\nabla_{\nu}l_{\mu} + m^{\mu}n^{\nu}\nabla_{\nu}\bar{m}_{\mu}) \\
	\alpha &=& -\frac{1}{2}(n^{\mu}\bar{m}^{\nu}\nabla_{\nu}l_{\mu}+ m^{\mu}\bar{m}^{\nu}\nabla_{\nu}\bar{m}_{\mu}) \\
	\beta &=& -\frac{1}{2}(n^{\mu}m^{\nu}\nabla_{\nu}l_{\mu} + m^{\mu}m^{\nu}\nabla_{\nu}\bar{m}_{\mu})
\eea
and the Weyl scalars are defined as
\bea
	\psi_0 &=& C_{\alpha \beta \mu \nu} l^{\alpha}m^{\beta}l^{\mu}m^{\nu} \\
	\psi_1 &=& C_{\alpha \beta \mu \nu} l^{\alpha}n^{\beta}l^{\mu}m^{\nu} \\
	\psi_2 &=& C_{\alpha \beta \mu \nu} l^{\alpha}m^{\beta} \bar m^{\mu}n^{\nu}\\
	\psi_3 &=& C_{\alpha \beta \mu \nu} l^{\alpha}n^{\beta}\bar m^{\mu} n^{\nu} \\
	\psi_4 &=& C_{\alpha \beta \mu \nu} n^{\alpha}\bar{m}^{\beta}n^{\mu}\bar{m}^{\nu} 
\eea
For further use, remember that under a type III rotation of the spin frame 
\bea
l^\mu \rightarrow A^{-1} l^\mu, \qquad n^\mu \rightarrow A n^\mu,\qquad m^\mu \rightarrow e^{i \theta} m^\mu
\eea
with $A,\theta$ arbitrary real numbers, the $\psi_4$ scalar transforms as \cite{1985AmJPh..53.1013C} 
\bea
\psi_4 \rightarrow A^2 e^{-2 i \theta} \psi_4.\label{tpsi4}
\eea

For the Kerr black hole, we use Kinnersley's tetrad
\bea
	\hat{l}^{\mu}  &=&  \Delta^{-1}((\hat{r}^2+a^2), \Delta, 0, a)  \\
	\hat{n}^{\mu}  &=& (2\Sigma)^{-1}((\hat{r}^2+a^2), -\Delta, 0, a) \\
	\hat{m}^{\mu} &=& (\sqrt{2}(\hat{r} + ia \cos{\theta}))^{-1}(ia \sin{\theta}, 0, 1, \frac{i}{\sin{\theta}})			 	
\eea
where we recall the definitions \eqref{defD}. With respect to this tetrad the spin coefficients are given by
\begin{align}
	\kappa &= \sigma = \lambda = \nu = \epsilon = 0,  \qquad\qquad &\beta =  -\frac{\bar{\rho}\cot{\theta}}{2 \sqrt{2}}  ,\nn\\
	\rho &= -\frac{1}{\hat{r}-ia\cos{\theta}} ,\qquad &\mu = \frac{\rho \Delta}{2 \Sigma} \nn\\
	\tau &=\frac{-ia\sin{\theta}}{\sqrt{2} \Sigma}  ,\qquad\qquad\qquad 
	&\varpi = \frac{ia\rho^2\sin{\theta}}{\sqrt{2}} ,\nn \\
	\alpha &= \varpi - \bar{\beta}, \qquad\qquad\qquad\qquad\qquad
	&\gamma = \mu + \frac{(\hat{r}-M)}{2 \Sigma}.\label{Kerrspins}
\end{align}
The Kerr solution is type D and out of the 5 Newman-Penrose scalars of the background only $\psi_2$ does not vanish for Kerr \cite{1973JMP....14.1453W}. It is given by
\be
\psi_2 = -\frac{M}{(\hat r-i a \cos\theta)^3} . 
\ee 

In the Poincar\'e NHEK limit \eqref{NHEKlimit}, we perform a tetrad rotation $\hat l^\mu = M \lambda^{2/3} L^\mu$, $\hat n^\mu =\frac{1}{M \lambda^{2/3}} N^\mu$, $\hat m^\mu = M^\mu$. The resulting Kinnersley tetrad is given by
\bea
	L^{\mu}  &=&  (\frac{1}{R^2}, 1, 0, -\frac{1}{R})  ,\nn\\
	N^{\mu}  &=& \frac{1}{2M^2(1+\cos^{2}{\theta})}(1,-R^2,0,-R), \label{KinNHEK}\\
	{M}^{\mu} &=& (\sqrt{2}M(1 + i \cos{\theta}))^{-1}(0, 0, 1, \frac{i(1+\cos^2{\theta})}{2\sin{\theta}})		\nn.	 	
\eea
The spin coefficients read as
\begin{align}
	\kappa &= \sigma = \lambda = \nu = \epsilon = \mu = \rho = 0,  \qquad\qquad &\beta = \frac{\cot{\theta}}{2 \sqrt{2} M(1+i \cos\theta)} ,\nn\\
	\tau &= \frac{-i\sin{\theta}}{M \sqrt{2} (1+\cos^{2}{\theta})} ,\qquad\qquad\qquad 
	&\varpi = \frac{i \sin{\theta}}{M \sqrt{2} (1-i\cos{\theta})^2},\label{spinNHEK} \\
	\alpha &= \varpi - \bar{\beta}, \qquad\qquad\qquad\qquad\qquad
	&\gamma = \frac{R}{2 M^2(1+\cos^2{\theta})}.\nn
\end{align}
and all Newman-Penrose scalars are vanishing except
\bea
\psi_2 =  -\frac{1}{M^2 (1-i \cos\theta)^3}\label{psi2NHEK}
\eea

In the near-NHEK limit \eqref{chgtnear}, we perform a tetrad rotation $\hat l^\mu = \frac{M \lambda}{\kappa} l^\mu$, $\hat n^\mu =\frac{\kappa}{M \lambda} n^\mu$, $\hat m^\mu = m^\mu$. The Kinnersley tetrad is given by
\bea
	l^{\mu}  &=&  (\frac{1}{r(r+2\kappa)}, 1, 0, -\frac{(r+\kappa)}{r(r+2\kappa)}) , \nn\\
	n^{\mu}  &=& \frac{1}{2M^2(1+\cos^{2}{\theta})}(1,-r(r+2\kappa),0,-(r+\kappa)), \label{KinnearNHEK} \\
	{m}^{\mu} &=& (\sqrt{2}M(1 + i \cos{\theta}))^{-1}(0, 0, 1, \frac{i(1+\cos^2{\theta})}{2\sin{\theta}})	.\nn		 	
\eea
All spin coefficients are identical to \eqref{spinNHEK} except for 
\bea
\gamma &=& \frac{(r+\kappa)}{2 M^2(1+\cos^2{\theta})}.
\eea
We also have \eqref{psi2NHEK}.

\subsection{Master equation of Teukolsky}

The Teukolsky master equation unifies the description of the linearized dynamics of various fields around a Kerr black hole or, more generally, a type D spacetime in a single partial differential equation. All gravitational perturbations are encoded in either $\delta \psi_0$ or $\delta \psi_4$ which are invariant under linearized diffeomorphisms. The only perturbations with $\delta\psi_0=\delta\psi_4=0$ are deformations which change $M$ or $J$, which introduce acceleration (given by the C-metric), NUT charge \cite{1973JMP....14.1453W} or introduce boundary gravitons (asymptotic symmetries).

It was shown by Teukolsky \cite{1973ApJ...185..635T} that linearized gravitational perturbations of vacuum Petrov type D spacetimes satisfy\footnote{The intermediate conventions (metric signature, sign of $\psi_0$, $\psi_4$ and sign of spin coefficients) differ from Teukolsky but signs combine so that this final equation is exactly identical \cite{Merlin:2016boc}.} 
\begin{eqnarray}
		&&\hspace{-2cm} [(D-3 \epsilon + \bar{\epsilon}-4\rho-\bar{\rho})(\Delta - 4 \gamma + \mu) - \nn\\
		 && (\delta + \overline{\varpi}-\bar{\alpha}-3\beta-4\tau)(\bar{\delta} + \varpi-4 {\alpha}) - 3\psi_2]\delta\psi_0 = 4 \pi \mathcal  T_0, \label{eqp0}\\
		&&\hspace{-2cm}  [(\Delta+3 \gamma - \bar{\gamma}+4\mu+\bar{\mu})(D + 4 \epsilon - \rho)-\nn\\
		 && (\bar{\delta} - \bar{\tau}+\bar{\beta}+3\alpha+4\varpi)(\delta - \tau+4\beta) - 3\psi_2 ] \delta\psi_4 = 4 \pi \mathcal T_4,\label{eqp4}
%		 \label{eqn:generalTeukolsky}
	\end{eqnarray}
for a Newman-Penrose tetrad with $l^{\mu}$, $n^{\nu}$ along the two principle null directions. Denoting projections onto the Kinnersley tetrad as $T_{lm} = l^{\mu} m^{\nu}T_{\mu \nu}$, etc, the sources in these perturbation equations are related to the stress-energy tensor by\footnote{Note the minus sign typo in front of $\bar\mu$ in \cite{Porfyriadis:2014fja} which is harmless because $\mu=0$ in NHEK.}\footnote{Compared to \cite{Merlin:2016boc}, $\mathcal T_0$ is their $-T_{+2}/(8\pi \Sigma)$ defined in (A10) and $\mathcal T_4$ is their $-\rho^4 T_{-2}/(8\pi \Sigma)$ defined in (A11). The identity can be checked using the spin coefficients \eqref{Kerrspins}.}
\begin{eqnarray}
		\mathcal T_0 &=& (\delta +\overline{\varpi}-\bar{\alpha}-3\beta-4\tau)[(D-2\epsilon-2\bar{\rho})T_{lm}-(\delta + \overline{\varpi}-2\bar{\alpha}-2\beta)T_{ll}] \nn\\ &+& (D-3\epsilon + \bar{\epsilon}-4\rho-\bar{\rho})[(\delta + 2\overline{\varpi}-2\beta)T_{lm}-(D-2\epsilon+2\bar{\epsilon}-\bar{\rho})T_{mm}] ,\\
\mathcal T_4 &=& (\Delta -\bar{\gamma}+\bar{\mu}+3\gamma+4\mu)[(\bar{\delta}-2\bar{\tau}+2\alpha)T_{n\bar{m}}-(\Delta + \bar{\mu}-2\bar{\gamma}+2\gamma)T_{\bar{m}\bar{m}}]\nn \\ &+& (\bar{\delta}+3\alpha + \bar{\beta}+4\varpi-\bar{\tau})[(\Delta + 2\bar{\mu}+2\gamma)T_{n\bar{m}}-(\bar{\delta}+2\alpha+2\bar{\beta}-\bar{\tau})T_{nn}].
	 	\label{eqn:generelTeukolskySource}
\end{eqnarray}

For the Kerr metric in Boyer-Lindquist coordinates  $(\hat{t},\hat{r},\theta,\hat{\phi})$, the Teukolsky equation \cite{1973ApJ...185..635T} for a general spin $s$ field is given by\footnote{The sign convention for $T_{(s)}$ is defined so that the radial derivative terms have the same sign as the source terms.}
\begin{eqnarray}
4\pi \Sigma T_{(s)} &=& -[\frac{(\hat{r}^2+a^2)^2}{\Delta}-a^2 \sin^2{\theta}]\frac{\partial^2 \Psi_{(s)}}{\partial \hat{t}^2} - \frac{4Ma\hat{r}}{\Delta}  \frac{\partial^2 \Psi_{(s)} }{\partial \hat{t} \partial \hat{\phi}}\nn \\ &-& 2s[\hat{r}-\frac{M(\hat{r}^2-a^2)}{\Delta}+ia\cos{\theta}]\frac{\partial \Psi_{(s)} }{\partial \hat{t}} + \Delta^{-s}\frac{\partial}{\partial \hat{r}} (\Delta^{s+1}\frac{\partial\Psi_{(s)} }{\partial \hat{r}}) \nn \\ &+& \frac{1}{\sin{\theta}} \frac{\partial}{\partial \theta}(\sin{\theta} \frac{\partial \Psi_{(s)} }{\partial \theta}) + [\frac{1}{\sin^{2}{\theta}}-\frac{a^2}{\Delta}]\frac{\partial^2 \Psi_{(s)} }{\partial \hat{\phi}^2} + 2s[\frac{a(\hat{r}-M)}{\Delta}+\frac{i \cos{\theta}}{\sin^{2}{\theta}}]\frac{\partial \Psi_{(s)} }{\partial \hat{\phi}} \nn \\ &-& (s^2 \cot^{2}{\theta}-s)\Psi_{(s)} \, . \label{Master}
\end{eqnarray}
The equations \eqref{eqp0}-\eqref{eqp4} precisely reduce to \eqref{Master} for the spin $s=-2$ case with $T_{(-2)} =-2 \mathcal T_4 \rho^{-4}$ and $\Psi_{(-2)}  = \rho^{-4}\delta\psi_4$ and for the spin $2$ case with $T_{(2)} =-2 \mathcal T_0$ and $\Psi_{(2)}=\delta\psi_0$. These two scalars contain the complete information about the perturbing field \cite{1973ApJ...185..635T}. They are related to each other via the Teukolsky-Starobinsky identity \cite{Chandrasekhar:1985kt} such that it suffices to consider one field to reconstruct the entire metric perturbation. The $s=-2$ field is convenient as it readily relates to asymptotic outgoing gravitational waves. 

For the near-NHEK geometry \eqref{nearNHEKGEO}, the spin $-2$ Teukolsky equation \eqref{eqp4} reads explicitly as 
\begin{eqnarray}
-8\pi M^2(1 + \cos^2{\theta}) \eta^{-4} \mathcal T_4 &=& r(r+2\kappa) \frac{\partial^2 (\eta^{-4}\delta\psi_4)}{\partial r^2}  -2(r+\kappa) \frac{{\partial} (\eta^{-4}\delta\psi_4)}{\partial r}  \nn \\ 
&&\hspace{-3.5cm}- \frac{1}{r(r+2\kappa)}\frac{\partial^2 (\eta^{-4}\delta\psi_4)}{\partial t^2}+ \frac{2(r+\kappa)}{ r(r+2\kappa)} \frac{\partial^2 (\eta^{-4}\delta\psi_4)}{\partial t \partial \phi} - \frac{4(\kappa +r)}{r(r+2\kappa)}\frac{\partial (\eta^{-4}\delta\psi_4)}{\partial t} \nn 
\\ && \hspace{-3.5cm}  +\frac{\partial^2 (\eta^{-4}\delta\psi_4)}{\partial \theta^2}  +\cot{\theta} \frac{\partial (\eta^{-4}\delta\psi_4)}{\partial \theta}  +(-\frac{\kappa^2}{r(r+2\kappa)}+\frac{1}{\sin^2{\theta}}-2+\frac{\sin^2{\theta}}{4}) \frac{\partial^2 (\eta^{-4}\delta\psi_4)}{\partial \phi^2}  \nn \\ 
&&\hspace{-3.5cm}- (-\frac{4 \kappa^2}{r(r+2\kappa)}+2i\cos{\theta}+\frac{4i \cos{\theta}}{\sin^2{\theta}})\frac{\partial (\eta^{-4}\delta\psi_4)}{\partial \phi}  + (-\frac{4}{\sin^2{\theta}}+2)(\eta^{-4}\delta\psi_4)
\label{eqn:nearNHEKmaster}
\end{eqnarray}
with 
\be
\eta = -\frac{1}{1-i\cos{\theta}}\label{defeta}
\ee

For the NHEK geometry \eqref{NHEKGEO}, equation \eqref{eqp4} reads explicitly as 
\begin{eqnarray}
-8\pi M^2(1 + \cos^2{\theta}) \eta^{-4}\mathcal  T_4 &=& R^2 \frac{\partial^2 (\eta^{-4}\delta\psi_4)}{\partial R^2}  -2R \frac{{\partial} (\eta^{-4}\delta\psi_4)}{\partial R} - \frac{1}{R^2}\frac{\partial^2 (\eta^{-4}\delta\psi_4)}{\partial T^2} \nn \\ 
&&\hspace{-2cm}  + \frac{2}{ R} \frac{\partial^2 (\eta^{-4}\delta\psi_4)}{\partial T \partial \Phi} -\frac{4}{R}\frac{\partial (\eta^{-4}\delta\psi_4)}{\partial T} +  \frac{\partial^2 (\eta^{-4}\delta\psi_4)}{\partial \theta^2} + \cot{\theta} \frac{\partial (\eta^{-4}\delta\psi_4)}{\partial \theta} \nn \\ 
&&\hspace{-2cm}  +(\frac{1}{\sin^2{\theta}}-2+\frac{\sin^2{\theta}}{4}) \frac{\partial^2 (\eta^{-4}\delta\psi_4)}{\partial \Phi^2}  \nn \\ 
&&\hspace{-2cm}  -(2i \cos{\theta}+\frac{4i \cos{\theta}}{\sin^2{\theta}})\frac{\partial (\eta^{-4}\delta\psi_4)}{\partial \Phi}  + (-\frac{4}{\sin^2{\theta}}+2)(\eta^{-4}\delta\psi_4)\label{eqn:NHEKmaster}
\end{eqnarray}
where $\eta$ is also defined as \eqref{defeta}.

For a description on how the metric can be reconstructed from $\delta\psi_4$ see \cite{Dias:2009ex} which expands the original results of \cite{1975PhLA...54....5C,1975PhRvD..11.2042C}.

\subsection{Separation of variables}

\paragraph{Kerr equation} The Teukolsky master equation in Kerr \eqref{Master} can be separated as 
\be
\Psi_{(s)}= \frac{1}{\sqrt{2 \pi}} \int_{-\infty}^{\infty} \d \hat{\omega} \sum_{lm} \hat R_{lm\hat{\omega}}(\hat{r})_{} S_{lm\hat \omega}(\theta)e^{im\hat{\phi}}e^{-i\hat{\omega}\hat{t}}. 
\label{eqn:KerrDecomposition}
\ee 
Here, the spin-weighted spheroidal harmonics $S_{lm\hat \omega}(\theta)$ satisfy
\be
	\frac{1}{\sin{\theta}} \frac{\d}{\d \theta}(\sin{\theta} \frac{\d S_{lm\hat \omega}}{\d \theta}) + [ (a \hat{\omega})^2 \cos^2{\theta} - 2a \hat{\omega} s \cos{\theta} -(\frac{m^2 + 2ms \cos{\theta} + s^2}{\sin^2{\theta}}) + \cE_{lm\hat\omega} ] S_{lm\hat \omega} = 0 	
	\label{eqn:AngularTeukolsy}
\ee
where $\cE_{lm\hat\omega}=\cE_{lm\hat\omega}[a,s]$ is the separation constant\footnote{Our $\cE_{lm\hat\omega}$ is $E$ in \cite{1974ApJ...193..443T}. The separation constant $\lambda_{s\ell m \omega}$ defined in \cite{Merlin:2016boc} is also defined in \eqref{lam} in our notation. The separation constant $K$ defined in \cite{Amsel:2009ev} relates to ours as $K = \cE_{lm \hat\omega}+\frac{m^2}{4}$.}. We have $-l \leq m \leq l$ and $l \geq |s|$. We adopt the convention to keep the dependence in the black hole parameters $M,a$ and the spin $s$ implicit. %\footnote{while \cite{teukolsky1973perturbations} on the other hand used $A = \cE_{lm}-s(s+1)$.}
The spheroidal harmonics are normalized according to 
\bea
\int_{-1}^1d\cos\theta (S_{lm\hat \omega}(\theta) )^2=1. \label{convC}
\eea
The radial equation is given by
\be
	\Delta^{-s} \frac{\d}{\d \hat{r}}(\Delta^{s+1}\frac{\d \hat R_{lm \hat{\omega}}}{\d \hat{r}}) - V(\hat{r})\hat R_{lm \hat{\omega}}(\hat{r}) = T_{l m \hat{\omega}}(\hat{r})
	\label{eqn:RadialTeukolsky}
\ee
with potential
\bea
	V(\hat{r})  &=& -\frac{(K_{m \hat \omega})^2-2si(\hat{r}-M)K_{m \hat\omega}}{\Delta}- 4si \hat{\omega} \hat{r}+ \lambda_{\ell m \hat\omega} ,\\
	K_{m \hat\omega}  &\equiv& (\hat{r}^2 + a^2)\hat{\omega} -ma, \\
	\lambda_{\ell m \hat\omega} & \equiv& \cE_{lm\hat\omega} -2am\hat{\omega}+a^2\hat{\omega}^2-s(s+1).\label{lam}
\eea
The source of the radial equation is defined as 
\bea
T_{l m \hat{\omega}}(\hat{r}) = \frac{1}{(2\pi)^{3/2}} \int_{-\infty}^\infty d\hat t \int_{-1}^1d\cos\theta \int_0^{2\pi} d\hat \phi e^{i(\hat\omega \hat t - m \hat\phi)}(4\pi \Sigma)T_{(s)}S_{lm\hat \omega}(\theta). 
\eea
or equivalently one decomposes
\bea
(4\pi \Sigma) T_{(s)} =  \frac{1}{\sqrt{2 \pi}} \int_{-\infty}^{\infty} \d \hat{\omega} \sum_{lm} T_{lm\hat{\omega}}(\hat{r})S_{lm\hat \omega}(\theta)e^{im\hat{\phi}}e^{-i\hat{\omega}\hat t}.
\eea

\paragraph{near-NHEK equation} 

Let us now turn to Teukolsky's equation $s=-2$ in near-NHEK spacetime \eqref{eqn:nearNHEKmaster}. It is separated similarly as 
\bea
(1-i\cos{\theta})^4 \delta\psi_4 =  \frac{1}{\sqrt{2 \pi}} \int_{-\infty}^{\infty} \d \omega \sum_{lm} R_{lm\omega}(r)_{} S_{lm}(\theta)e^{im \phi}e^{-i\omega t}
\label{eqn:nearNHEKDecomposition},\\
-8\pi M^2(1 + \cos^2{\theta}) \eta^{-4} \mathcal T_4 =  \frac{1}{\sqrt{2 \pi}} \int_{-\infty}^{\infty} \d \omega \sum_{lm} T_{lm\omega}(r)_{} S_{lm}(\theta)e^{im \phi}e^{-i\omega t}.
\eea
The $s=2$ equation can be separated similarly but we will omit the explicit formulae.  The extremal spheroidal harmonics $S_{lm}(\theta)$ are now defined from \eqref{eqn:AngularTeukolsy} as\footnote{The extremal spheroidal harmonics for $s=0$ are built in Mathematica as \textsf{SpheroidalPS}. The $s=-2$ case can be solved using the spectral decomposition method, see appendix A of \cite{Hughes:1999bq}. A online package is available on \cite{warb} based on the resummation method outlined in \cite{Shah:2012gu}.}
\bea
S_{lm}(\theta) = S_{lm\hat \omega}(\theta) |_{\hat \omega = \frac{m}{2a}} \label{Slm}
\eea
where the separation constant can now be written as $ \cE_{lm}$ instead of $\cE_{lm \hat\omega}$\footnote{The separation constant $K_\ell$ for  $S_{lm}(\theta)$ defined in \cite{Porfyriadis:2014fja} relates to ours as $K_\ell = \cE_{lm}+\frac{m^2}{4}$. The separation constant $\Lambda_{lm}^{(s)}$ defined in \cite{Dias:2009ex} relates to ours as $\Lambda_{lm}^{(s)} = \cE_{lm}-s(s+1)$. Note that $\cE_{lm} = \frac{7m^2}{4}+h(h-1)$ where $h$ is defined in \eqref{defh}.}, namely 
\be
	\frac{1}{\sin{\theta}} \frac{\d}{\d \theta}(\sin{\theta} \frac{\d S_{lm}}{\d \theta}) + [ \frac{m^2}{4} \cos^2{\theta} - ms \cos{\theta} -(\frac{m^2 + 2ms \cos{\theta} + s^2}{\sin^2{\theta}}) + \cE_{lm} ] S_{lm} = 0.  	
	\label{eqn:extAngularTeukolsy}
\ee
Only these specific harmonics occur because the NHEK spacetime is spanned only by corotating modes ($\hat \omega = \Omega_{ext} m=\frac{m}{2a}$) up to small deviations.  For $m$ small and any $s$ we have $\mathcal E_{lm} = l (l+1) + O(m)$. The equation is invariant under $\theta \mapsto \pi - \theta$ together with either $m \mapsto -m$ or $s \mapsto -s$. We therefore have $\mathcal E_{lm}^{(-s)}= \mathcal E_{lm}^{(s)}= \mathcal E_{l,-m}^{(s)} $. The radial separation function now satisfies the $s=-2$ or $s=2$ case of
\be
	(r(r+2\kappa))^{-s} \frac{\d}{\d r}((r(r+2\kappa))^{s+1}\frac{\d R_{lm\omega}}{\d r}) - V(r)R_{lm\omega}(r) = T_{l m \omega}(r)
	\label{eqn:nearNHEKRadial}
\ee
with potential\footnote{Note that our convention for $V(r)$ differs from \cite{Porfyriadis:2014fja} by an overall minus sign, so that it has the interpretation of a standard potential.} 
\be
V(r)  = -\frac{3}{4}m^2 - s(s+1) + \cE_{lm} - 2ism + \frac{(mr+\kappa n)(\kappa (2si-n)+r(2si-m) )}{r(r+2\kappa)} \label{Vr}
\ee
where
\be
n = m +\frac{\omega}{\kappa}. 
\ee

\paragraph{NHEK equation} 
Let us finally turn to Teukolsky's equation $s=-2$ in NHEK spacetime \eqref{eqn:NHEKmaster}. It is separated similarly as 
\bea
(1-i\cos{\theta})^4\delta \psi_4 &=& \frac{1}{\sqrt{2 \pi}} \int_{-\infty}^{\infty} \d \Omega \sum_{lm} R_{lm\Omega}(R)_{} S_{lm}(\theta)e^{im \Phi}e^{-i\Omega T}
\label{eqn:NHEKDecomposition},\\
-8\pi M^2(1 + \cos^2{\theta}) \eta^{-4} \mathcal T_4 &=&  \frac{1}{\sqrt{2 \pi}} \int_{-\infty}^{\infty} \d \Omega \sum_{lm} T_{lm\Omega}(R)_{} S_{lm}(\theta)e^{im \Phi}e^{-i\Omega T}.\label{decT}
\eea
The spheroidal harmonics are defined as \eqref{Slm}. The radial separation function now satisfies the $s=-2$ case of
\be
R^{-2s} \frac{\d}{\d R}(R^{2s+2}\frac{\d R_{lm \Omega}}{\d R}) - V(R)R_{lm \Omega}(R) = T_{l m \Omega}(R) \label{eqn:NHEKradial}
\ee
with potential\footnote{Our sign convention for $V(R)$ is again the opposite of the one chosen in \cite{Porfyriadis:2014fja}.}
\be
V(R)  = -\frac{7}{4}m^2 + \cE_{lm} - s(s+1) - \frac{2 \Omega (m-is)}{R} - \frac{ \Omega^2}{R^2}.\label{defV}
\ee
Formally, the NHEK equation is obtained from the near-NHEK equation upon substituting $(r,t,\phi) \mapsto (R,T,\Phi)$ and setting $\kappa=0$.

\section{Detailed taxonomy of (near-)NHEK orbits}

\subsection{Timelike NHEK orbits}
\label{formulae_NHEK}
\subsubsection{$\ell = \ell_*$, $E = 0$ Circular$_*$ (ISCO)}

\bea
R&=& R_0, \\
\Phi &=& \Phi_0 - \frac{3}{4} R_0 T\label{ISCOeq}
\eea
where $R_0 > 0$ and $\Phi_0 \in [0,2\pi]$.

\subsubsection{$\ell = \ell_*$, $E > 0$ Plunging$_*$$(E)$}

\bea
T &=&T_0  +\frac{1}{R}\sqrt{1+\frac{2 \ell_* R}{E}} ,\\
\Phi &=& \Phi_0 + \frac{3}{4}\sqrt{1+\frac{2 \ell_* R}{E}}-2\ \text{arctanh} \sqrt{1+\frac{2 \ell_* R}{E}}-i \pi
\eea
where $T_0 \in \mathbb R$ and  $\Phi_0 \in [0,2\pi]$.

\subsubsection{$\ell>\ell_*$, $E>0$ Plunging$(E,\ell)$ or $E<0$ Osculating$(E,\ell)$ }

\bea
T &=& T_0+\frac{F(R)}{2E R},\\
\Phi &=& \Phi_0- \log \frac{2E+2\ell R+F(R)}{R}+\frac{\sqrt{3}\ell}{2 \sqrt{\ell^2-\ell_*^2}}\log(\sqrt{3(\ell^2-\ell_*^2)} F(R) +3R (\ell^2-\ell_*^2) +4E \ell)\nn
\eea
where $F(R)=\sqrt{4E^2+8E \ell R+3(\ell^2-\ell_*^2)R^2}$. 

\subsubsection{$\ell>\ell_*$, $E = 0$ Marginal$(\ell)$}

\bea
T &=& T_0+\frac{2\ell}{\sqrt{3(\ell^2-\ell_*^2)}R},\nn\\
\Phi &=& \Phi_0 + \frac{\sqrt{3}\ell}{2\sqrt{\ell^2-\ell_*^2}}\log R.\label{margl}
\eea
It is marginally plunging or, equivalently, marginally osculating.

\subsection{Timelike near-NHEK orbits}
\label{formulae_nearNHEK}
\subsubsection{$\ell = \ell_*$, $e =0$ Plunging$_*$($e=0$)}

\bea
t &=& t_0-\frac{1}{2\kappa}\log \left( r(r+2\kappa) \right),\\
\phi &=& \phi_0 +\frac{3}{4\kappa} r -\frac{1}{2}\log (1+\frac{2\kappa}{r}).\label{Plungingstare0}
\eea

\subsubsection{$\ell = \ell_*$, $e > 0$ Plunging$_*$($e$)}

\bea
t &=& t_0+ \frac{1}{\kappa}\text{arccosh} \frac{r+\kappa(1+\frac{\kappa \ell_*}{e})}{\sqrt{r(r+2\kappa)}},\\
\phi &=&  \phi_0+\frac{3}{4e}F - \frac{1}{2}\log (1+\frac{2\kappa}{r}) +\log\frac{F+\ell_* \kappa -e}{F+\ell_* \kappa + e}\label{Plungingstare}
\eea
where $F=\sqrt{(e+\ell_* \kappa)^2+ 2 \ell_* e r}$. Note that for $e>0$, the argument of the $\text{arccosh}$ is bigger than $1$ and therefore $t$ is real.

Note that for $e =  \kappa \ell_*$ the orbit reduces to
\bea
t &=& t_0+\frac{1}{2\kappa}\log \frac{\sqrt{1+\frac{r}{2 \kappa}} +1}{\sqrt{1+\frac{r}{2 \kappa}} -1},\\
\phi &=& \phi_0 +\frac{3}{2} \sqrt{1+\frac{r}{2\kappa}} -\frac{1}{2}\log \frac{\sqrt{1+\frac{r}{2 \kappa}} +1}{\sqrt{1+\frac{r}{2 \kappa}} -1}.
\eea

\subsubsection{$\ell > \ell_*$, $e =-\frac{\sqrt{3}\kappa}{2}\sqrt{\ell^2-\ell_*^2}$ Circular($\ell$)}

\bea
r &=& r_0 =\frac{2\kappa \ell}{\sqrt{3(\ell^2-\ell_*^2)}}-\kappa,\\
\phi &=& \phi_0 -\frac{3}{4}(r_0+\kappa)t.\label{circ1}
\eea
Note that $r_0 \geq (\frac{2}{\sqrt{3}}-1)\kappa$.

\subsubsection{$\ell > \ell_*$, Osculating($e,\ell$) or Plunging($e,\ell$)}

\bea
t &=& t_0-\frac{1}{2\kappa} \log (1+\frac{2\kappa}{r}) +\frac{1}{2\kappa} \log \frac{(3\kappa \ell^2-4\kappa M^2-4 e \ell )r+2(\kappa^2 \ell^2 -2 e^2 -4\kappa^2 M^2+(e-\kappa \ell)F)}{(3\kappa \ell^2-4\kappa M^2+4 e \ell )r+4(e+\kappa \ell)^2 - 2 (e+\kappa \ell)F},\nn\\
\phi &=&\phi_0-\frac{1}{2}\log(r(r+2\kappa))+\frac{\sqrt{3}\ell}{2\sqrt{\ell^2-\ell_*^2}}\log(3(\ell^2-\ell^2_*)(r+\kappa)+\sqrt{3(\ell^2-\ell^2_*)}F+4 e \ell) \nn\\
&& +\frac{1}{2}\log ( (7\ell^2-4M^2)r(r+2\kappa)+16 e \ell r+8 (e+\ell \kappa)^2 -4(e+\ell(r+\kappa))F)\label{gen7}
\eea
where $F=\sqrt{3(\ell^2-\ell_*^2)r(r+2\kappa)+8 e \ell r+ 4 (e+\kappa \ell)^2}$. 

For $e < -\frac{\sqrt{3}\kappa}{2}\sqrt{\ell^2-\ell_*^2}$ the orbits are osculating. For $e > -\frac{\sqrt{3}\kappa}{2}\sqrt{\ell^2-\ell_*^2}$ they are plunging. For  $e = -\frac{\sqrt{3}\kappa}{2}\sqrt{\ell^2-\ell_*^2}$ the orbits are circular and given by \eqref{circ1}. These orbits were considered in \cite{Hadar:2016vmk}.

\subsubsection{$\ell > \ell_*$, $e =-\kappa \ell$ Second Circular($\ell$)}

\bea
r &=& r_0 =\frac{2\kappa (\ell^2+4M^2)}{3(\ell^2-\ell_*^2)},\\
\phi &=& \phi_0 -\frac{3\ell^2+4M^2}{3(\ell^2-\ell_*^2)}\kappa t.
\eea
Note that $r_0 \geq \frac{2}{3}\kappa$. 

\subsection{Conformal transformations}
\label{formulae_ct}

First, the inverse transformation of \eqref{glob} defined for $R>0$, $1+R^2(1-T^2)>0$ is given by
\bea
y &=& \frac{R^2(1+T^2)-1}{2R},\nn\\
\tau &=& \arctan \frac{2 R^2 T}{1+R^2(1-T^2)}, \label{globinverse} \\
\varphi &=& \Phi+\log \frac{R^2 + (1+R T)^2}{\sqrt{1+2R^2 (1-T^2)+R^4(1+T^2)^2}}.\nn
\eea
We can then deduce that under a global translation $\tau \rightarrow \bar \tau = \tau + \zeta$, the Poincar\'e coordinates $(R,T,\Phi)$ are transformed to new Poincar\'e coordinates $(\bar R,\bar T,\bar \Phi)$ as
\bea
\bar R &=& \frac{R^2(1+T^2)-1 + (1+R^2 (1-T^2))\cos\zeta - 2 R^2 T \sin\zeta }{2R},\nn\\
\bar T &=& \frac{2R^2 T \cos\zeta + (1+R^2 (1-T^2)) \sin\zeta}{2R}\frac{1}{\bar R},\label{barc}\\
\bar \Phi &=& \Phi+\log \frac{\cos\frac{\zeta}{2} R - \sin\frac{\zeta}{2} (1+RT)}{\cos\frac{\zeta}{2} R - \sin\frac{\zeta}{2} (-1+RT)}.\nn
\eea

%\be
%\frac{R \cos{\frac{\zeta}{2}} + (1+RT)\sin{\frac{\zeta}{2}}}{R}
%\ee

The inverse coordinate transformation of \eqref{NHEKtonearNHEK} defined for $R > 0$ and $RT<-1$ is given by
\bea
 r &=& \kappa (-R T - 1),\nn \\
 t &=& \frac{1}{\kappa}\log \frac{R}{\sqrt{R^2 T^2 -1}},\label{nearNHEKtoNHEK} \\
 \phi &=& \Phi + \frac{1}{2} \log\frac{-RT -1}{-RT + 1}.\nn 
\eea

%\be
%\xi = \sqrt{\frac{r}{R}} 
%\ee

We now define 

\bea
\bar r &=& \kappa (-\bar R \bar T - 1),\nn \\
\bar t &=& \frac{1}{\kappa}\log \frac{\bar R}{\sqrt{\bar R^2 \bar T^2 -1}}, \\
\bar \phi &=& \bar\Phi + \frac{1}{2} \log\frac{-\bar R \bar T -1}{-\bar R \bar T + 1}. \nn
\eea

Substituting \eqref{barc} and \eqref{NHEKtonearNHEK} we obtain the transformation from near-NHEK to near-NHEK coordinates which corresponds to a global time translation $\tau \rightarrow \bar \tau = \tau + \zeta$ as 
\bea
 r &=& (\bar r + \kappa) \cos \zeta -\sqrt{\bar r (\bar r+2 \kappa)} \sin\zeta \sinh \kappa \bar t  - \kappa ,\nn\\
 t &=& \frac{1}{2\kappa}\log \frac{\sqrt{\bar r (\bar r+2 \kappa)} (\cosh \kappa \bar t + \cos\zeta \sinh \kappa \bar t)+\sin\zeta (\bar r+\kappa)}{\sqrt{\bar r (\bar r+2 \kappa)} (\cosh \kappa \bar t - \cos\zeta \sinh \kappa \bar t)-\sin\zeta (\bar r+\kappa)},\label{tautr}\\
\phi &=& \bar \phi + \frac{1}{2}\log\frac{ r}{ r +2\kappa} + \log\frac{e^{\kappa \bar t} \sqrt{\bar r+2\kappa} \cos \frac{\zeta}{2}+\sqrt{\bar r}\sin\frac{\zeta}{2}}{e^{\kappa \bar t} \sqrt{\bar r} \cos \frac{\zeta}{2}+\sqrt{\bar r+2\kappa}\sin\frac{\zeta}{2}}.\nn
\eea
The two near-NHEK patches admit an overlap in the range $-\frac{\pi}{2} < \zeta < \frac{\pi}{2}$. For $\zeta = -\pi/2$, \eqref{tautr} reduces to the change of coordinates of \cite{Hadar:2016vmk} for $\chi = 0$, 
\bea
r &=& \sqrt{ \bar r (  \bar r + 2\kappa)} \sinh \kappa \bar t - \kappa,\\
t &=& \frac{1}{2 \kappa} \log \frac{\sqrt{ \bar r (  \bar r + 2\kappa)} \cosh \kappa \bar  t - ( \bar r+\kappa)}{\sqrt{ \bar r (  \bar r + 2\kappa)} \cosh \kappa   \bar t + (  \bar r+\kappa)},\\
\phi &=&  \bar  \phi + \frac{1}{2}\log \frac{( \bar r+\kappa) \sinh \kappa  \bar t + \kappa \cosh \kappa \bar  t}{( \bar r+\kappa) \sinh \kappa  \bar t - \kappa \cosh \kappa  \bar t}.
\eea

In the following we will enumerate the conformal transformations relating the Circular$_*$ (ISCO) orbit \eqref{ISCOeq} of NHEK to either NHEK or near-NHEK orbits, and  the conformal transformations relating the Circular$(\ell)$ orbit \eqref{circ1} of near-NHEK to either NHEK or near-NHEK orbits. 

\subsubsection{ISCO to NHEK orbits}

\begin{itemize}
\item Circular$_*$ (ISCO) $\Leftrightarrow$ Plunging$_*$$(E)$. 

We denote the initial NHEK coordinates as $(T,R,\Phi)$ and the new NHEK  coordinates as $(\bar T,\bar R,\bar \Phi)$. We can map the $RT \leq -1$ patch to the $\bar R \bar T \geq 1$ patch via (see \cite{Hadar:2015xpa})
\bea
T &=& - \frac{\bar R^2 \bar T}{\bar R^2 \bar T^2-1},\nn\\
R&=&\frac{\bar R^2 \bar T^2-1}{\bar R}, \label{NHEKdiff}\\
\Phi &=&\bar \Phi + \log \frac{\bar R \bar T+1}{\bar R \bar T-1}.\nn
\eea
The Circular$_*$ orbit is then mapped to the Plunging$_*$$(E)$ orbit with parameters
\bea
E &=& \frac{2\ell_*}{R_0},\qquad \ell= \ell_*, \\
\bar T_0 &=& 0,\qquad \bar \Phi_0 = \Phi_0.
\eea
Generic orbits are obtained after the shift 
\bea
\bar T \rightarrow \bar T + \bar T_0.
\eea
The two orbits are therefore related by a real $SL(2,\mathbb R) \times U(1)$ transformation.

\end{itemize}

\subsubsection{ISCO to near-NHEK orbits}
\label{sec:map3}

\begin{itemize}
\item Circular$_*$ (ISCO) $\Leftrightarrow$ Plunging$_*$$(e=0)$. 

These orbits are simply related by the defining change of coordinates between the NHEK patch and the near-NHEK patch \eqref{NHEKtonearNHEK}. 
The parameters of the orbits are related as 
\bea
R_0=\frac{1}{\kappa} e^{\kappa t_0},\qquad \Phi_0=\phi_0-\frac{3}{4}.\label{mapp1}
\eea 
This parametrizes a generic orbit. 

\item Circular$_*$ (ISCO) $\Leftrightarrow$ Plunging$_*$$(e)$.

We can reach the orbit \eqref{Plungingstare} (with barred coordinates) starting from the orbit Plunging$_*$$(e=0)$  (with unbarred coordinates) by applying a shift as $\tau \rightarrow \bar \tau= \tau + \zeta$ followed by a near-NHEK time shift $\bar t \rightarrow \bar t  -  \frac{i \pi}{\kappa}$. 

The diffeomorphism between two near-NHEK patches of same parameter $\kappa$ which is exactly a $\tau \rightarrow \bar \tau= \tau + \zeta$ translation is given by \eqref{tautr}. We then obtain the orbit \eqref{Plungingstare}  (with barred coordinates)  with $0 < \zeta < \pi$ and
\bea
e &=& \kappa^2 \ell_* \sin\zeta e^{-\kappa t_0},\qquad \ell=\ell_*, \\
\bar t_0 &=& \frac{i \pi}{\kappa}+ \frac{1}{\kappa}\log(\frac{\sin\zeta}{1+\cos\zeta}),\qquad \bar\phi_0 = \phi_0 - \frac{3}{4}(\frac{\cot \zeta}{\kappa} e^{\kappa t_0}+1).\label{mapp2}
\eea
It is then necessary to perform an imaginary time shift $\bar t \rightarrow \bar t -  \frac{i \pi}{\kappa}$. All energies can be reached by using simply $\zeta = \pm\frac{\pi}{2}$ (for $\zeta <0$ the energy $e$ is positive after an initial time shift $t_0 \rightarrow t_0 +  \frac{i \pi}{\kappa}$).

\end{itemize}

\subsubsection{near-NHEK circular to NHEK orbits}
\label{conf:3}
\begin{itemize}

\item Circular$(\ell)$ $\Leftrightarrow$Marginal$(\ell)$

We start from the Circular$(\ell)$ orbits. We first apply the inverse coordinate transformation of (\ref{NHEKtonearNHEK}) given in \eqref{nearNHEKtoNHEK}. We then apply the $PT$ flip $T \rightarrow -T $ and $\Phi \rightarrow - \Phi$. We obtain the Marginal$(\ell)$ orbits \eqref{margl} with parameters
\bea
E &=& 0,\qquad \ell>\ell_*, \qquad T_0 = 0,\\
\Phi_0 &=& -\phi_0+\frac{3\log \kappa}{4\kappa}(r_0+\kappa)+\frac{1}{2}\log\frac{r_0}{r_0+2\kappa}-\frac{3(r_0+\kappa)}{8\kappa}\log ( r_0(r_0+2\kappa)).\nn
\eea
We can always shift $T_0$ to be nonzero.

\item Marginal$(\ell)$ $\Leftrightarrow$ Plunging$(E,\ell)$ or Osculating$(E,\ell)$ 

Starting from the Marginal$(\ell)$ orbit \eqref{margl}, we can apply the shift $\tau \rightarrow \bar \tau= \tau + \zeta$ expressed as a NHEK to NHEK diffeomorphism \eqref{barc} or, equivalently,  
\bea
 R &=& \frac{\bar R^2(1+\bar T^2)-1 + (1+\bar R^2 (1-\bar T^2))\cos\zeta +2 \bar R^2 \bar T \sin\zeta }{2\bar R},\nn\\
 T &=& \frac{2\bar R^2 \bar T \cos\zeta - (1+\bar R^2 (1-\bar T^2)) \sin\zeta}{2\bar R}\frac{1}{R}\label{barc2},\\
\Phi &=& \bar \Phi+\log \frac{\cos\frac{\zeta}{2} \bar R +\sin\frac{\zeta}{2} (1+\bar R \bar T)}{\cos\frac{\zeta}{2} \bar R + \sin\frac{\zeta}{2} (-1+\bar R \bar T)}. \nn
\eea
Assuming $\zeta \neq 0$, we obtain the Plunging$(E,\ell)$ or Osculating$(E,\ell)$ orbits in barred coordinates with 
\bea
E &=& \frac{\sqrt{3(\ell^2-\ell_*^2)}}{2}(\sin\zeta+T_0 (\cos\zeta -1)),\qquad  \ell>\ell_*, \nn\\
\bar{T}_0&=& \frac{-\cos\zeta +T_0 \sin\zeta}{\sin\zeta + T_0(\cos\zeta -1)},\label{EPEl}\\ 
\bar{\Phi}_0 &=&\Phi_0+\log(2\ell+\sqrt{3(\ell^2-\ell_*^2)}) - \frac{\sqrt{3}\ell}{2\sqrt{\ell^2-\ell_*^2}} \log( 6 (\ell^2-\ell_*^2) (\cos \frac{\zeta}{2}-T_0 \sin \frac{\zeta}{2})^2 ) .\nn
\eea

\end{itemize}

\subsubsection{near-NHEK circular to near-NHEK orbits}

\begin{itemize}

\item Circular$(\ell)$ $\Leftrightarrow$ Osculating$(e,\ell)$ or Plunging$(e,\ell)$

We consider a $T \rightarrow T - \chi$ translation followed by a $\tau \rightarrow  \tau - \frac{\pi}{2}$ translation. It leads to another near-NHEK patch $(\bar t,\bar r,\bar \phi)$ given by (see also \cite{Hadar:2016vmk})
\bea
r &=& \sqrt{\bar r(\bar r+2\kappa)} (\sinh \kappa \bar t + \chi \cosh \kappa \bar t) - \chi (\bar r + \kappa) - \kappa ,\nn \\
t &=& \frac{1}{\kappa} \log \frac{\sqrt{\bar r(\bar r+2\kappa)} \cosh \kappa \bar t - (\bar r + \kappa)}{\sqrt{r ( r + 2\kappa)}}\label{eq:1}, \\
\phi &=& \bar \phi - \frac{1}{2}\log \frac{\sqrt{\bar r(\bar r+2\kappa)} - (\bar r + \kappa) \cosh \kappa \bar t + \kappa \sinh \kappa \bar t}{\sqrt{\bar r(\bar r+2\kappa)} - (\bar r + \kappa) \cosh \kappa \bar t - \kappa \sinh \kappa \bar t} \frac{ r+2\kappa}{ r}. \nn
\eea
Using this diffeomorphism with $\chi \neq \pm 1$ we can map the near-NHEK circular orbits \eqref{circ1} to the near-NHEK Osculating$(e,\ell)$ or Plunging$(e,\ell)$ orbits \eqref{gen7} with 
\bea
e &=&  \frac{1}{2}\sqrt{3(\ell^2-\ell_*^2)}\kappa \chi,\qquad  \ell>\ell_*, \label{defechi}\\
\bar{t}_0&=&-\frac{1}{2\kappa}\log \frac{1+\chi}{1-\chi},\nn\\ 
\bar{\phi}_0&=&\phi_0 - \frac{1}{2}\log(\ell^2+4M^2) -\frac{\sqrt{3}\ell}{4\sqrt{\ell^2-\ell_*^2}}\log (3(\ell^2-\ell_*^2)\kappa^2 (\ell^2+4M^2)). 
\eea
For $-1 <\chi < 1$, $\bar t_0$ is real, but for $|\chi| > 1$, $\bar t_0$ is complex, and therefore a complex shift of $\bar t$ is required as a final step.

Alternatively, one can use the diffeomorphism (\ref{tautr}) with $\zeta < 0$ to map near-NHEK circular orbits \eqref{circ1} to near-NHEK Osculating$(e,\ell)$ or Plunging$(e,\ell)$ orbits \eqref{gen7} with 
\bea
\frac{e}{\kappa} &=& - \frac{1}{2}\sqrt{3(\ell^2-\ell_*^2)} \cos\zeta,\qquad  \ell>\ell_*,\\
\bar{t}_0&=&0.\nn
%\bar{\phi}_0&=&\phi_0+\frac{\sqrt{3}\pi l i}{2\sqrt{l^2-l_*^2}}-\frac{\sqrt{3}l\log(l^2-l_*^2)}{2\sqrt{l^2-l_*^2}}-(\frac{\sqrt{3}l}{4\sqrt{l^2-l_*^2}}+\frac{1}{2})\log(l^2+4M^2)\nn\\&&+\frac{\sqrt{3}l}{4\sqrt{l^2-l_*^2}}\log\frac{1}{\kappa^2\sin^2\zeta}.
\eea
However, this map is limited to the range $|e| < \frac{1}{2}\sqrt{3(\ell^2-\ell_*^2)}\kappa$. 

\end{itemize}

\section{Emission from the ISCO orbit in NHEK}
\label{app:circ1}

The circular orbit is given in \eqref{ISCOeq} where we set $\Phi_0=0$. The covariant components of the stress-tensor \eqref{Tmunu} are then vanishing except for 
\bea
T_{\Phi\Phi} = \frac{m_0 R_0}{\sqrt{3}M} \delta(R-R_0)\delta(\theta-\frac{\pi}{2})\delta(\Phi-\tilde \Omega T)
\eea
where 
\be
\tilde{\Omega} = - \frac{3 R_0}{4}. \label{deftO}
\ee
From the NHEK limit \eqref{NHEKlimit}, the ISCO is located at $R_0=2^{1/3}$. As a result, only the frequency $\Omega = m \tilde{\Omega}$ will appear in the source \eqref{decT} and therefore also in the curvature perturbation \eqref{eqn:NHEKDecomposition} which then simplify to  

\bea
\delta\psi_4 &=& \frac{1}{(1-i\cos{\theta})^4}  \sum_{lm} R_{lm \tilde{\Omega}}(R) S_{lm}(\theta) e^{im(\Phi-\tilde{\Omega} T)}.
\label{eqn:circDecomposition}\\
4 \pi \mathcal T_4 &=& -\frac{1}{2 M^2(1+\cos^2{\theta})(1-i\cos{\theta})^4} \sum_{l,m} T_{lm\tilde{\Omega}}(R) S_{l m}(\theta) e^{im(\Phi-\tilde{\Omega} T)}. \label{T42}
\eea

The source term $\mathcal T_4$ is a particular combination of differential operators acting on projections of the stress-energy tensor. The general expression is given in  \eqref{eqn:generelTeukolskySource} in Appendix \ref{Teuk}. We find the same result as \cite{Porfyriadis:2014fja}: 
\begin{eqnarray}
	\mathcal T_4 &=& \frac{m_0 R^3_0}{64 \sqrt{3} M^7} \lbrace 144  \delta(R-R_0)\delta(\theta-\frac{\pi}{2})\delta(\Phi-\tilde{\Omega} T) + 16 R_0 \delta'(R-R_0)\delta(\theta-\frac{\pi}{2})\delta(\Phi-\tilde{\Omega} T)\nn\\ 
	&-&48i \delta(R-R_0)\delta'(\theta-\frac{\pi}{2} ) \delta(\Phi-\tilde{\Omega} T) - 21 \delta(R-R_0)\delta(\theta-\frac{\pi}{2})\delta'(\Phi-\tilde{\Omega} T)\nn\\ 
	&-&8i R_0\delta'(R-R_0)\delta'(\theta-\frac{\pi}{2})\delta(\Phi-\tilde{\Omega} T) - 3 R_0 \delta'(R-R_0)\delta(\theta-\frac{\pi}{2})\delta'(\Phi-\tilde{\Omega} T)\nn\\
	 &+&6i \delta(R-R_0)\delta'(\theta-\frac{\pi}{2})\delta'(\Phi-\tilde{\Omega} T) + 2R^2_0 \delta''(R-R_0)\delta(\theta-\frac{\pi}{2})\delta(\Phi-\tilde{\Omega} T) \nn\\
	  &-& 8\delta(R-R_0)\delta''(\theta-\frac{\pi}{2})\delta(\phi-\tilde{\Omega} T)  + \frac{9}{8}\delta(R-R_0)\delta(\theta-\frac{\pi}{2})\delta''(\Phi-\tilde{\Omega} T) \rbrace .
	\label{eqn:sourceNHEK}
\end{eqnarray}
The expression can be written more conveniently as a special case of the general formula
\be
\mathcal T_4 = \frac{m_0 R_0^3}{M^7}\times \hspace{-10pt}\sum_{\footnotesize \begin{array}{l} \left\lbrace i,j,k \in \mathbb{N} | \right. \\ \left. i+j+k \leq 2 \right \rbrace \end{array} } a_{ijk} R^{i}_0 \delta^{(i)}(R-R_0)\delta^{(j)}(\theta-\frac{\pi}{2})\delta^{(k)}(\Phi-\tilde{\Omega} T) 
\label{eqn:T4form}
\ee
where $a_{ijk}$ are pure numbers. Using the convention \eqref{convC}, the radial function $T_{lm\tilde{\Omega}}(R)$ is defined from \eqref{T42} as
\bea
T_{lm\tilde{\Omega}}(R) = -4 M^2\int_0^{2\pi} d\Phi e^{- i m( \Phi-\tilde \Omega T)}\int_{-1}^1 d\cos\theta S_{lm}(\theta) (1+\cos^2\theta)(1-i \cos\theta)^4 \mathcal T_4. 
\eea
For the generic form \eqref{eqn:T4form} we find 

\bea
T_{lm \tilde{\Omega}} &=&- \frac{4  m_0 R_0^3}{ M^5} \lbrace \delta(R-R_0) \left[ \cS(a_{000}+ima_{001}-m^2a_{002}) - \cS'(a_{010}+ima_{011}) + \cS''(a_{020})\right] \nn \\ &+&R_0 \delta'(R-R_0) \left[ \cS(a_{100}+ima_{101}) - \cS'(a_{110})\right] + R^2_0 \delta''(R-R_0) \left[ \cS(a_{200})\right]  \rbrace 
\label{eqn:genseparatedsource}
\eea
where
\bea 
\cS &=& S_{lm}(\pi/2) \\
\cS'&=& S_{lm}'(\pi/2)+4iS_{lm}(\pi/2) \\
\cS''&=& S_{lm}''(\pi/2)+8iS_{lm}'(\pi/2) -11 S_{lm}(\pi/2) 
\eea
and $'$ is the derivative with respect to $\theta$. In the case of \eqref{eqn:sourceNHEK}, we find 
\be
T_{lm \tilde{\Omega}}(R) = a_0 \delta(R-R_0) + a_1 R_0 \delta'(R-R_0) + a_2 R^2_0 \delta''(R-R_0)
\ee
with  
\begin{eqnarray}
a_0 &=&- \frac{m_0 R^3_0}{16 \sqrt{3} M^5}[(40+ 3im - \frac{9}{8}m^2)S_{lm}(\frac{\pi}{2})+(6m-16i)S_{lm}'(\frac{\pi}{2})-8S_{lm}''(\frac{\pi}{2})]\nn \\
a_1 &=& - \frac{m_0 R^3_0}{16 \sqrt{3} M^5}[(-3im-16)S_{lm}(\frac{\pi}{2})+8iS_{lm}'(\frac{\pi}{2})] \\
a_2 &=&- \frac{m_0 R^3_0}{8 \sqrt{3} M^5} S_{lm}(\frac{\pi}{2})\nn
\end{eqnarray}
in agreement with \cite{Porfyriadis:2014fja} up to the global sign which differs\footnote{The global sign difference originates from the sign of the stress-tensor in Teukolsky equations \eqref{Master}. This global sign difference leads to phase shift which does not modify the amplitude or energy fluxes of \cite{Porfyriadis:2014fja}, which have been confirmed independently numerically \cite{Gralla:2015rpa}. }.

Following Green's function methods, the general solution which is ingoing at the horizon and sourced by $\mathcal T_4$ is given by 
\begin{equation}
R_{lm \tilde{\Omega}}(R) = \frac{R_0^{2s}}{W}(\cX \Theta(R_0-R) \cW^{\text{in}}(R) + \cZ \Theta(R-R_0) \cM^{\text{D}}(R)+\cY \cW^{\text{in}}(R) ) + a_2 \delta(R-R_0)
\label{eqn:NHEKsolution}
\end{equation}
Here, $\cY$ is an arbitrary coefficient, and the Wronskian $W$ and $\cX$, $\cZ$ coefficients are given by
\begin{eqnarray}
	W &=& 2im\tilde{\Omega} \frac{\Gamma(2 h)}{\Gamma(h-im-s)}, \label{defW}\\
	\cX &=& R_0 \cM^{\text{D}\prime}(R_0)(2sa_2-a_1-2a_2) + \cM^{\text{D}}(R_0) (a_0-2sa_1-2sa_2 + 4s^2a_2+a_2V(R_0)),\nn\\
	\cZ &=&  R_0 \cW^{\text{in}\prime}(R_0)(2sa_2-a_1-2a_2) + \cW^{\text{in}}(R_0) (a_0-2sa_1-2sa_2 + 4s^2a_2+a_2V(R_0))\nn
\end{eqnarray}
where the potential is defined in \eqref{defV} with $\Omega = m \tilde \Omega$ and $s=-2$ is understood.

If one attaches this solution to the asymptotically flat region, matching with \eqref{Rexo} requires that the coefficient multiplying the homogenous solution be given by $\cY = R_0^{-2s} W A$. One then finds 
\bea
B= \frac{\cZ}{R_0^{4}W}\sqrt{2\pi}\delta(\Omega - m \tilde \Omega).
\eea
where the delta function originates because only one frequency is present. The asymptotic behavior of the solution is then given by \eqref{apsi4} where $B$ is now known.  Note that the factors of $R_0$ cancel out in the final expression. Also note that 
\bea
\delta \psi_4 = O(\lambda^{1/3}).
\eea
The power of $1/3$ originates from the NHEK limit which was fine-tuned to include the ISCO in the NHEK region, see \eqref{NHEKlimit}. 

At $\hat x \rightarrow \infty$, 
\bea
\delta\psi_4 (\hat x \rightarrow \infty) \rightarrow \frac{1}{2} (\ddot h_+-i \ddot h_\times)
\label{eqn:asymptoticmetricperturbations}
\eea
The energy and angular momentum flux in gravitational waves is given in terms of the Landau-Lifschitz pseudo-tensor as
\bea
(\frac{dE}{d\hat t} )^\infty &=& \frac{1}{16 \pi} \int_{0}^{2\pi} d\hat \phi \int_{0}^{\pi} \sin\theta d\theta \hat r^2 \left\{ (\dot h_+)^2 + (\dot h_\times)^2 \right\} ,\\
(\frac{dJ}{d\hat t} )^\infty &=& -\frac{1}{16 \pi} \int_{0}^{2\pi} d\hat \phi \int_{0}^{\pi} \sin\theta d\theta \hat r^2 \left\{ \dot h_+ \frac{\p h_+}{\p \hat\phi} + \dot h_\times \frac{\p h_\times}{\p \hat\phi} \right\} . 
\eea
where $\dot {}$ denotes a $\hat t$ derivative. Since $ \frac{\p h_+}{\p \hat\phi} = -2M \dot h_+$ at leading order in $\lambda \rightarrow 0$, we have $(\frac{dJ}{d\hat t} )^\infty = -2 M (\frac{dE}{d\hat t} )^\infty $.

In the case of a single mode $m$, the energy and angular momentum flux are given by 
\bea
(\frac{dE}{d\hat t} )^\infty =\int_{0}^{2\pi} d\hat \phi \int_{0}^{\pi} \sin\theta d\theta \hat r^2 \frac{\delta \psi_4^* \delta \psi_4}{4\pi \hat \omega^2} = \frac{2 M^8}{ m^2}|B \mathcal K|^2 = O(\lambda^{\frac{4}{3} \text{Re}(h)}).\label{dEdt}
\eea
The energy flux is suppressed by the factor $\lambda^{\frac{4}{3} \text{Re}(h)}$, as already noticed in \cite{Gralla:2015rpa}. The dominant modes are the complex modes with $\text{Re}(h)=\frac{1}{2}$. The formula \eqref{dEdt} exactly matches with Eq. (77) of \cite{Gralla:2015rpa} after using the ISCO value $R_0=2^{1/3}$ where their $x_0=R_0 \eps^{2/3}$ and $\epsilon$ is our $\lambda$. The flux of energy at the horizon is given by \cite{1974ApJ...193..443T}
\bea\label{dEHdt}
(\frac{dE}{d\hat t} )^H =  \frac{128 \hat \omega \hat k (\hat k^2+4\eps^2)(\hat k^2+16 \eps^2)(2M r_+)^5}{|C|^2}|Z_{hole}|^2
\eea
where $\eps = \frac{\lambda}{2M}+o(\lambda)$ can be neglected compared to $\hat k = \hat \omega - \frac{m}{2M}=O(\lambda^{2/3})$. After matching the conventions of \cite{1974ApJ...193..443T} we find 
\bea
Z_{hole} &=& \lambda^{-\frac{4}{3}}(M^2)R_0^{2s} \frac{\cX+\cY}{W} (-2 i \Omega)^{i m +s}, \label{Zhole} \\
|C|^2 &=& (m^2+(h-2)^2)(m^2+(h-1)^2)(m^2+h^2)(m^2+(h+1)^2). 
\eea 
We checked that the formula \eqref{dEHdt} with \eqref{Zhole} exactly matches with Eq. (76) of \cite{Gralla:2015rpa} after using the ISCO value $R_0=2^{1/3}$ and $\frac{R^{2s}_0}{W} \cY = A$.

\section{Relevant integrals and QNM overtone summation}
\label{formulae_integrals}

In this appendix, we provide further details about several integrals which have been used in the main text and about the QNM overtone summations.  \\

\begin{itemize}
	
\item Integrals involving an exponential of an exponential:
\be
\int_{-\infty}^{\infty} \d z e^{-pz} e^{-y e^{-z}} = y^{-p} \Gamma(p)
\label{eqn:integral1}
\ee
This is 3.331 p. 340 of \cite{Gradshteyn} for 
\bea
\text{Re}(p) > 0 ,\qquad \text{Re}(y) > 0 
\eea
By analytic continuation, the first inequality relaxes to the condition that $p$ shouldn't be a non-positive integer.

\item Integrals involving an exponential of a hyperbolic function: 
\bea
\int^{\infty}_{0} &\d z& e^{-p z} (\sinh{\frac{z}{2}})^{-2h}e^{-y \coth{\frac{z}{2}}} =  (\frac{y}{2})^{-h}\Gamma(h+p)W_{-p,-h{\boxed{\footnotesize +}}\frac{1}{2}}(2y). \label{eqn:integral2} 
%\\ &=&  (\frac{y}{2})^{-h-\frac{1}{2}}\frac{\Gamma(h+p)}{2}(W_{-p+\frac{1}{2},-h}(2y)+(-p-h)W_{-p-\frac{1}{2},-h}(2y)) \nn
\eea
The first equality is 3.547 (10) p. 386 of \cite{Gradshteyn} up to a minus sign correction, which is emphasized here in the box.
%\footnote{This minus sign was found by K. Fransen and will be submitted as an erratum of \cite{Gradshteyn}.}.  
%The second identity uses the property of Whittaker functions that \cite[Eq.~13.15.9]{NIST:DLMF}
%\be
%W_{-p,-h+\frac{1}{2}}(2y) = \frac{1}{\sqrt{2y}}(W_{-p+\frac{1}{2},-h}(2y)+(-p-h)W_{-p-\frac{1}{2},-h}(2y))
%\ee
The conditions for \eqref{eqn:integral2} to hold are 
\bea
	\text{Re}{(h+p)} &>& 0 ,\qquad
	\text{Re}{(y)}> 0 .
	\label{eqn:conditionsint2}
\eea

\item Integrals involving arbitrary powers of hyperbolic functions:
\bea
\int^{\infty}_{0} &\d z&  e^{-p z} (\coth{\frac{z}{2}}+\chi)^{-y}(\sinh{z}+\chi(\cosh{z}-1))^{-h}  \label{eqn:integral3} \\	&=& (1+\chi)^{-y-h}2^{h} B(h+p, 1-h+y)  {}_2F_1(h+y, h+p,1+y+p, -\frac{1-\chi}{1+\chi} )  \nn
\eea
if
\bea
\text{Re}(h+p) > 0 , \qquad \text{Re}(1-h+y) > 0,\qquad \chi > -1.
\eea
\underline{Proof:}	
\begin{eqnarray*}
	\int^{\infty}_{0} &\d z&  e^{-p z} (\coth{\frac{z}{2}}+\chi)^{-y}(\sinh{z}+\chi(\cosh{\kappa t}-1))^{-h} \\
	&=& 2^{-h} \int^{\infty}_{0} \d z e^{-pz} (\coth{\frac{z}{2}}+\chi)^{-y-h}( \sinh{\frac{z}{2}})^{-2h} \\
	&=&  2^{h} \int^{\infty}_{0} \d z e^{(-p-h)z} (1+\chi + (1-\chi)e^{-z})^{-y-h}( 1-e^{-z})^{y-h} .
\end{eqnarray*}	
If $\chi > -1$, this can be written as	
\begin{eqnarray*}		
	\int^{\infty}_{0} &\d z&  e^{-p z} (\coth{\frac{z}{2}}+\chi)^{-y}(\sinh{z}+\chi(\cosh{z}-1))^{-h} \\
	&=& (1+\chi)^{-y-h}2^{h} \int^{\infty}_{0} \d z e^{(-p-h)z} (1+\frac{(1-\chi)}{(1+\chi)}e^{-z})^{-y-h}( 1-e^{-z})^{y-h}  \\
	&=& (1+\chi)^{-y-h}2^{h} B(h+p, 1-h+y) \\ &\times& {}_2F_1(h+y, h+p,1+y+p, -\frac{1-\chi}{1+\chi} )  
\end{eqnarray*}
using integral 3.312 (3) page 337 of \cite{Gradshteyn} under the conditions
\bea
\text{Re}(h+p) > 0 ,\qquad \text{Re}(1-h+y) > 0 ,\qquad 
|\text{arg}(1+\frac{1-\chi}{1+\chi})| < \pi 
\eea
where the last condition is satisfied in particular for $\chi > -1$.	

Using the integral expression for a hypergeometric function in two variables $F_2$, 9.184 (2) on page 1030, along with 3.312 (3) page 337 of \cite{Gradshteyn} one can also write
\begin{eqnarray*}
	& & \int^{\infty}_{0} \d z e^{(-p-h)z} (1+\chi + (1-\chi)e^{-z})^{-y-h}( 1-e^{-z})^{y-h}  \\ &=& - \chi(1+y+h)B(h+p,1-h+y)F_2(1+y+h;p+h,1;1+y+p,2;-(1-\chi),-\chi)  \\ &+& B(h+p,1-h+y){}_2F_1(h+y,h+p,1+y+p,-(1-\chi)) 
\end{eqnarray*}
if 
\bea
\text{Re}(h+p)>0,\qquad 
\text{Re}(1+y-h)>0,\qquad 
|\text{arg}(2-\chi)| < \pi 
\eea
such that this should hold for $\chi < -1$.

\item Integrals involving arbitrary powers of a polynomial:

\bea
&&\int^{\infty}_{\tan{\frac{\zeta}{2}}} \d \bar{T} e^{- p \bar{T}} (\frac{\bar{T}\cos{\zeta}-\frac{1}{2}(1-\bar{T}^2)\sin{\zeta}}{(\cos{\frac{\zeta}{2}}+\sin{\frac{\zeta}{2}}\bar{T})^2})^{-y} (\bar{T} \cos{\zeta}-\frac{1}{2}(1-\bar{T}^2)\sin{\zeta})^{-h} \nn \\ &=& (\tan{\frac{\zeta}{2}})^{y} \Gamma(1-h-y)p^{h-1}e^{-\frac{p}{2}(\tan{\frac{\zeta}{2}}-\cot{\frac{\zeta}{2}})}  W_{y,\frac{1}{2}-h}(\frac{2p}{\sin\zeta}) \label{eqn:integral4}
\eea
if $\sin{\zeta} > 0$.

\underline{Proof:}
\bea
&&\int^{\infty}_{\tan{\frac{\zeta}{2}}} \d \bar{T} e^{-p \bar{T}} (\cos{\frac{\zeta}{2}}+\sin{\frac{\zeta}{2}}\bar{T})^{2y} (\bar{T} \cos{\zeta}-\frac{1}{2}(1-\bar{T}^2)\sin{\zeta})^{-h-y} \nn 
\\ &=& (\frac{\sin\zeta}{2})^{-y-h}(\sin\frac{\zeta}{2})^{2y}\int_{\tan\frac{\zeta}{2}}^{\infty}d\bar{T}e^{-p\bar{T}}(\bar{T}-\tan\frac{\zeta}{2})^{-h-y}(\bar{T}+\cot\frac{\zeta}{2})^{-h+y}\nn\\&=&(\frac{\sin\zeta}{2})^{-y-h}(\sin\frac{\zeta}{2})^{2y}e^{-p\tan\frac{\zeta}{2}}\int_{0}^{\infty}dx e^{-p x}x^{-h-y}(x+\frac{2}{\sin\zeta})^{-h+y}\nn\\
&=&(\tan{\frac{\zeta}{2}})^{y} \Gamma(1-h-y)p^{h-1}e^{-\frac{p}{2}(\tan{\frac{\zeta}{2}}-\cot{\frac{\zeta}{2}})}  W_{y,\frac{1}{2}-h}(\frac{2p}{\sin\zeta}).
\eea
In the last step we have used the integral representation of Whittaker function 
\be
W_{k,m}(z)=\frac{e^{-z/2}z^k}{\Gamma(1/2-k+m)}\int_0^{\infty}t^{-k-1/2+m}(1+\frac{t}{z})^{k-1/2+m}e^{-t}dt
\ee
and
%upon setting $\bar T = \frac{2}{\sin\zeta}e^x - \cot \frac{\zeta}{2}$. The last equality follows from 3.331 (4) page 340 of \cite{Gradshteyn} if 
\bea
\text{Re}(p) > 0, \qquad \text{Re}(1-h-y)> 0, \qquad  \sin{\zeta} > 0.
\eea

\item Integrals involving the exponential of an inverse:
\bea
\int^{\infty}_{0} \d \bar{T} e^{- p \bar{T}} e^{\frac{- y }{\bar{T}}} \bar{T}^{-2h}  = 2(\frac{y}{p})^{\frac{1}{2}-h}K_{1-2h}(2 \sqrt{p y})
\label{eqn:integral5}
\eea
which is valid for 
\bea
\text{Re}(y) > 0 ,\qquad 
\text{Re}(p) > 0 .
\eea
From 3.471(9) page 370 \cite{Gradshteyn} for 
\bea
\text{Re}{(\alpha \pm \mu +\frac{3}{2})} > 0,\qquad 
\text{Re}{(s+\frac{q}{2})}>0 ,\qquad 
q > 0.
\eea

%\end{itemize}

%\section{QNM overtone summation}
%\label{formulae_sums}
%In this appendix, the overtone summations are calculated. 

%\begin{itemize}
\item Summation involving a hypergeometric function: 	
	\bea
	 \sum^{\infty}_{N=0} & & \frac{(-1)^Nx^{N}}{N!} \frac{\Gamma(1-c_+)}{\Gamma(1-c_+-N)} {}_2F_1(c_-, -N,1-c_+-N, z) \nn \\
	&=& (1-z x)^{-c_-}(1-x)^{-c_+}	
	\label{eqn:sum3first}
	\eea
\underline{Proof:}	
	\begin{eqnarray*}
		\sum^{\infty}_{N=0} & & \frac{(-1)^Nx^{N}}{N!} \frac{\Gamma(1-c_+)}{\Gamma(1-c_+ -N)} {}_2F_1(c_-, -N,1-c_+ -N, z) \\ &=&
	  \sum^{\infty}_{N=0} \frac{\Gamma(c_+ +N)}{N! \Gamma(c_+)} {}_2F_1(c_-, -N,1-c_+-N, z)x^N \\
		&=& \sum^{\infty}_{N=0} \frac{(c_+)_{N}}{N!} {}_2F_1(c_-, -N,1-c_+-N, z)x^N \\
	&=& \sum^{\infty}_{N=0} \sum^{N}_{k=0} (-1)^{k} \binom{N}{k}\frac{(c_+)_{N}(c_-)_k}{N!(1-c_+-N)_k} z^{k} x^{N}  \\
	&=& \sum^{\infty}_{N=0} \sum^{N}_{k=0}  \frac{(c_-)_k (c_+)_{N-k}}{k!(N-k)!} (zx)^{k} (x)^{N-k}\\
	&=& \sum^{\infty}_{N=0} \sum^{\infty}_{k=0}  \frac{(c_-)_k (c_+)_{N}}{k!N!} (zx)^{k} x^{N}  \\
	&=& (1-zx)^{-c_-}(1-x)^{-c_+}  
	\end{eqnarray*}
		where the last line uses
	\be
	(1+x)^{c_+} = \sum^{\infty}_{N=0} (-c_+)_N \frac{(-x)^{N}}{N!}
	\ee

\item After using the property of the hypergeometric function  \cite[Eq.~15.8.1]{NIST:DLMF}
\be
{}_2F_1(a,b,c;z) = (1-z)^{-a}  {}_2F_1(a,c-b,c; \frac{z}{z-1}) 
\ee
we also have
	\bea
	 \sum^{\infty}_{N=0} & & \frac{(-1)^Nx^{N}}{N!} \frac{\Gamma(1-c_+)}{\Gamma(1-c_+-N)} (1-z)^{-c_-} {}_2F_1(c_-, 1-c_+,1-c_+-N, \frac{z}{z-1}) \nn \\
	&=& (1-z x)^{-c_-}(1-x)^{-c_+}	
	\label{eqn:sum3}
	\eea
		
	\item Involving one Whittaker function:
	\bea
	\sum_{N=0}^{\infty}\frac{(-a)^N}{N!} W_{h+N,h-\frac{1}{2}}(z)=z^h e^{-\frac{z}{2}}(1-a)^{-2h}e^{-\frac{z a}{1-a}}\label{oneW}
	\eea

\end{itemize}

%\bibliography{refs}
\providecommand{\href}[2]{#2}\begingroup\raggedright\endgroup

\end{document}